\documentclass[twocolumn]{aastex631}

\graphicspath{{./}{figures/}}
\usepackage{graphicx}
\usepackage[caption=false]{subfig}

\begin{document}

\title{X-Ray Constraints on the Hot Gaseous Corona of Edge-on Late-type Galaxies in Virgo}


\author[0000-0001-9062-8309]{Meicun Hou}
\affiliation{Kavli Institute for Astronomy and Astrophysics, Peking University, Beijing 100871, China}
\email{houmc@pku.edu.cn}

\author[0000-0002-7875-9733]{Lin He}
\affiliation{School of Astronomy and Space Science, Nanjing University, Nanjing 210023, China}
\affiliation{Key Laboratory of Modern Astronomy and Astrophysics (Nanjing University), Ministry of Education, Nanjing 210023, China}

\author[0009-0009-9972-0756]{Zhensong Hu}
\affiliation{School of Astronomy and Space Science, Nanjing University, Nanjing 210023, China}
\affiliation{Key Laboratory of Modern Astronomy and Astrophysics (Nanjing University), Ministry of Education, Nanjing 210023, China}

\author[0000-0003-0355-6437]{Zhiyuan Li}
\affiliation{School of Astronomy and Space Science, Nanjing University, Nanjing 210023, China}
\affiliation{Key Laboratory of Modern Astronomy and Astrophysics (Nanjing University), Ministry of Education, Nanjing 210023, China}

\author[0000-0003-2206-4243]{Christine Jones}
\affiliation{Smithsonian Astrophysical Observatory, Cambridge, MA 02138, USA}

\author[0000-0002-9478-1682]{William Forman}
\affiliation{Smithsonian Astrophysical Observatory, Cambridge, MA 02138, USA}

\author[0000-0002-3886-1258]{Yuanyuan Su}
\affiliation{Department of Physics and Astronomy, University of Kentucky, Lexington, KY 40506, USA}

\author[0000-0002-6593-8820]{Jing Wang}
\affiliation{Kavli Institute for Astronomy and Astrophysics, Peking University, Beijing 100871, China}

\author[0000-0001-6947-5846]{Luis C. Ho}
\affiliation{Kavli Institute for Astronomy and Astrophysics, Peking University, Beijing 100871, China}
\affiliation{Department of Astronomy, School of Physics, Peking University, Beijing 100871, China}

\begin{abstract}
We present a systematic study of the putative hot gas corona around late-type galaxies (LTGs) residing in the Virgo cluster, 
based on archival {\it Chandra} observations. 
Our sample consists of 21 nearly edge-on galaxies representing a star formation rate (SFR) range of ($0.2-3\rm~M_\odot~yr^{-1}$) a stellar mass ($M_*$) range of $(0.2-10) \times 10^{10}\rm~M_{\odot}$, the majority of which have not been explored with high-sensitivity X-ray observations so far.
Significant extraplanar diffuse X-ray (0.5--2 keV) emission is detected in only three LTGs, which are also the three galaxies with the highest SFR. 
A stacking analysis is performed for the remaining galaxies without individual detection,
dividing the whole sample into two subsets based on SFR, stellar mass, or specific SFR.
Only the high-SFR bin yields a significant detection, which has a value of $L\rm_X \sim3\times10^{38}\rm~erg~s^{-1}$ per galaxy.
The stacked extraplanar X-ray signals of the Virgo LTGs are consistent with the empirical $L\rm_X - SFR$ and $L\rm_X - M_*$ relations found among highly inclined disk galaxies in the field, but appear to be systematically lower than that of a comparison sample of simulated cluster star-formation galaxies identified from the Illustris-TNG100 simulation.  
The apparent paucity of hot gas coronae in the sampled Virgo LTGs might be understood as the net outcome of the long-lasting effect of ram pressure stripping exerted by the hot intra-cluster medium and in-disk star-forming activity acting on shorter timescales.
A better understanding of the roles of environmental effects in regulating the hot gas content of cluster galaxies invites sensitive X-ray observations for a large galaxy sample.
\end{abstract}

\keywords{Virgo Cluster(1772) --- X-ray astronomy(1810) --- Late-type galaxies(907) --- Disk galaxies(391) --- Circumgalactic medium(1879)}

\section{Introduction} \label{sec:intro}

The presence of a hot gas corona (sometimes called hot gas {\it halo}) around present-day disk galaxies, including our own Galaxy, has long been theorized \citep{Spitzer1956,White1978} and is now understood to be the manifestation of the interplay between galactic disks and their large-scale environment, an indispensable physical component of the so-called {\it galactic ecosystem}.  

There are two main origins, external and internal, for this hot galactic corona. 
On the one hand, the formation of baryonic structures, which in the $\Lambda$CDM paradigm follows the gravitational collapse of dark matter, is continuously supplied by accretion of the intergalactic medium (IGM). 
Consequently, a hot gas corona (or hot circumgalactic medium) forms around disk galaxies, especially the massive ones, as the relic of the infallen IGM heated to X-ray-emitting temperatures by accretion shocks and gravitational compression \citep{White1978,White1991,Benson2010}.
On the other hand, once a disk of cooled and condensed gas is in place and begins to form stars, a substantial fraction of energy and mass released by the stars, mediated by stellar winds and supernovae (SNe), may produce a vertical outflow or ``fountain'' of metal-enriched, hot gas back into the halo \citep{Cox1974,Bregman1980,Chevalier1985}, in the meantime entraining and evaporating cool gas originally situated in the disk and/or halo \citep{Strickland2000,LiM2017}. 
A hot corona thus forms as the result of this so-called ``disk-halo interaction'' \citep{Cox2005}, provided successive star formation in the disk.
While the amount of externally accreted hot gas is expected to be a strong function of the host galaxy mass, the amount of the internally supplied hot gas is expected to be heavily dependent on the star formation history of the host galaxy.

To date, most observational searches for hot gas coronae around disk galaxies have followed such expectations, leading to some degree of success with contemporary X-ray telescopes, in particular {\it Chandra} and {\it XMM-Newton}. 
Early ROSAT observations analyzed by \citet{Benson2000} lacked a good sensitivity for the low-intensity, extended X-ray emission and did not yield significant detection of X-ray coronae around three massive spirals, NGC\,2841, NGC\,4594 and NGC\,5529. 
A positive {\it Chandra} detection of extraplanar X-ray emission from a moderately massive, highly-inclined disk galaxy NGC\,5746 was first reported by \citet{Pedersen2006}, but was later found to be mistaken \citep{Rasmussen2009}, which underscores the challenge in unambiguously detecting the tenuous hot corona.  
Based on {\it Chandra} and {\it XMM-Newton} observations, extraplanar X-ray emission out to a vertical height of $\sim20$ kpc from the disk was detected in NGC\,2613 \citep{Li2006} and NGC\,4594 \citep{Li2007}, two of the most massive edge-on spirals within a distance of $\sim$25 Mpc. 
This was followed by the detection of the X-ray coronae in more distant, more massive spirals, starting to place useful constraints on their hot circumgalactic medium and the baryon budget, out to a galactocentric distance of tens of kpc \citep{Anderson2011,Anderson2016,Dai2012,Bogdan2013a,Bogdan2013b,Li2016,LiJT2017}. 
A stacking analysis of six massive spiral further pushes the X-ray detection out to a mean galactocentric distance $\sim$100 kpc \citep{Li2018}. 


In parallel, extraplanar hot gas has also been found in a small sample of moderate-size, late-type galaxies typically with active star formation \citep[e.g.,][]{Strickland2004,Grimes2005,Tullmann2006,Li2008}. 
These studies utilizing {\it Chandra} observations found a positive correlation between the extraplanar X-ray luminosity and the in-disk star formation rate (SFR), providing compelling evidence for SF-driven activity as the dominant powering source for the observed hot coronae, although the very details of this process remain largely uncertain.  
\citet{Li2013a} and follow-up studies \citep{Li2013b,Li2014,Wang2016,Jiang2019}, conducted an in-depth investigation of the X-ray properties of hot coronae and their correlations with host galaxy properties, on the largest sample to date of 53 highly-inclined disk galaxies, which have the advantage of minimal contamination by the typically X-ray-bright disk.
Most recently, the eROSITA all-sky X-ray survey discovered a pair of giant shells, i.e., the so-called eROSITA bubbles, with a physical extent of $\sim15$ kpc above and below the Galactic plane \citep{Predehl2020}, providing direct evidence for extraplanar X-ray emission in our own Galaxy.


These aforementioned X-ray constraints on the hot coronae have been used to confront idealized hydrodynamical simulations of disk galaxies \citep[e.g.,][]{Toft2002,Rasmussen2009,Crain2010,Bogdan2013b,Li2014}, and more recently, large-scale galaxy formation simulations \citep[e.g.,][]{Bogdan2015,Chadayammuri2022,Comparat2022},
which assume various and progressively improved recipes of active galactic nucleus (AGN) feedback and stellar feedback. 
It is generally recognized from such comparison that the total amount and spatial distribution of the extraplanar hot gas in the simulated disk galaxies sensitively depend on the assumed AGN feedback and stellar feedback. 
Indeed, the two influential cosmological simulations, EAGLE \citep{Schaye2015} and Illustris-TNG \citep{Pillepich2018}, predict drastically different amounts of coronal hot gas around normal galaxies \citep{Oppenheimer2020}, presumably owing to the distinct AGN/stellar feedback modes adopted in these simulations (e.g., Illustris-TNG assumes thermal and kinetic modes of feedback, while EAGLE assumes only thermal mode; \citealp{Truong2021}). 
This strongly indicates that X-ray measurements of the hot galactic corona, while proven to be a great observational challenge, continue to offer a fundamental test for the ever-improving galaxy formation models.

A further, perhaps equally important factor, in determining the content of the hot coronae around normal galaxies, is environmental effects.
In particular, in a dense environment such as galaxy groups and clusters, ram pressure stripping (RPS) can effectively remove gas from individual galaxies, at a rate that depends on the galaxy's position and velocity relative to the intra-cluster medium \citep[ICM;][]{Gunn1972}. 
Manifestation of the RPS effect is particularly remarkable in gas-rich disk galaxies, with lopsided gas disks and multi-phase long filaments (also known as RPS tails) reaching out to tens of kpc as the most commonly noted features (see recent review by \citealp{Boselli2022}). 
A hot gas corona, if it exists, should also be susceptible to the RPS effect, which is proven by extensive observations and  numerical simulations \citep[e.g.,][]{Roediger2015,Vijayaraghavan2015}. 
In the meantime, ram pressure may also act to halt galactic outflows and produce pressure-confined hot coronae with temporarily enhanced diffuse X-ray luminosity \citep{Brown2000}. 
Moreover, any cool gas stripped from the galaxy can be heated up to X-ray-emitting temperatures before finally mixing with the hot ICM, which may also boost the X-ray luminosity in certain regions of the halo \citep{Wezgowie2012}.
Ram pressure can also drive mass flows, which can increase the dense gas fraction in the disk regions and thus enhance star-forming activity. The subsequent massive stellar wind and SNe can act to replenish the hot gas corona \citep{Zhu2023}. 

The Virgo cluster, the nearest galaxy cluster and the host of $\sim$1600 known member galaxies \citep{Kim2014}, offers a unique opportunity to study the environmental effects on the hot gas content in galaxies spanning a wide range of global properties, such as morphological type, stellar mass and SFR.
A disturbed X-ray morphology has been seen in a number of Virgo spirals \citep[e.g.,][]{Tschoke2001,Machacek2004,Wezgowie2011,Wezgowie2012,Ehlert2013} and ellipticals \citep[e.g.,][]{Biller2004,Randall2004,Machacek2006,Randall2008,Kraft2011,Kraft2017,Paggi2017,Wood2017,Su2019}, providing strong
evidence for on-going RPS in these galaxies. 
Recently, \citet{Hou2021} conducted a {\it Chandra} survey of the hot gas content in a sizable sample of low- to intermediate-mass Virgo early-type galaxies (ETGs). They revealed a general paucity of hot gas around these galaxies, which might be understood as an efficient gas stripping by the Virgo ICM.  
A logical next-step would be to obtain a census of the putative hot gas coronae around late-type, star-forming galaxies in Virgo, which would have important implications on the dense environment of galaxy clusters in general. 

We are thus motivated to carry out a systematic study of the putative hot gas coronae around Virgo late-type galaxies (LTGs), based on archival {\it Chandra} observations. {\it Chandra}'s superb angular resolution and sensitivity for point source detection is crucial for minimizing the contamination of stellar objects and/or background AGNs to the diffuse X-ray emission. 

The rest of the paper is organized as follows. Section 2 describes our sample selection and data reduction. The analysis of diffuse X-ray emission in the Virgo LTGs are presented in Section 3. The implications of our results, in a close comparison with field LTGs and LTGs generated in cosmological simulations, are discussed in Section 4, followed by a summary in Section 5. Throughout this work, we adopt a uniform distance of 16.5 Mpc (1$''$ corresponds to 80 pc; \citealp{Mei2007}) for all sources and galaxies in Virgo. The line-of-sight depth of Virgo may introduce an uncertainty of $\lesssim 10\%$ in the derived luminosities, which should have little effect on our results and conclusion. 

\section{Data Preparation} \label{sec:data}
\subsection{Sample Selection and Galaxy Properties}\label{subsec:sample}

\begin{figure}\centering
\includegraphics[scale=0.4, angle=0]{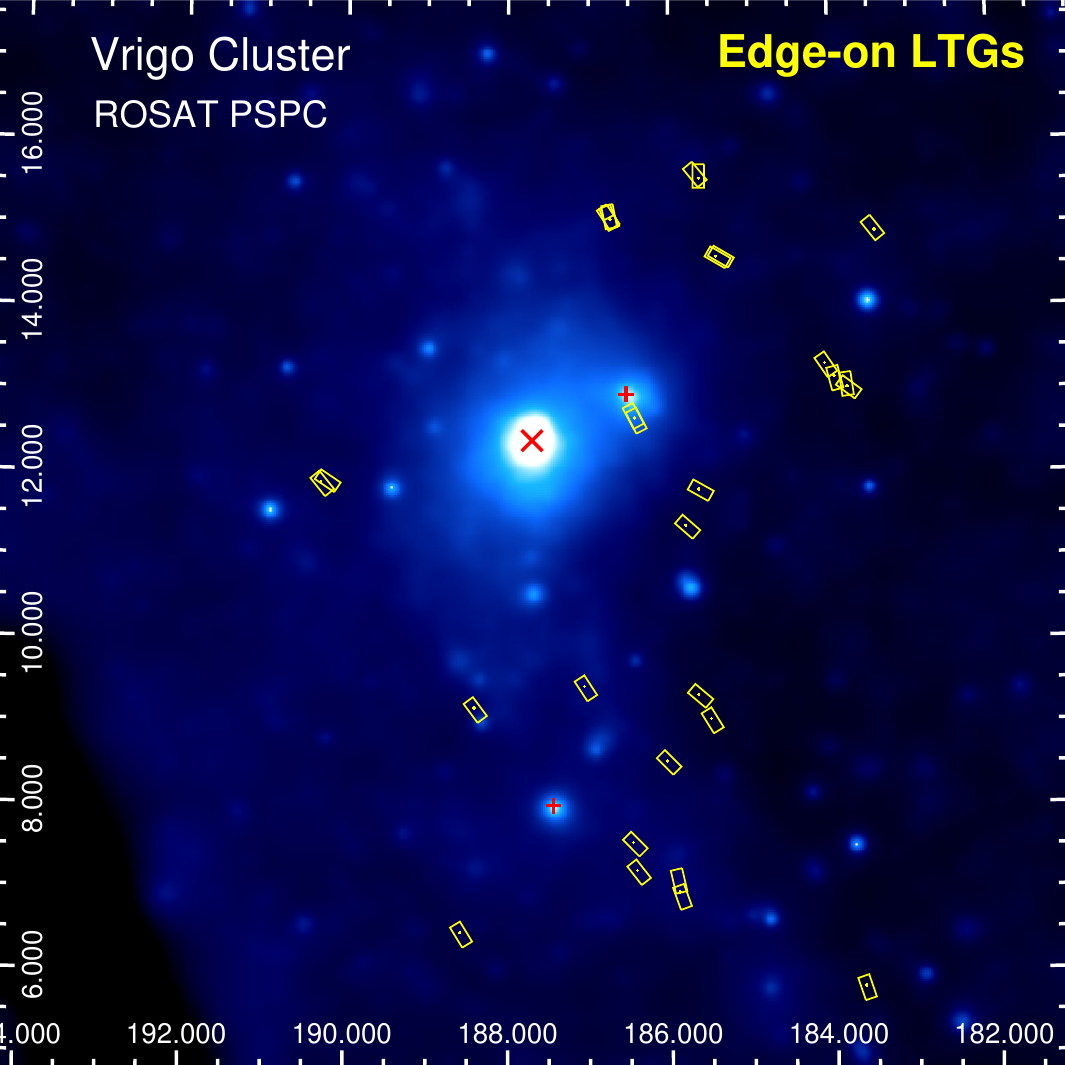}
\includegraphics[scale=0.5, angle=0]{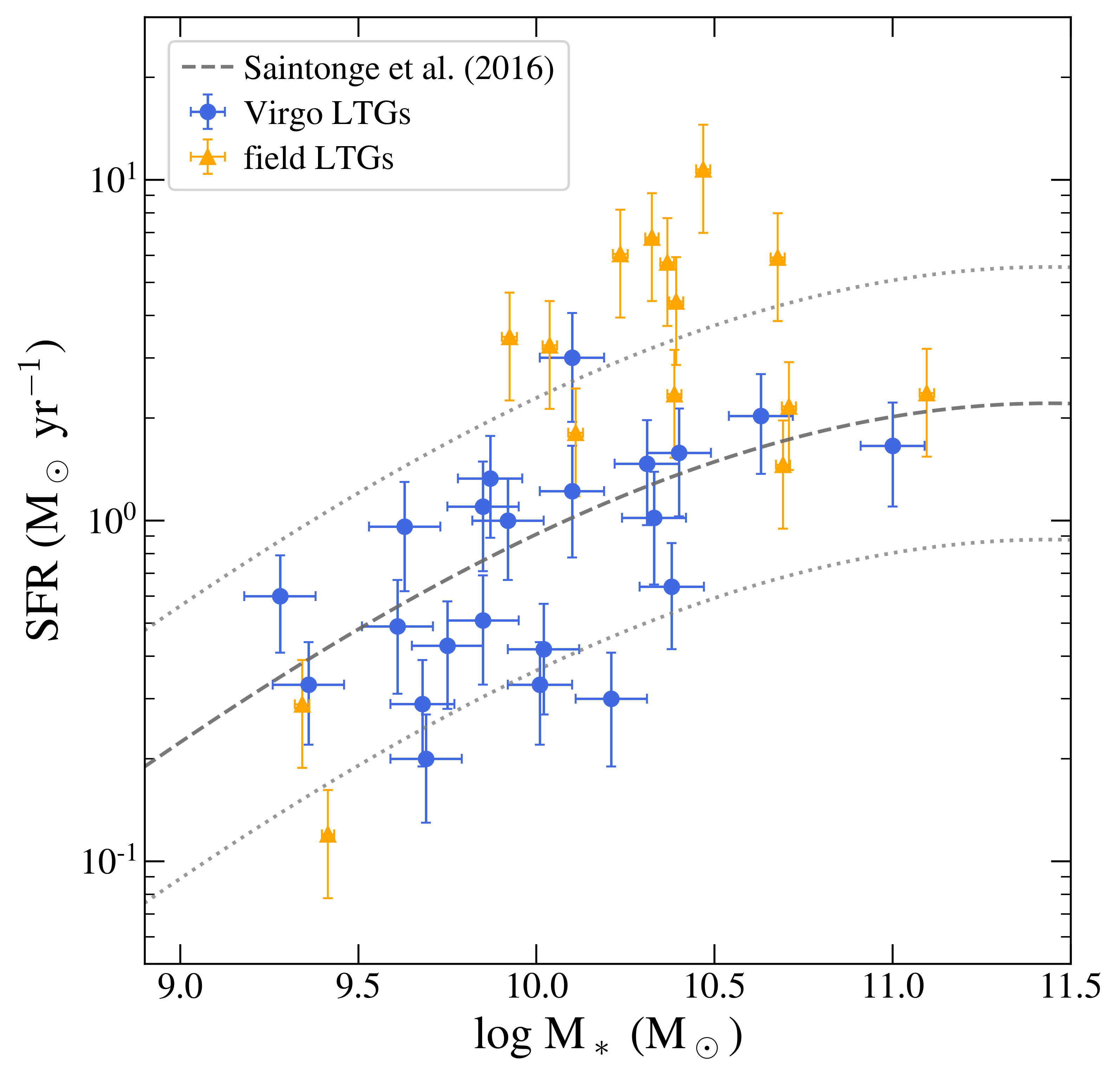}
\caption{{\it Top}: An illustration of the spatial distribution of the 21 edge-on late-type galaxies against the ROSAT all-sky survey X-ray image of the Virgo cluster \citep{Bohringer1994}. The individual {\it Chandra}/ACIS fields are outlined by the yellow boxes, with the center of each galaxy marked by a dot. The center of M87 is marked by a red cross, while the centers of M49 and M86 are marked by a red `+'. 
That most sample galaxies appear to the west of the cluster center is likely a mere coincidence, depending on observers' choices of Chandra targets.
{\it Bottom}: Distribution of the Virgo edge-on late-type galaxies (blue symbols) in the SFR--M$_*$ plane. The dashed and dotted lines indicate the position of the galaxy main sequence and the $\pm 0.4$ dex scatter \citep{Saintonge2016}.
The field late-type galaxies discussed in Section~\ref{subsec:field} are also plotted for comparison in orange triangles.
}
\label{fig:mosaic}
\end{figure} 

We started from a parent sample of 75 Virgo cluster LTGs defined by \citet{Soria2022}, which consists of 52 galaxies with new observations from a {\it Chandra} large program aiming primarily at detecting and quantifying ultra-luminous X-ray sources and nuclear X-ray sources, and an addition of 23 galaxies with sufficiently deep archival {\it Chandra} observations. 
This sample spans a large range in galaxy morphology, including both early-type spirals (i.e., with a substantial bulge), disk-dominated spirals, as well as irregular galaxies. In this work, we focus on a subsample of nearly edge-on disk galaxies and defer a thorough study of the diffuse X-ray emission of the entire sample to a subsequent work.
Specifically, we selected 22 galaxies with a disk inclination angle $\gtrsim 75 \degr$, which corresponds to the major-to-minor axis ratio less than $ 0.32$, according to the \citet{Hubble1926} formula for oblate spheroids with an intrinsic axis ratio of 0.2. The semi-major axis ($a$), semi-minor axis ($b$) of galaxies and the position angle (PA) are adopted from NASA/IPAC Extragalactic Database (NED\footnote{\url{http://ned.ipac.caltech.edu/}}), which is determined from the 2MASS Ks band isophote of 20.0 mag arcsec$^{-2}$ \citep{Jarrett2000}. 
One galaxy, NGC\,4402, was then discarded due to its proximity to the massive elliptical galaxy M86, the extended hot gas corona of which casts substantial contamination. 
This results in a final sample of 21 nearly edge-on Virgo LTGs.
The top panel of Figure~\ref{fig:mosaic} illustrates the locations of the sample galaxies inside Virgo.
Among them, seven galaxies have more than one ACIS observations, for which we incorporate all the available data to maximize the signal-to-noise ratio (S/N) for the putative diffuse emission. The total on-target exposure of these galaxies ranges between 10--48 ks. The remaining galaxies have a single exposure of $\sim$ 8--15 ks. 


The stellar mass and SFR of the sample LTGs are adopted from \citet{Soria2022}, which are calculated based on Wide-field Infrared Survey Explorer \citep[WISE;][]{Wright2010} photometry ($W1- W2$ color and $W1$ band luminosity for stellar mass and $W3$ magnitude for SFR, based on the calibrated relations in \citealp{Cluver2014,Cluver2017}). 
The SFR of our sample ranges from $0.2-3.0~\rm M_\odot~yr^{-1}$ and the stellar mass ranges from $(0.2-10.0) \times 10^{10}~\rm M_\odot$, which occupy a regime of normal SF galaxies with intermediate stellar masses and apparently follow the general trend of the so-called galaxy main sequence \citep{Saintonge2016}, as illustrated by the bottom panel of Figure~\ref{fig:mosaic}.
We stress that the majority of this sample have not been explored with high-sensitivity X-ray observations so far.
The basic information of the 21 Virgo edge-on LTGs are listed in Table~\ref{tab:info}.

\begin{deluxetable*}{lccccccccccc}
\tabletypesize{\footnotesize}
\tablecaption{Basic Information of the Virgo Edge-on Late-type Galaxies
\label{tab:info}}
\tablehead{
\colhead{Galaxy} & 
\colhead{Type} & 
\colhead{R.A.} & 
\colhead{Decl.} & 
\colhead{PA} & 
\colhead{$a$} &
\colhead{$b$} &
\colhead{log $M_{\ast}$} &
\colhead{SFR} & 
\colhead{ObsID} &
\colhead{Exp.} & 
\colhead{$L_{\rm X, corona}$} 
}
\colnumbers
\startdata
NGC\,4192 & SAB(s)ab & 183.451188 & 14.900334 & -30.0 & 313.3 & 75.2 & $10.63 \pm 0.09$ & $2.03 \pm 0.66$ & 19390       & 14.9 & $4.12^{+1.08}_{-1.08}$ \\ 
NGC\,4197 & Sd       & 183.660636 &  5.805730 &  35.0 &  78.2 & 15.6 & $ 9.92 \pm 0.10$ & $1.00 \pm 0.33$ & 19420       &  8.0 & $<$ 2.03 \\ 
NGC\,4206 & SA(s)bc  & 183.820030 & 13.023982 &   2.5 & 170.3 & 34.1 & $10.02 \pm 0.10$ & $0.42 \pm 0.15$ & 16993 19431 & 10.0 & $<$ 1.88 \\ 
NGC\,4216 & SAB(s)b  & 183.976833 & 13.149389 &  19.5 & 317.8 & 57.2 & $11.00 \pm 0.09$ & $1.66 \pm 0.56$ & 19391       &  9.5 & $4.51^{+1.28}_{-1.26}$ \\ 
NGC\,4222 & Sd       & 184.093833 & 13.307056 &  60.0 &  60.4 &  9.7 & $ 9.61 \pm 0.10$ & $0.49 \pm 0.18$ & 19432       &  8.0 & $<$ 1.20 \\ 
NGC\,4302 & Sc       & 185.427000 & 14.598306 & -2.0  & 214.5 & 25.7 & $10.40 \pm 0.09$ & $1.58 \pm 0.55$ & 19392 19397 & 22.0 & $<$ 4.44 \\ 
NGC\,4307 & Sb       & 185.523681 &  9.043633 &  25.0 &  91.7 & 16.5 & $10.38 \pm 0.09$ & $0.64 \pm 0.22$ & 19405       & 14.7 & $<$ 2.42 \\ 
NGC\,4312 & SA(rs)ab & 185.630667 & 15.537917 & -7.5  & 170.3 & 44.3 & $ 9.68 \pm 0.09$ & $0.29 \pm 0.10$ & 7083  19414 &  9.9 & $<$ 3.92 \\ 
NGC\,4313 & SA(rs)ab & 185.660601 & 11.800933 & -40.0 & 127.6 & 33.2 & $10.01 \pm 0.09$ & $0.33 \pm 0.11$ & 19416       &  8.0 & $<$ 4.40 \\ 
NGC\,4316 & Scd      & 185.676000 &  9.332472 & -70.0 &  87.2 & 17.4 & $10.10 \pm 0.09$ & $1.22 \pm 0.44$ & 19400       & 14.9 & $<$ 2.04 \\ 
NGC\,4330 & Scd      & 185.821880 & 11.367970 &  59.0 & 170.3 & 27.2 & $ 9.85 \pm 0.10$ & $0.51 \pm 0.18$ & 19419       &  8.0 & $<$ 3.22 \\ 
NGC\,4343 & SA(rs)b  & 185.911250 &  6.954083 & -50.0 &  77.3 & 23.2 & $10.33 \pm 0.09$ & $1.02 \pm 0.37$ & 7129  4687  & 40.9 & $<$ 1.20 \\ 
NGC\,4356 & Scd      & 186.060542 &  8.535861 &  40.0 &  83.6 & 13.4 & $10.21 \pm 0.10$ & $0.30 \pm 0.11$ & 19435       & 14.9 & $<$ 1.28 \\ 
IC\,3322A & SB(s)cd  & 186.427333 &  7.216667 & -25.0 &  77.4 & 10.8 & $ 9.85 \pm 0.10$ & $1.10 \pm 0.39$ & 19401       & 13.9 & $<$ 1.51 \\ 
NGC\,4388 & SA(s)b   & 186.444780 & 12.662086 &  89.5 & 190.3 & 60.9 & $10.10 \pm 0.09$ & $3.01 \pm 1.06$ & 12291 1619  & 47.6 & $1.79^{+0.28}_{-0.29}$ \\ 
IC\,3322  & SAB(s)cd & 186.475417 &  7.554778 & -25.0 &  87.1 & 20.9 & $ 9.36 \pm 0.10$ & $0.33 \pm 0.11$ & 19424       & 14.5 & $<$ 2.29 \\ 
NGC\,4419 & SB(s)a   & 186.735167 & 15.047385 & -50.0 & 118.0 & 37.8 & $10.31 \pm 0.09$ & $1.47 \pm 0.50$ & 25186 4016  & 35.2 & $<$ 2.20 \\ 
          &          &            &           &       &       &      &                  &                 & 5283  19394 &      &          \\ 
NGC\,4445 & Sab      & 187.066366 &  9.436197 & -75.0 &  64.1 & 11.5 & $ 9.69 \pm 0.10$ & $0.20 \pm 0.07$ & 19433       & 14.9 & $<$ 1.54 \\ 
NGC\,4522 & SB(s)cd  & 188.415462 &  9.175018 &  35.0 &  84.0 & 18.5 & $ 9.75 \pm 0.10$ & $0.43 \pm 0.15$ & 19428       &  7.8 & $<$ 2.07 \\ 
NGC\,4532 & IBm      & 188.580542 &  6.467694 & -20.0 &  88.2 & 26.5 & $ 9.28 \pm 0.10$ & $0.60 \pm 0.19$ & 19407       &  8.0 & $<$ 3.37 \\ 
NGC\,4607 & SBb      & 190.301667 & 11.886639 &   5.0 &  81.2 & 13.0 & $ 9.63 \pm 0.10$ & $0.96 \pm 0.34$ & 19403 19437 & 15.4 & $<$ 1.50 \\ 
\enddata
\tablecomments{ (1) Name of Virgo galaxies; (2) morphological galaxy type, adopted from NED; (3)-(4): celestial coordinates of the galactic center (J2000); (5) position angle, in units of degree; (6)-(7): semi major-axis ($a$) and semi minor-axis ($b$), in units of arcsecond; (8) logarithmic stellar mass of host galaxies, in units of $\rm M_\odot$, adopted from \citet{Soria2022} and derived from WISE $W1$ and $W2$ fluxes as calibrated by \citet{Cluver2014}; (9) SFR in units of $\rm M_\odot~yr^{-1}$, adopted from \citet{Soria2022} and derived from WISE $W3$ flux as calibrated by \citet{Cluver2017}; (10) {\it Chandra} observation ID. For galaxies with multiple observations, all observation ID are listed; (11) {\it Chandra} effective exposure, in units of ks; (12) 0.5--2 keV luminosity of detected diffuse X-ray hot gas corona emission and $3\sigma$ upper limit of non-detections, in units of ${10^{39}\rm~erg~s^{-1}}$.
}
\end{deluxetable*}

\subsection{X-ray Data Reduction}\label{subsec:X-ray data}
We downloaded and reprocessed the archival data using CIAO v4.15 and the calibration files CALDB v4.10.4, following the standard procedure\footnote{\url{http://cxc.harvard.edu/ciao/}} for all selected ACIS fields. 
The aimpoint of all relevant observations are placed on the S3 CCD, and we only included data from the S3 and S2 CCDs for better spatial resolution and sensitivity.
We examined the light curve of each observation 
and filtered the time intervals contaminated by an abnormally high particle background. 
The cleaned exposure of each galaxy is listed in Table~\ref{tab:info}.
For each observation, we produced counts and exposure maps on the native pixel scale ($0 \farcs 492 {\rm~pixel^{-1}}$) in the 0.5--2, 2--8 and 0.5--8 keV bands. 
The exposure maps were weighted by a fiducial incident spectrum, an absorbed power-law with a photon-index of 1.7 and an absorption column density $N_{\rm H}$ = $10^{21} {\rm~cm^{-2}}$, which is typical for X-ray binaries and also suitable for the ICM and cosmic X-ray background sources across the field-of-view (FoV) with a relatively hard intrinsic spectrum.
Adopting a softer incident spectrum characteristic of the coronal hot gas as the weight does not affect our main conclusions described in Section~\ref{sec:discussion}.
We then produced point-spread function (PSF) maps, at a given enclosed count fraction (ECF; 50\% and 90\%), using the same spectral weighting as for the exposure maps. 
We also generated the corresponding instrumental background maps from the ``stowed background'' data, normalizing with the 10--12 keV count rate.
For galaxies with multiple observations, we calibrated their relative astrometry by matching the centroid of commonly detected point sources, using the CIAO tool {\it reproject\_aspect}. 
We then reprojected the count and exposure maps of individual observations to a common tangential point, i.e., the optical center of the target galaxy, to produce combined images.

Following the procedures detailed in \citet{Hou2017} and \citet{Jin2019}, we performed source detection across the FoV of each galaxy in each of the three energy bands. 
Specifically, we used CIAO tool {\it wavdetect}, supplying the exposure map and 50\%-ECF PSF map and adopting a false detection probability of  $10^{-6}$. 
Point sources detected in any of the three energy bands were considered.
A total of 548 independent point-like sources are detected in and around the 21 Virgo edge-on LTGs, at a typical limiting 0.5--8 keV luminosity (assuming that the source is located at the distance of Virgo) of $\sim 10^{38}\rm~erg~s^{-1}$, which is similar to that achieved for the Virgo ETGs studied by \citet{Hou2021}.
These sources are mainly composed of AGNs and active galaxies in the cosmic background, AGNs and XRBs associated with the individual galaxies, as well as possible intra-cluster X-ray sources (Z. Hu et al. in preparation).
In particular, when adopting a matching radius of 2$\arcsec$, an X-ray source is found at the nucleus of 8 edge-on LTGs, which might be tracing a low-luminosity AGN.

To study the diffuse X-ray emission, we removed pixels falling within two times the 90\% ECR around each detected point source.
These pixels were consistently removed from the counts, exposure, and instrumental background maps, and no artificial counts were introduced to fill these pixels in the following quantitative analysis.

\section{Analysis and results} \label{sec:analysis}
\subsection{Detection of Diffuse X-ray Emission} \label{subsec:diffuse}

\begin{figure*}\centering
\includegraphics[scale=0.34, angle=0]{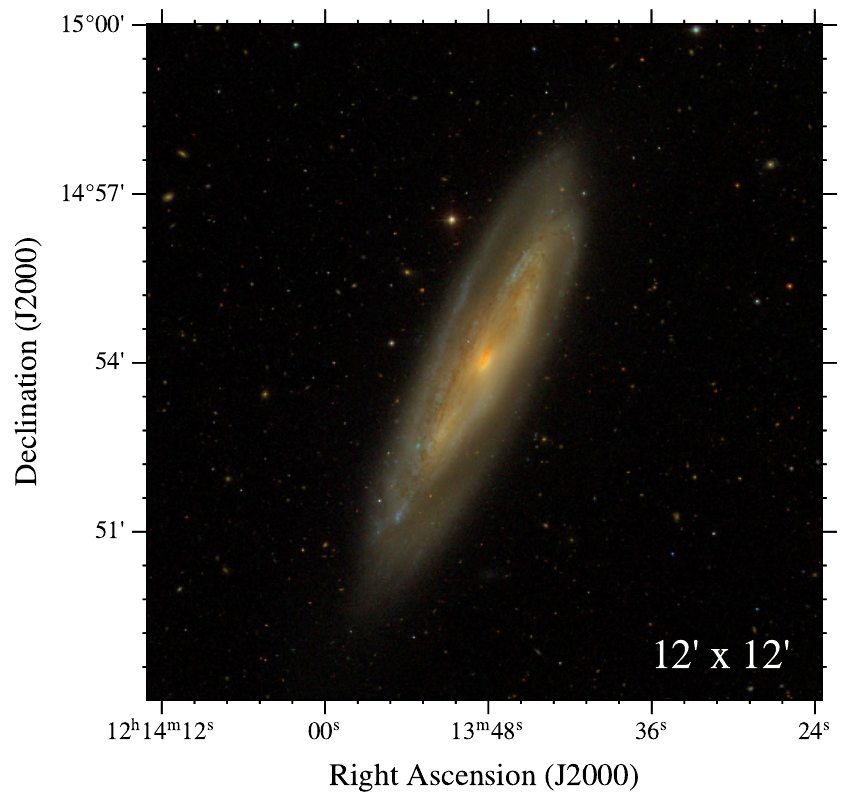}
\includegraphics[scale=0.35, angle=0]{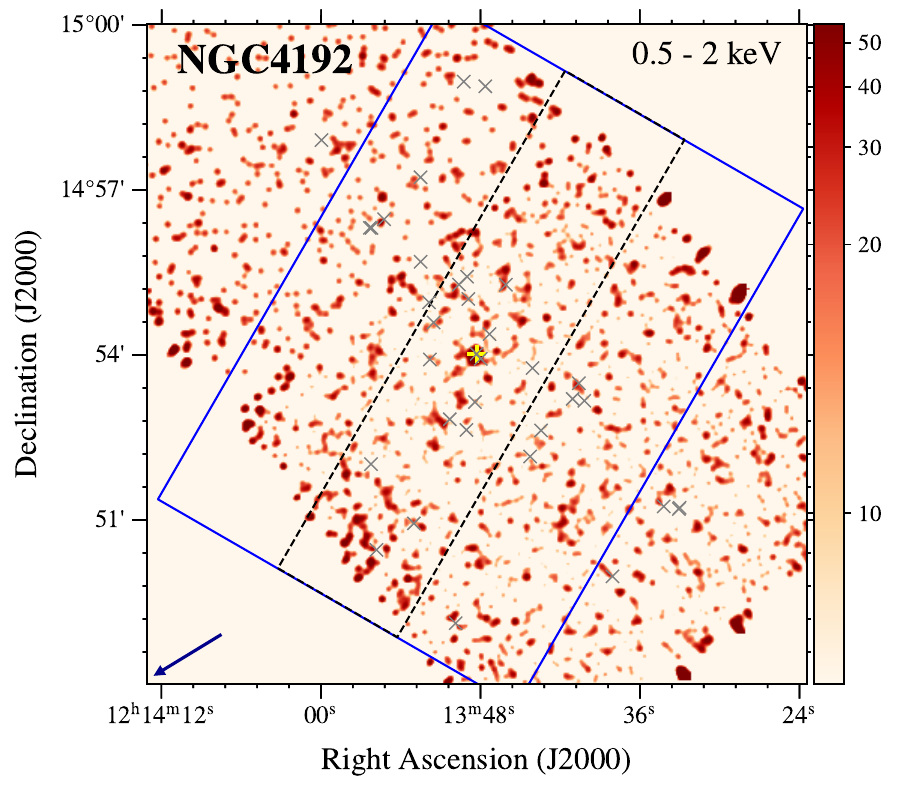}
\includegraphics[scale=0.3, angle=0]{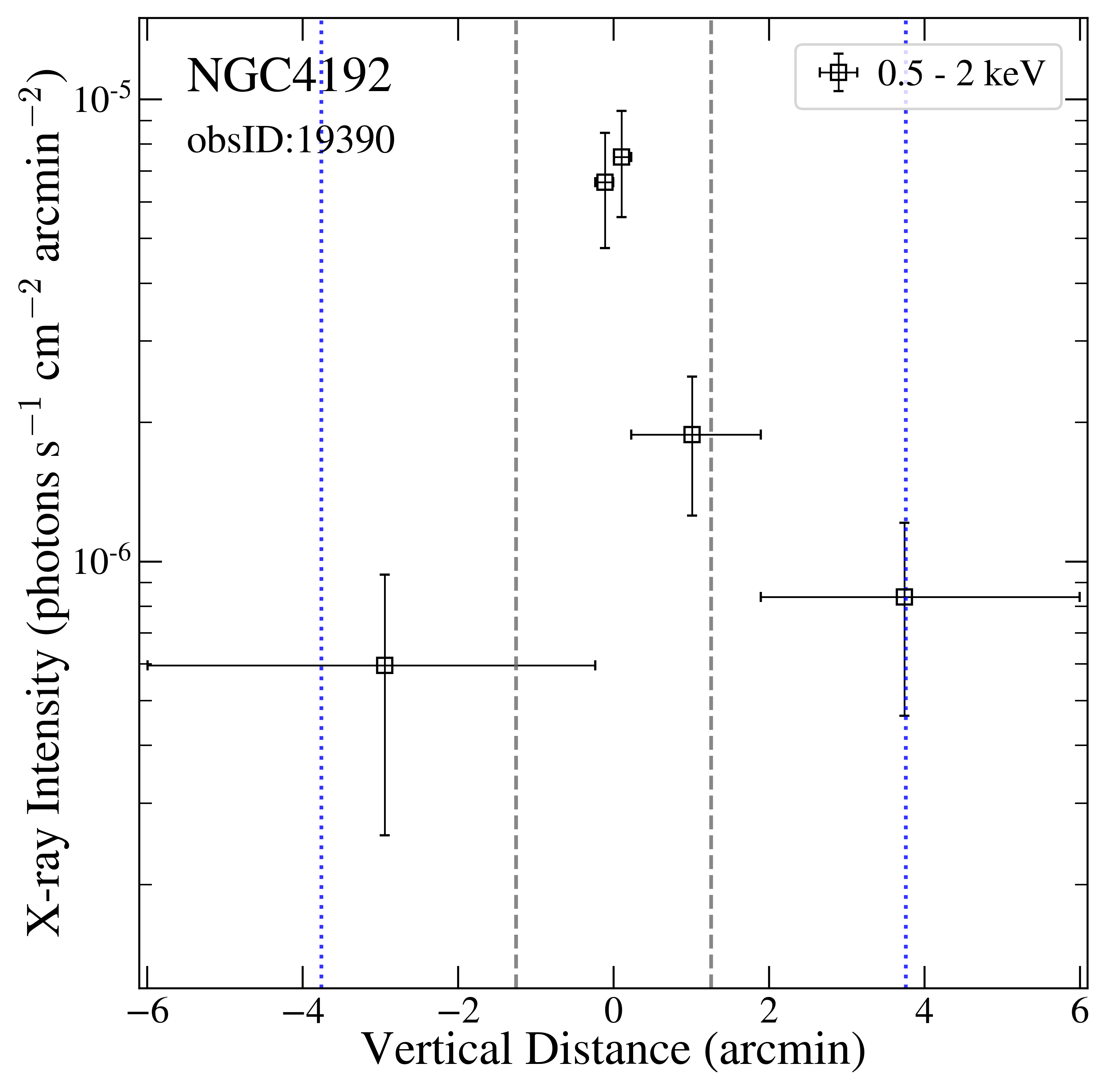} 
\\
\includegraphics[scale=0.34, angle=0]{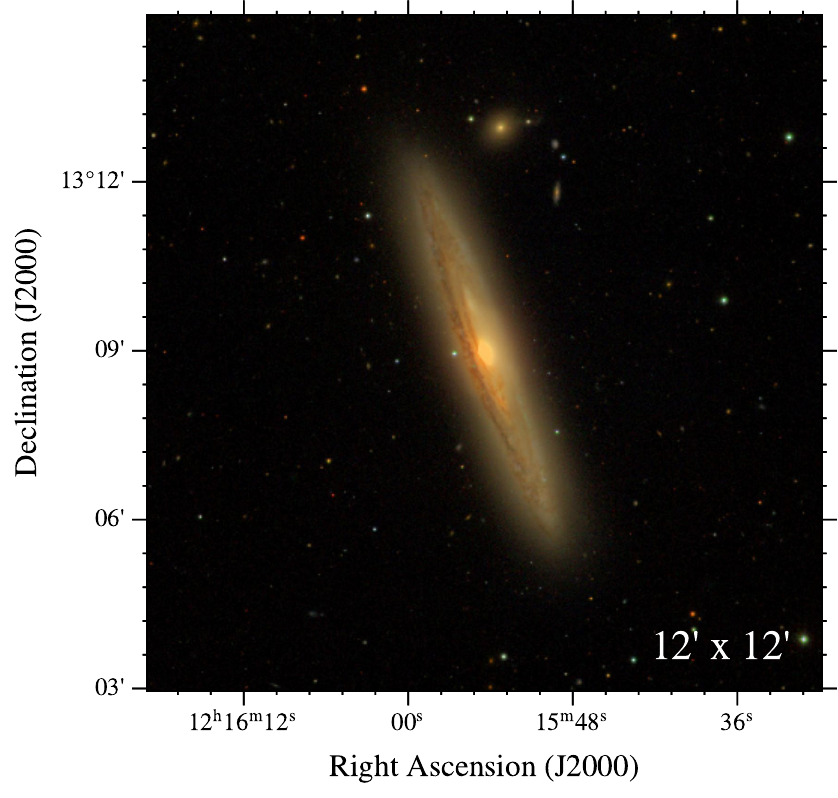}
\includegraphics[scale=0.35, angle=0]{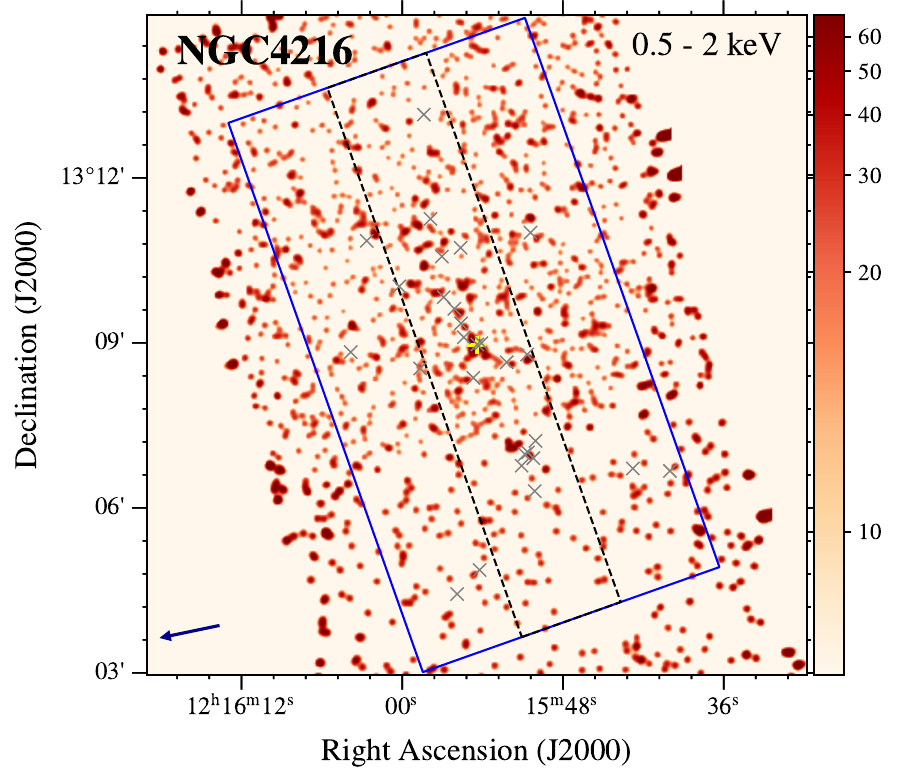}
\includegraphics[scale=0.3, angle=0]{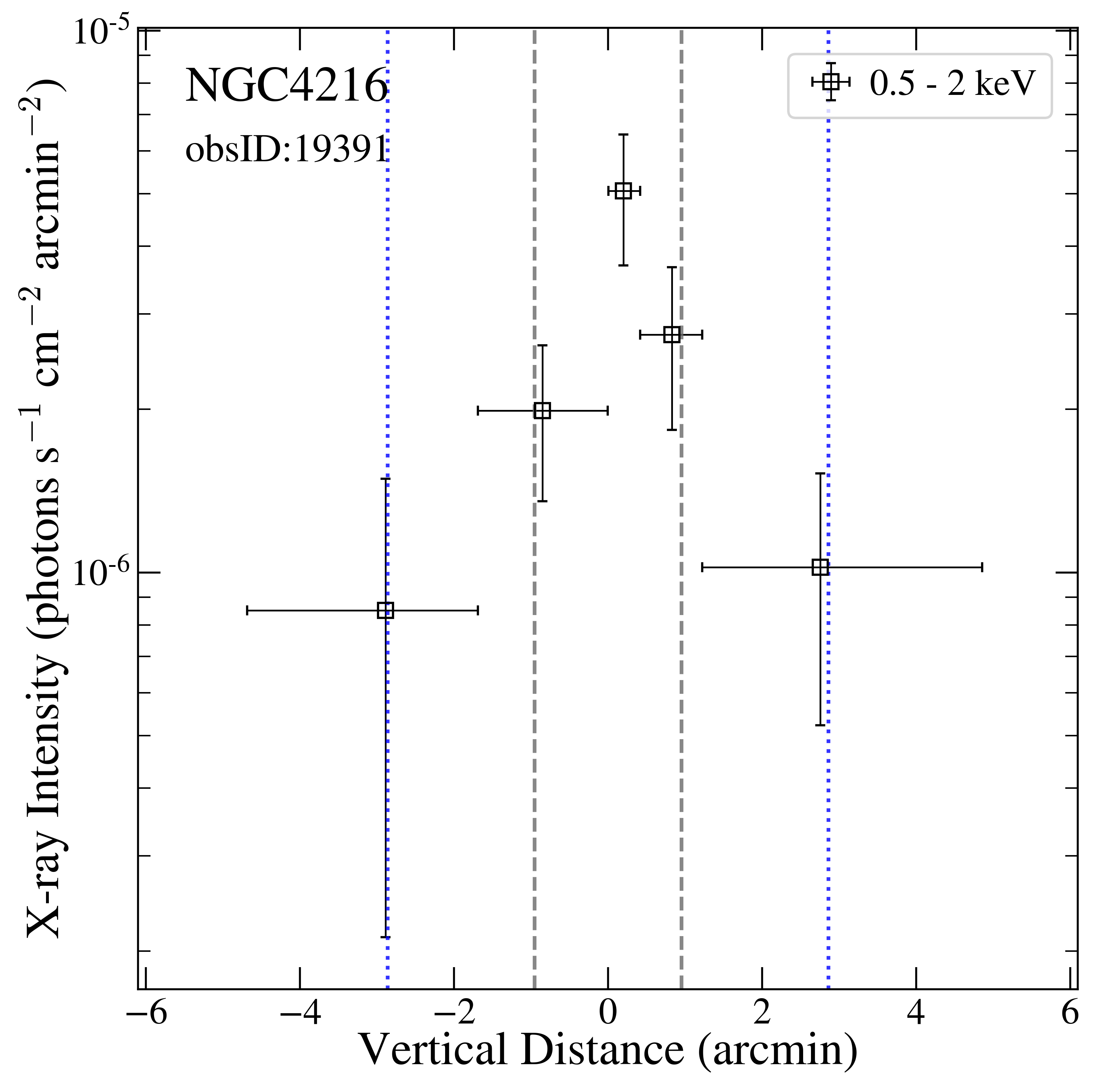} 
\\
\includegraphics[scale=0.34, angle=0]{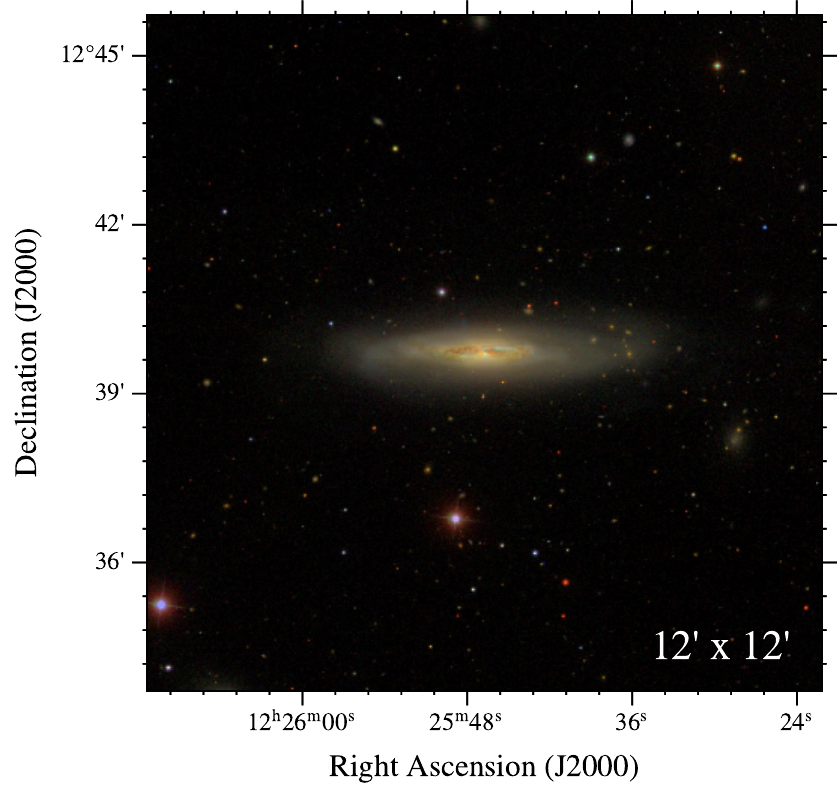}
\includegraphics[scale=0.35, angle=0]{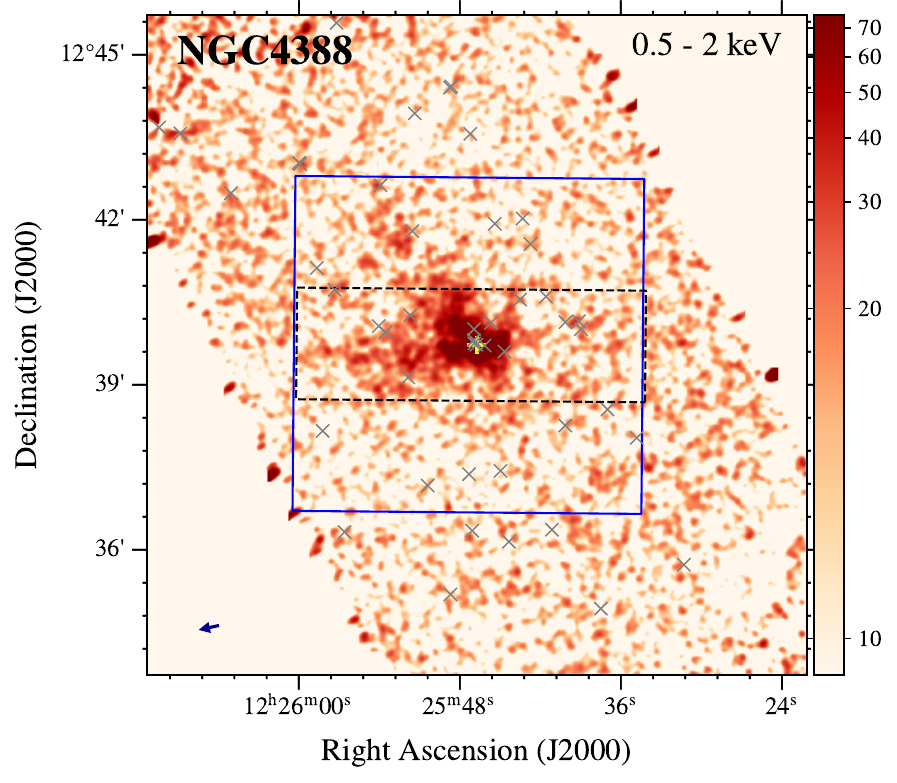}
\includegraphics[scale=0.3, angle=0]{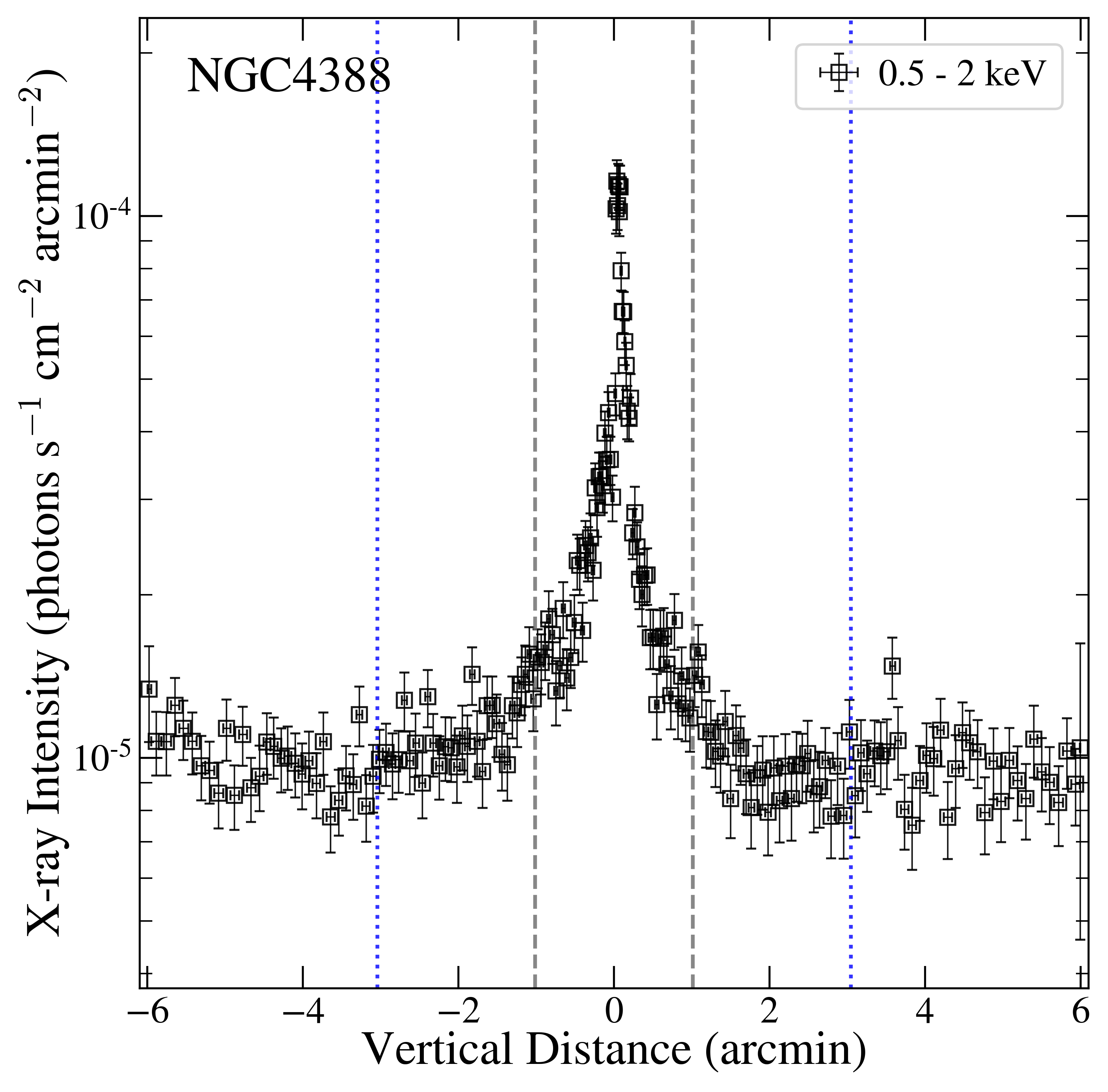}
\caption{{\it Left}: The SDSS $gri$-color composite image in size of $12' \times 12'$. {\it Middle}: X-ray flux map in 0.5--2 keV band, corrected with exposure and particle background and smoothed by a Gaussian kernel of 6 pixels, in unit of $10^{-6}\rm~ph~cm^{-2}~s^{-1}$. The solid blue rectangle represents the X-ray corona region and the dashed black rectangle marks the subtracted galactic disk region. The yellow plus marks the position of the galaxy center and the gray crosses mark the positions of detected X-ray point sources. The dark blue arrow in the lower left corner shows the direction toward the center of the Virgo cluster (i.e. center of M87) and the length of the arrow represents the relative distance to the cluster center. {\it Right}: X-ray intensity profile along the minor axis of each galaxy in the 0.5--2 keV band. The positive direction is the side closer to M87, as indicated in the middle panel. The pair of dashed black lines and dotted blue lines mark the position of 1 and 3 times the galaxy's semi-minor axis. NGC\,4192 and NGC\,4216 have single observation with the obsID labeled, while NGC\,4388 have multiple observations.
}
\label{fig:image}
\end{figure*}

We first examine the flux maps of the 21 Virgo edge-on LTGs in the 0.5--2 keV band, which are produced by subtracting the corresponding instrumental background maps and dividing by the exposure maps. Due to low surface brightness nature of most galaxies, these maps are smoothed by a Gaussian kernel of 6 pixels for a better visualization. 
From the flux maps, extended features along the galactic disk and/or  extraplanar are visible around a few galaxies, while the rest, majority of galaxies show no obvious diffuse emission.

Since we are mainly interested in the hot gas corona, we uniformly quantify the S/N of the 0.5--2 keV extraplanar X-ray emission for each galaxy, as follows.
The hot gas corona region is defined as a rectangle in the size of $2a \times 6b$, where $a$ is the semi-major axis of the host galaxy and $b$ is the semi-minor axis, with the galactic disk region (a rectangle in size of $2a \times 2b$) and detected X-ray point sources (pixels within two times of the 90\% ECR) masked (as shown in Figure \ref{fig:image}).
The corresponding local background region is defined as a larger rectangle with the disk and corona region excluded. 

Next, we employ the CIAO tool {\it aprate}, which applies the Bayesian approach for Poisson statistics at the low-count regime, to calculate the net photon flux and bounds for each given galaxy.
In this calculation the local background and effective exposure 
are taken into account.
A galaxy is regarded as a firm detection of diffuse emission if the 3$\sigma$ lower limit of its net photon flux given by {\it aprate} is greater than zero.
We then assume a 0.3 keV, solar-abundance thermal plasma model \citep[apec;][]{Smith2001} subject to the Galactic foreground absorption to convert the observed photon flux into the unabsorbed 0.5--2 keV luminosity. The conversion factor is $1.6\times 10^{41}(D/\rm Mpc)^2\rm~erg~s^{-1}/(ph~cm^{-2}~s^{-1})$, only weakly dependent on the assumed gas temperature.
We have tested and found that even the assumed gas temperature is increased to 1.0 keV, the conversion factor would only increase by a factor of 1.2, which does not affect our conclusions.
The X-ray luminosity and the 1$\sigma$ errors are reported for galaxies with detection of diffuse X-ray emission (as listed in Table~\ref{tab:info}). 
For those galaxies regarded as non-detections, we report the 3$\sigma$ upper limit of the extraplanar X-ray luminosity. 

Somewhat surprisingly,
only three galaxies (NGC\,4192, NGC\,4216 and NGC\,4388) are found to have S/N $> 3$, indicating significant extraplanar X-ray emission.
To our knowledge, this is the first report of significant extended X-ray emission from 
NGC\,4192 and NGC\,4216, although the S/N is only moderate ($\sim 4$). Diffuse X-ray emission in and around NGC\,4388 has been extensively studied by ROSAT \citep{Matt1994,Colbert1998}, Chandra \citep{Iwasawa2003,Yi2021} and XMM-Newon \citep{Wezgowie2011} observations.
Unfortunately, the limited S/N of these three galaxies does not permit a meaningful X-ray spectral analysis for the extraplanar hot gas, for which we also adopt the above photon flux-to-luminosity conversion.
In the case of NGC\,4192, whose corona region slightly exceeds the ACIS FoV, we have assumed a spatially uniform corona surface brightness to correct for the derived X-ray hot gas corona luminosity.

The optical image, the X-ray flux map, and the X-ray intensity profile along the minor axis of these three galaxies are shown in Figure \ref{fig:image}.
The images of the remaining galaxies without individual detection are given in the Appendix.

\subsection{Stacking Non-detections} \label{subsec:stack}

The majority of our sample of edge-on LTGs do not have detectable diffuse X-ray emission based on the X-ray photometric results in Section~\ref{subsec:diffuse} . 
Thus, we perform a stacking analysis for the undetected edge-on LTGs. 
We first remove NGC\,4302 and NGC\,4343 from stacking because they each have a neighboring massive galaxy, which might introduce undesired contamination. 
This results in a total of 16 galaxies for the stacking analysis.
We use all available observations for the stacking to maximize the S/N of the putative diffuse emission. Only NGC\,4419 has a relative long exposure compared to the average, but our test confirms that the inclusion of all four observations of NGC\,4419 does not dominate the stacking signal.

\begin{figure*}\centering
\includegraphics[scale=0.4, angle=0]{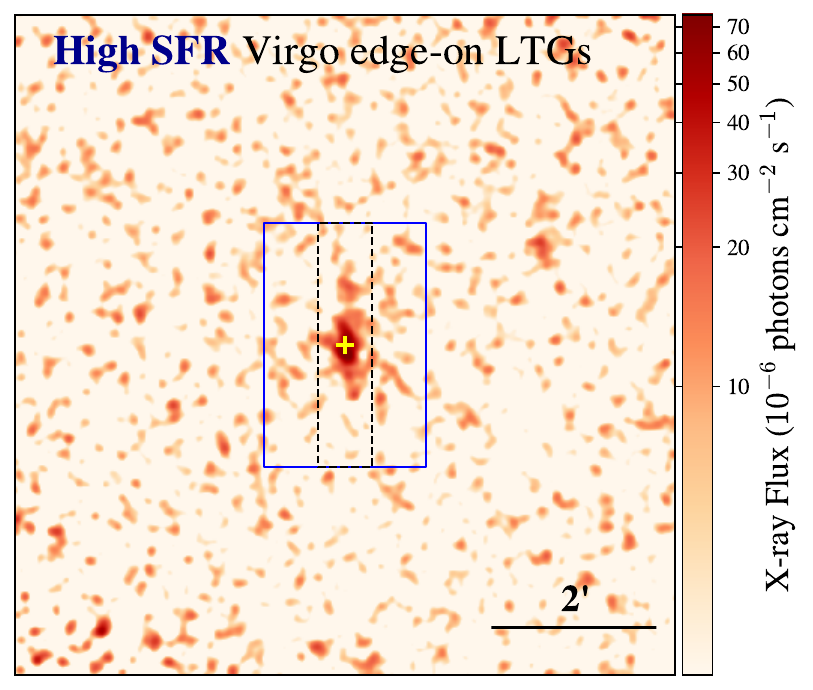}
\includegraphics[scale=0.4, angle=0]{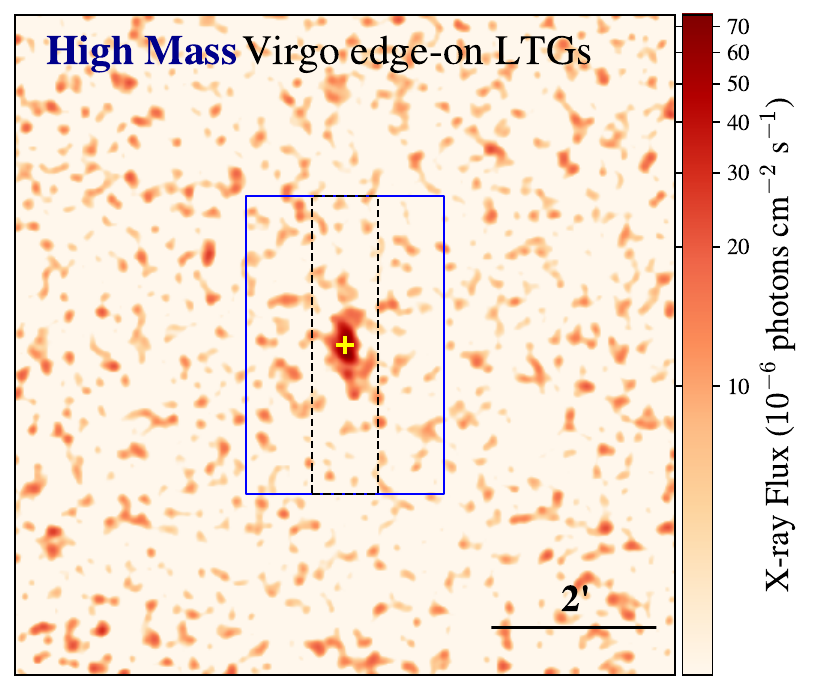}
\includegraphics[scale=0.4, angle=0]{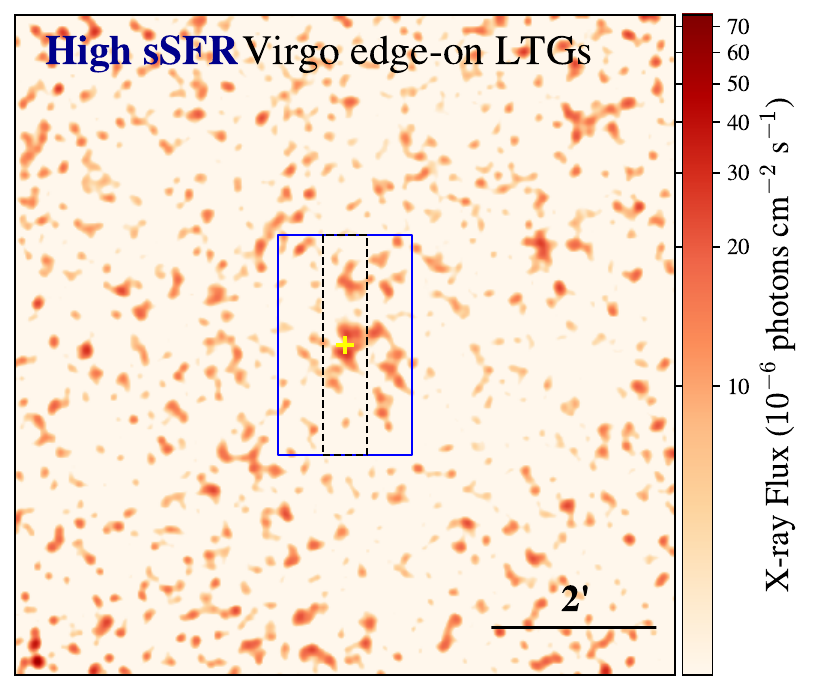}
\\
\includegraphics[scale=0.4, angle=0]{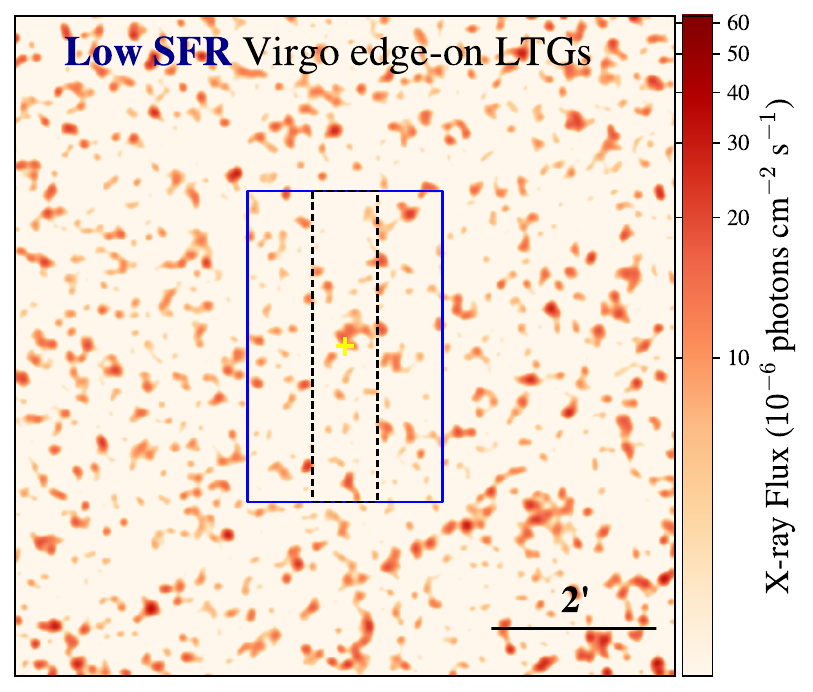}
\includegraphics[scale=0.4, angle=0]{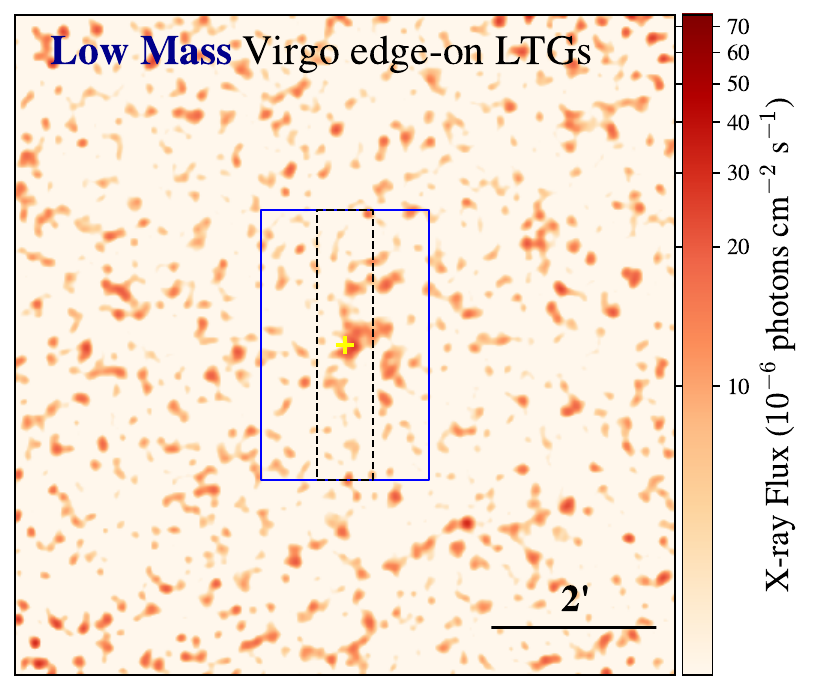}
\includegraphics[scale=0.4, angle=0]{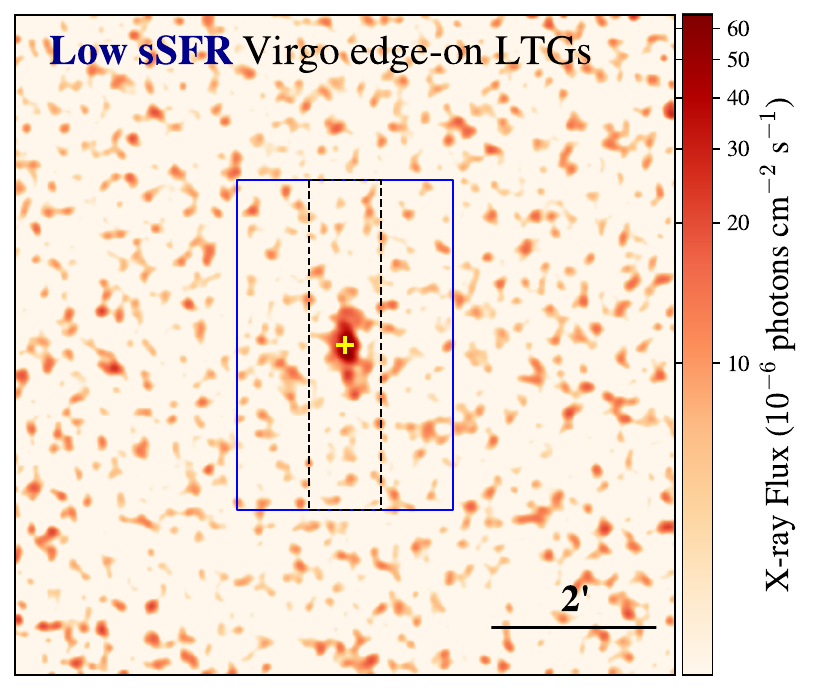}
\\
\caption{The stacked 0.5--2 keV flux map of individually undetected Virgo edge-on LTGs, divided into two bins based on SFR ({\it left panels}), stellar mass ({\it middle panels}), and specific SFR ({\it right panels}). 
The maps have a size of $8' \times 8'$ and units of $10^{-6}\rm~ph~cm^{-2}~s^{-1}$. 
The flux maps are smoothed with a Gaussian kernel of 6 pixels.
We align the major-axis of the galaxy disk and let the side closer to M87 at the right-hand side.
The solid blue rectangle represents the averaged X-ray hot gas corona region and the dashed black rectangle marks the subtracted galactic disk region. The yellow plus marks the position of the stacked galaxy center.
}
\label{fig:stackimage}
\end{figure*}

To probe the potential dependence of diffuse X-ray emission on the SFR and the stellar mass, we divide the undetected LTGs into two subsets according to their SFR (high-SFR and low-SFR), stellar masses (high-mass and low-mass), and specific SFR (high-sSFR and low-sSFR), respectively.
We then produce stacked counts maps for these two SFR bins, two mass bins, and two sSFR bins by reprojecting each galaxy to the center of the map and rotating each galaxy to align the major-axis of the galaxy disk. The detected point sources are also masked when stacking. 
The exposure maps and instrumental background maps are stacked in the same way. 
These stacked flux maps are shown in Figure \ref{fig:stackimage}, again smoothed for a better visualization. 

We quantify the S/N and photon flux of the stacked signal, in the same way as done for the individual galaxies, by extracting 0.5--2 keV counts from a rectangular region, which has a length and a width equaling to the mean major axis (2$\bar{a}$) and 3 times the mean minor axis (6$\bar{b}$) of the galaxies in a given subset. The disk region, which is defined as a rectangle in size of 2$\bar{a}\times2\bar{b}$, is similarly excluded.
This corresponds to a typical height of 3.9--6.3 kpc (these value are determined by the sSFR subsets, that is, galaxies in the two sSFR subsets have very different sizes)
from the disk plane.
The local background is calculated from a larger rectangle with the diffuse emission region excluded. 
{\it Aprate} reports S/N $> 3$ for the subset of high-SFR only, for which we report the derived X-ray luminosity and the 1$\sigma$ error. For the other five subsets, there is no significant extraplanar X-ray emission, and we derive the 3$\sigma$ upper limit of the average extraplanar X-ray luminosity.
The input and results of the stacking analysis are summarized in Table~\ref{tab:subset}. 

We note that the Virgo LTGs have a range of angular size. We have tested to divide the undetected LTGs into two subsets according to their semi-major axis. The large-galaxy and small-galaxy subsets have a mean semi-major axis of $123.4^{+46.9}_{-36.3}$ arcsec and $75.6^{+8.4}_{-15.2}$ arcsec, respectively. The stacking of these two subsets results in no detection either, indicating that optimizing the galaxy size does not help with detecting the coronal emission. Moreover, both the sky and instrumental background, which together dominate the detected counts, are a function of the angular size rather than the physical, which makes a single scaling for both the source and background signals problematic. Thus we do not perform any rescaling of the stacked galaxies.

\begin{deluxetable*}{cccccccc}
\tabletypesize{\footnotesize}
\tablecaption{Stacking results of various Virgo edge-on LTG subsets}
\label{tab:subset}
\tablehead{
\colhead{Subset} & \colhead{\# of LTGs} & \colhead{Range} & \colhead{Median Value} & \colhead{$\bar{a}$} & \colhead{$\bar{b}$} & \colhead{S/N} & 
\colhead{$L_{\rm X}$}
}
\colnumbers
\startdata
High-SFR LTGs& 7 &  $0.6 \leq$ SFR $< 1.5$ &  1.0 & 88.8  & 19.7 & 3.8 & $2.7 \pm 0.7$ \\ 
Low-SFR LTGs&  9 &  $0.2 \leq$ SFR $< 0.6$ &  0.3 & 113.1 & 23.6 & 0.0 & $<$ 9.0  \\ 
High-mass LTGs& 7 & $9.9 <$ log $M_* < 10.4$   & 10.1 & 108.1 & 24.0 & 2.4 & $<$ 11.2 \\ 
Low-mass LTGs&  9 & $9.2 <$ log $M_* < 9.9$    &  9.7 & 98.1  & 20.3 & 1.7 & $<$ 9.5  \\ 
High-sSFR LTGs& 7 &  $-10.0 \leq$ sSFR $\leq -9.5$  &  $-9.8 $ & 80.0  & 16.3 & 2.6 & $<$ 11.2 \\ 
Low-sSFR LTGs&  9 &  $-10.7 \leq$ sSFR $\leq -10.1$ &  $-10.4$ & 120.0 & 26.3 & 0.3 & $<$ 8.4  \\ 
\enddata
\tablecomments{(1) Subsets of stacked Virgo edge-on LTGs; (2) number of stacked galaxies; (3) range of SFR or logarithmic stellar mass or specific SFR; SFR in units of $\rm M_\odot~yr^{-1}$, $M_*$ in units of $\rm M_\odot$, and sSFR in units of $\rm~yr^{-1}$; (4) median SFR or stellar mass or sSFR of stacked galaxies; (5)-(6) averaged semi-major axis ($\bar{a}$) and averaged semi-minor axis ($\bar{b}$) of stacked galaxies, in units of arcsecond; (7) signal-to-noise ratio of the stacked signal; 
(8) 0.5--2 keV luminosity or 3$\sigma$ upper limit of the stacked signal in units of ${10^{38}\rm~erg~s^{-1}}$.
}
\end{deluxetable*}


\section{Discussion} \label{sec:discussion}
The putative diffuse X-ray emission from a hot gas corona are undetected in most Virgo edge-on LTGs , with an upper limit of $L_{X,0.5-2} \sim (1-6) \times 10^{39}\rm~erg~s^{-1}$. The only three galaxies (NGC\,4192, NGC\,4216 and NGC\,4388) with significant extraplanar X-ray emission also have its 0.5--2 keV luminosity within this range. This is illustrated in Figure~\ref{fig:lxsfrM}, which plots the extraplanar X-ray luminosity versus SFR in the left panel and the stellar mass in the right panel. This low detection fraction is partly owing to the moderate {\it Chandra} exposure for most objects, but could also reflect an intrinsic deficiency of diffuse hot gas in the halo of Virgo LTGs. 
This situation is quite similar to the case of Virgo ETGs with a similar range of stellar mass as the LTGs, for which \citet{Hou2021} found a general paucity of diffuse hot gas.

In a cluster environment, multiple physical processes may affect the content of the hot gas corona. 
Sufficiently strong ram pressure is certainly able to remove a pre-existing hot coronal gas that is on-average less gravitationally bound than the disk gas \citep{Bekki2009}. 
On the other hand, ram pressure can act to halt or confine an on-going hot gas outflow (for instance, driven by the in-disk SF activity), which otherwise could easily escape from the lower halo region, decreasing the observable amount of hot gas there \citep{Brown2000}.  
In the meantime, cool gas stripped from the galaxy can turn into hot gas by thermal conduction with the ICM, temporarily raising the extraplanar X-ray luminosity \citep{Wezgowie2012}.
As a more indirect effect, ram pressure can also trigger in-disk star formation by a localized compression of the cold gas, which is clearly evidenced in Virgo disk galaxies \citep{Nehlig2016}, and in the stripped tail \citep[e.g.,][]{Vollmer2012,Vulcani2020},  also illustrated in simulations \citep[e.g.,][]{Zhu2023}. 
Subsequently, such enhanced SF activity can replenish the hot gas corona.
Furthermore, it has been suggested that ram pressure can also enhance AGN activity \citep{Poggianti2017}, which may in turn produce a hot gas outflow, although such a scenario is still hotly debated and can be relevant to not only LTGs but also ETGs.   
In reality, the above processes may conspire to determine the amount of hot gas around a cluster LTG.

We address below the deficiency of hot gas coronae around Virgo LTGs in close comparison with field LTGs (Section \ref{subsec:field}) and cluster LTGs in cosmological simulations (Section~\ref{subsec:TNG}), in the context of star formation rate, host galaxy mass and environment.

\subsection{Comparison with field LTGs} \label{subsec:field}

\begin{figure*}
\centering
\includegraphics[width=0.45\textwidth]{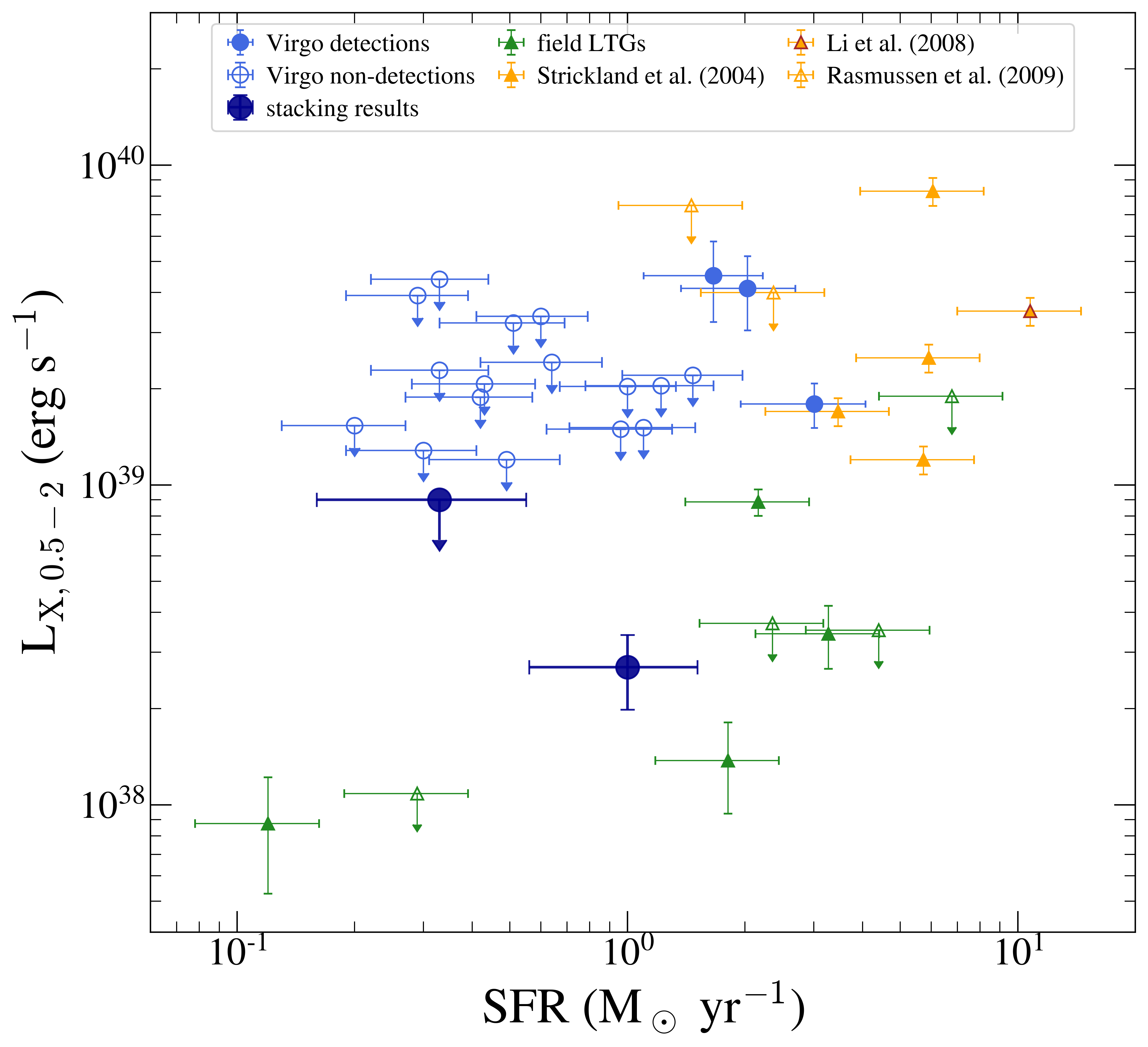}  
\includegraphics[width=0.45\textwidth]{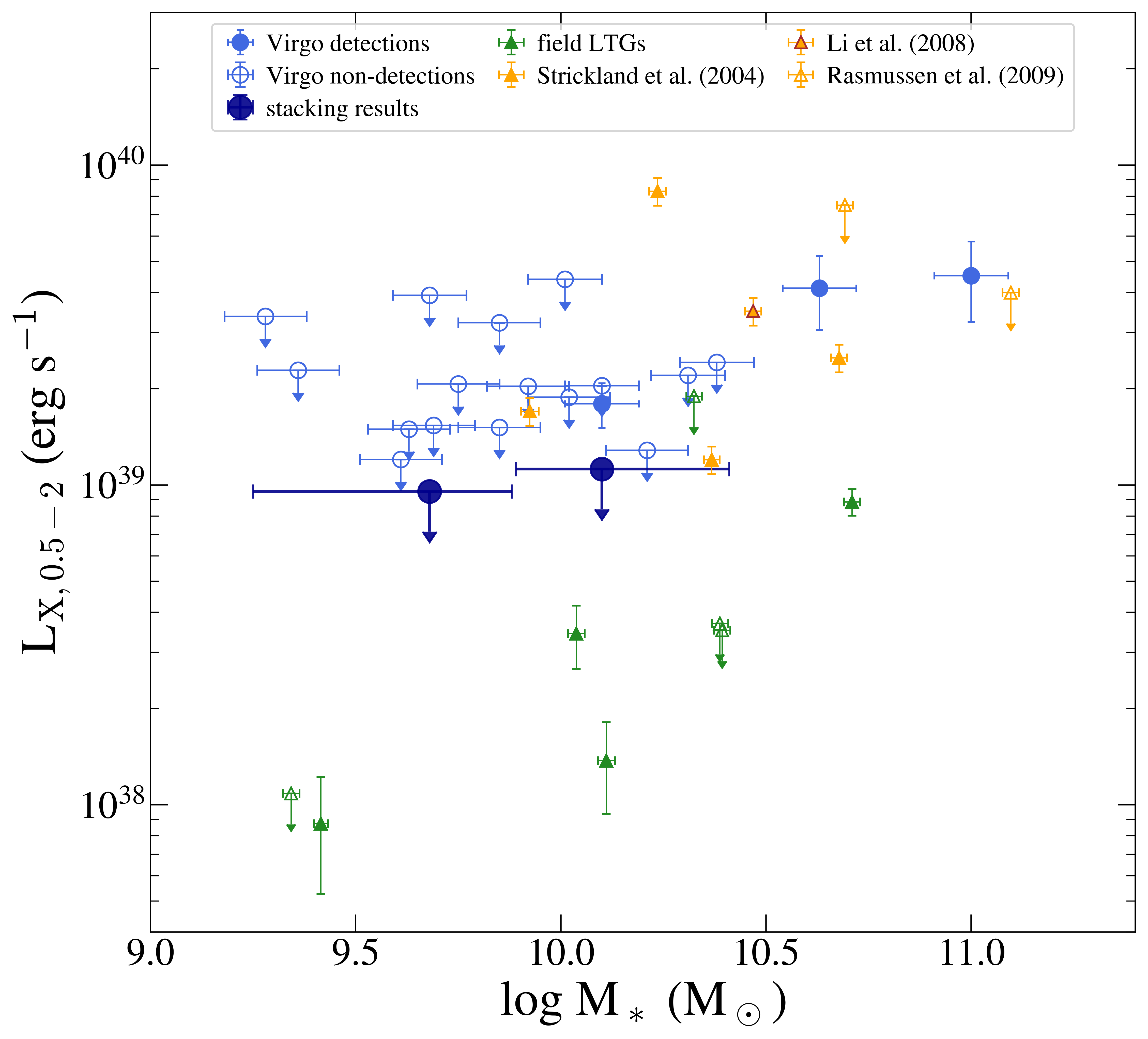} 
\includegraphics[width=0.45\textwidth]{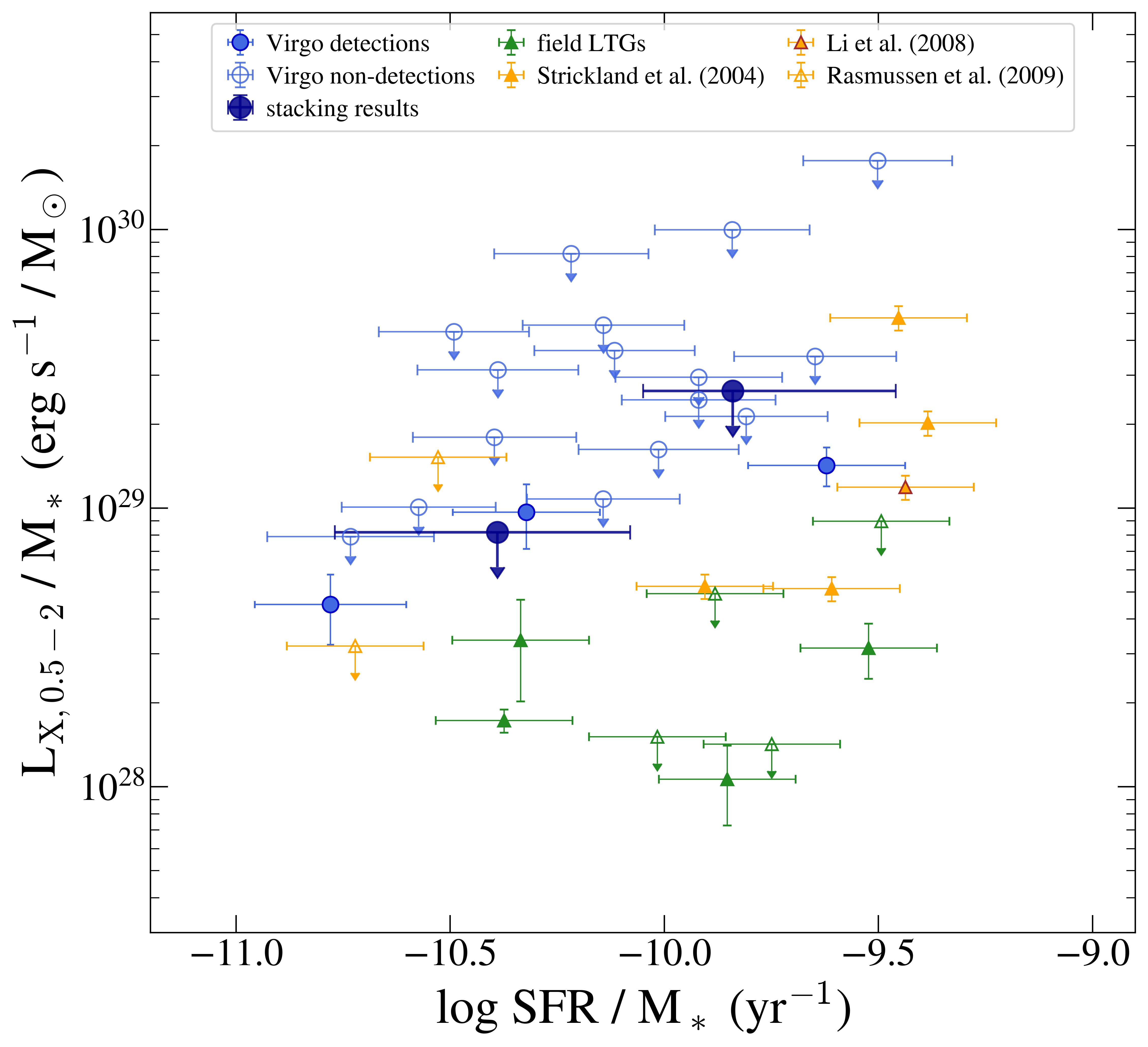} 
\caption{0.5--2 keV X-ray hot gas corona luminosity versus SFR ({\it top left}) and stellar mass ({\it top right}) of individual galaxies and the stacked bins. The {\it bottom panel} shows the X-ray luminosity normalized by stellar mass versus SFR normalized by stellar mass.
In all panels, the blue solid and open circles are for individual Virgo edge-on LTGs with detections and non-detections, respectively. The large dark blue circle are stacking results. Downward pointing arrows represent 3$\sigma$ upper limits.
The orange triangles are for field LTGs collected from \citet{Strickland2004}, \citet{Li2008}, and \citet{Rasmussen2009}, while additional field LTGs, originally studied by \citet{Li2013a}, are shown by the green triangles. See text in Sec. \ref{subsec:field} for details.
 Overall, both the Virgo individual detections of corona X-ray emission and the stacked signals are consistent with the positive correlations exhibited by the field LTGs.
}
\label{fig:lxsfrM}
\end{figure*}

A sample of field (i.e., not located inside clusters or massive groups) edge-on LTGs with existing X-ray hot gas corona measurements are plotted in Figure~\ref{fig:lxsfrM}.
These field LTGs are taken from previous work on the diffuse X-ray coronae of nearby disk galaxies. 
\citet{Strickland2004} performed the first systematic study of diffuse X-ray emission from 9 highly-inclined, actively star-forming galaxies using {\it Chandra} observations and 
provided 0.5--2 keV luminosity of the extraplanar hot gas for each galaxy. 
We adopt 4 galaxies (NGC\,891, NGC\,3079, NGC\,3628, and NGC\,4631) from this sample and neglect the other 5 galaxies with a distance $<$ 5 Mpc, such that the Chandra FoV does not cover a large enough  physical region for an equal comparison.
\citet{Li2008} derived the extraplanar 0.5--2 keV luminosity for the edge-on galaxy NGC\,5775, whereas \citet{Rasmussen2009} obtained upper limits for two 
highly-inclined disk galaxies (NGC\,5746 and NGC\,5170) with only a moderate SFR.  
These values (including upper limits) are plotted as oranges symbols in Figure~\ref{fig:lxsfrM}. 
The largest {\it Chandra} sample of highly inclined disk galaxies to date was presented by \citet{Li2013a}, which include both late-types and early-types. 
We select a subset of 8 galaxies ((NGC\,24, NGC\,3556, NGC\,3877, NGC\,4013, NGC\,4217, NGC\,4565, NGC\,4666, and NGC\,7090) from this sample, which have an inclination angle $\gtrsim 75\degr$ and a morphological type later than Sa, to be more consistent with our Virgo sample. 
A further complication arises because the diffuse X-ray luminosity provided by \citet{Li2013a} may include contribution from the star-forming disk.
For this reason, we re-derive the extraplanar X-ray luminosity (or upper limit) of these 8 galaxies in a way fully consistent with our treatment for the Virgo edge-on LTGs. 
It is noteworthy that potential contribution from unresolved stellar objects belonging to the host galaxy is generally negligible in the extraplanar region of interest here.
The resultant values are plotted as green symbols in Figure~\ref{fig:lxsfrM}, which include four detections and four upper limits. 
The stellar mass of all field LTGs are taken from \citet{Jiang2019} derived based on WISE $W1- W2$ color and $W1$ band luminosity. The SFR are calculated from WISE W3 magnitude as calibrated by \citet{Cluver2017}.
These 15 field LTGs are also plotted in the bottom panel of Figure~\ref{fig:mosaic}, half of which appear to lie systematically above the galaxy main sequence. This might be partially understood as a selection effect, for many of these field LTGs were targeted by Chandra because they are known to exhibit intensive star formation.  

As shown in the top panels of Figure~\ref{fig:lxsfrM}, the field LTGs exhibit a positive correlation between the corona X-ray luminosity and the SFR, or between the corona X-ray luminosity and the stellar mass, which was studied in detailed by \citet{Li2013b} and \citet{Wang2016}. 
The Virgo edge-on LTGs span a similar range in both SFR and $M_*$ as the field LTGs, but with a large fraction of galaxies at SFR $< 1\rm~M_\odot~yr^{-1}$ and log$(M_*/\rm~M_\odot) < 10.0$.
In fact the field sample contains only two galaxies (NGC\,24 and NGC\,7090) satisfying these two conditions.
The three Virgo LTGs with a significant detection roughly follow the positive correlations defined by the field LTGs.
They are in fact the three galaxies with the highest SFR,  and two of them (NGC\,4192 and NGC\,4216) have the highest stellar mass among the Virgo sample.  
This implies that either star formation or stellar mass is the primary factor responsible for the detected extraplanar X-ray emission.
The lower panel of Figure~\ref{fig:lxsfrM} further depicts $L\rm_X/M_*$ versus SFR/$M_*$.
The field LTGs show a previously known correlation \citep{Li2013b,Wang2016}. Interestingly, the three Virgo LTGs appear to lie above the field LTGs, that is, they seem to have a high specific X-ray luminosity for their specific SFR. This might be taken as a hint that the extraplanar X-ray emission is temporarily enhanced in these three galaxies due to ram pressure. In fact, at least in the case of NGC\,4388, evidence was found for extended X-ray emission related to RPS \citep{Wezgowie2011}.

As for the individually undetected Virgo LTGs. As upper limits, they lie significantly above the field LTGs, which is mainly owing to the shallow Chandra exposures. 
But interestingly, the six bins of stacking signals, in particular the high-SFR bin signifying a detection, are also broadly consistent with the trend of the field LTGs.
Apparently, the current data, which represent the largest X-ray sample to date of nearby edge-on LTGs, 
{\it suggest no significant difference in the amount of corona X-ray emission between Virgo and field LTGs}.
This situation is again similar to the case of low-to-intermediate mass Virgo ETGs, for which \citet{Hou2021} found no statistically significant difference in the amount of diffuse X-ray emission when compared to a sample of field ETGs of similar stellar masses.

The apparent paucity of hot gas coronae in the sampled Virgo galaxies as well as the similarity between the Virgo LTGs and ETGs might be taken as a hint that both in-disk star-forming activity and the subsequent stellar feedback, and the long-lasting effect of ram pressure exerted by the hot ICM act together to regulate the hot gas coronae for most present-day low-to-intermediate mass Virgo galaxies.
To find further clues on the possible role of environment effects, we turn to compare the observation with state-of-the-art cosmological simulations.

\subsection{Comparison with simulated cluster LTGs} \label{subsec:TNG}

We utilize data from IllustrisTNG, a set of cosmological, magneto-hydrodynamical simulations carried out using the advanced AREPO moving-mesh code \citep{Pillepich2018}. To select out simulated galaxies for a meaningful comparison with the Virgo LTGs, we make use of the TNG100 simulation \citep{Pillepich2018b,Springel2018,Nelson2019}, which evolves $1820^3$ gas cells in a cubic box volume of $(\rm 110.7~Mpc)^3$, with an average mass resolution of $\rm 1.4\times 10^6~M_{\odot}$ and $\rm 7.5\times 10^6~M_{\odot}$ for baryon and dark matter, respectively. With the combination of a large volume and intermediate resolution (best spatial resolution of 185 parsec), TNG100 is hence very suited for studying cluster galaxies.


First, we search for `Virgo-like' clusters, which is by definition a friends-of-friends (FoF) halo with $ 1.7\times 10^{14}\rm~M_{\odot}< M_{\rm halo} < 3.8\times 10^{14}\rm~M_{\odot}$ and the number of subhalos greater than 1000. This returns nine Virgo-like clusters in TNG100. Second, we select galaxies in these clusters by requiring their SFR larger than 0.1 $\rm M_{\odot}~yr^{-1}$ and their stellar mass log$(M_*/\rm~M_{\odot}~yr^{-1}) > 9.5$ to be compatible with the Virgo LTGs.
Lastly, we exclude probable ellipticals and S0 galaxies above the SFR cut, using the galaxy morphology catalog from \citet{Huertas-Company2019}, which assigns a probability of being late-type or early-type to TNG100 $z = 0$ galaxies with a machine learning algorithm, based on pseudo-$R$-band morphology.
We retain galaxies with a probability of being late-type greater than 0.5. 
This results in a final sample of 119 star-forming galaxies residing in a total of 9 Virgo-like clusters at redshift $z = 0$.
Hereafter these galaxies are referred to as simulated LTGs, although a minor fraction of these could be more like ETGs in terms of their morphology. 
The number of simulated LTGs per cluster ranges between 6 to 40.
Notably, these numbers are generally lower than the number of Virgo LTGs satisfying the same two conditions \citep[at least $\sim$70;][]{Soria2022}, possibly indicating an under-prediction of star-forming LTGs in TNG100 `Virgo-like' clusters.

\begin{figure}
\centering
\includegraphics[width=0.42\textwidth]{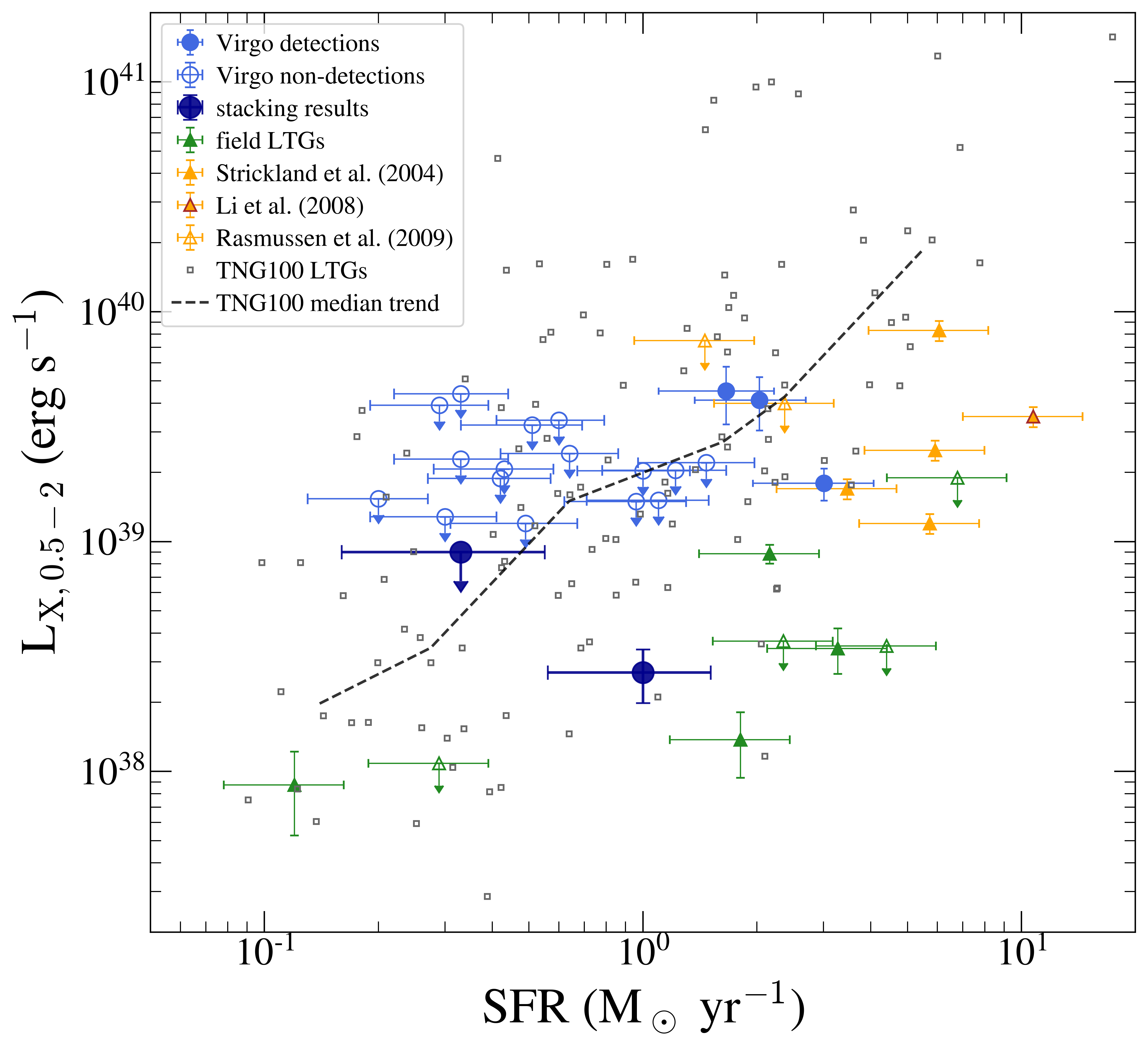} 
\includegraphics[width=0.42\textwidth]{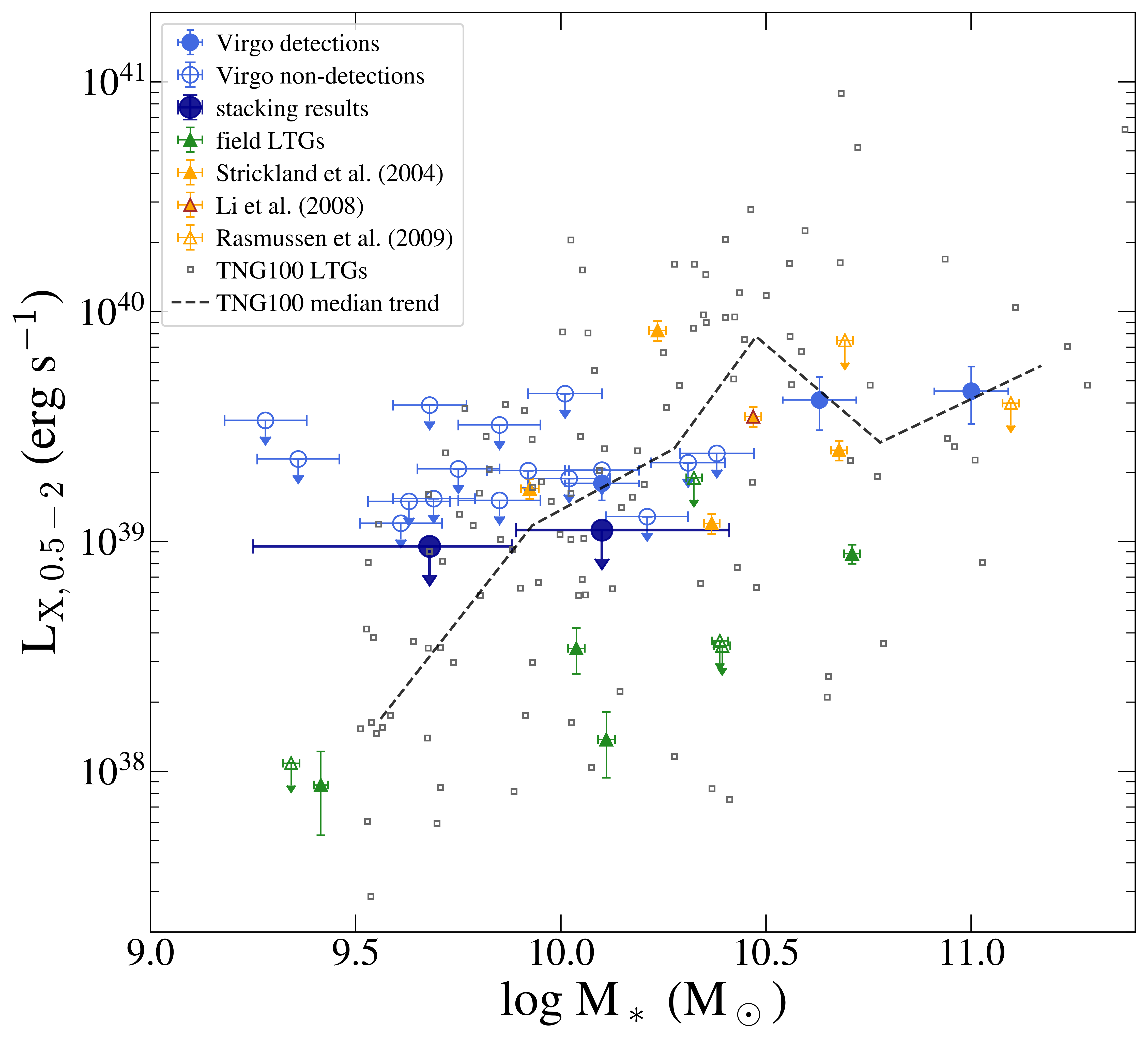} 
\includegraphics[width=0.42\textwidth]{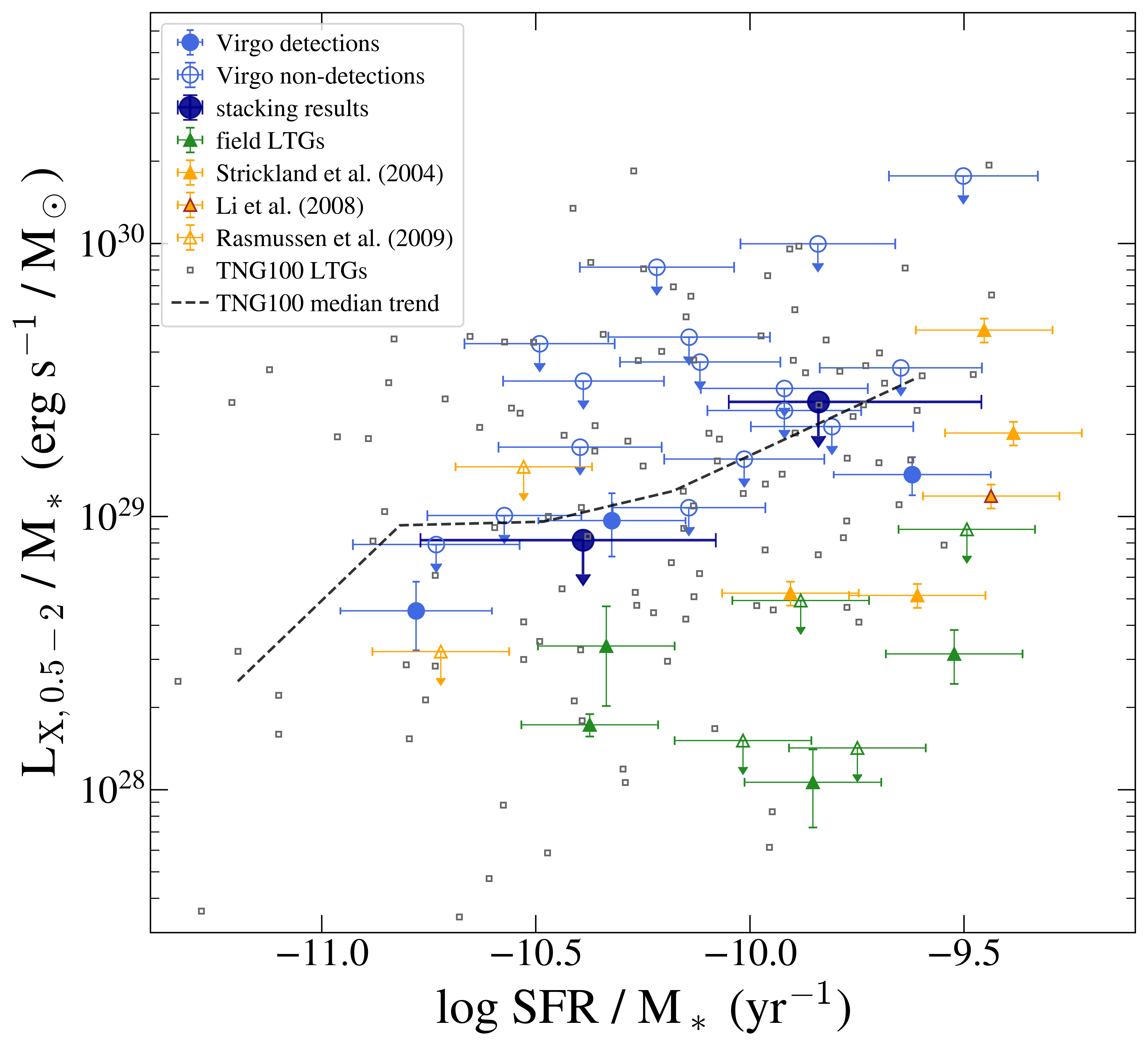} 
\caption{Similar to Figure \ref{fig:lxsfrM}, but adding the comparison with the TNG100 cosmological simulation. A total of 119 simulated LTGs from 9 ``Virgo-like'' clusters at redshift $z = 0$ identified in the TNG100 simulation are shown by open gray boxes. The dashed line represents the median trend of the TNG100 sample.}
\label{fig:lxTNG}
\end{figure}

\begin{figure*}
\centering
\includegraphics[scale=0.4, angle=0]{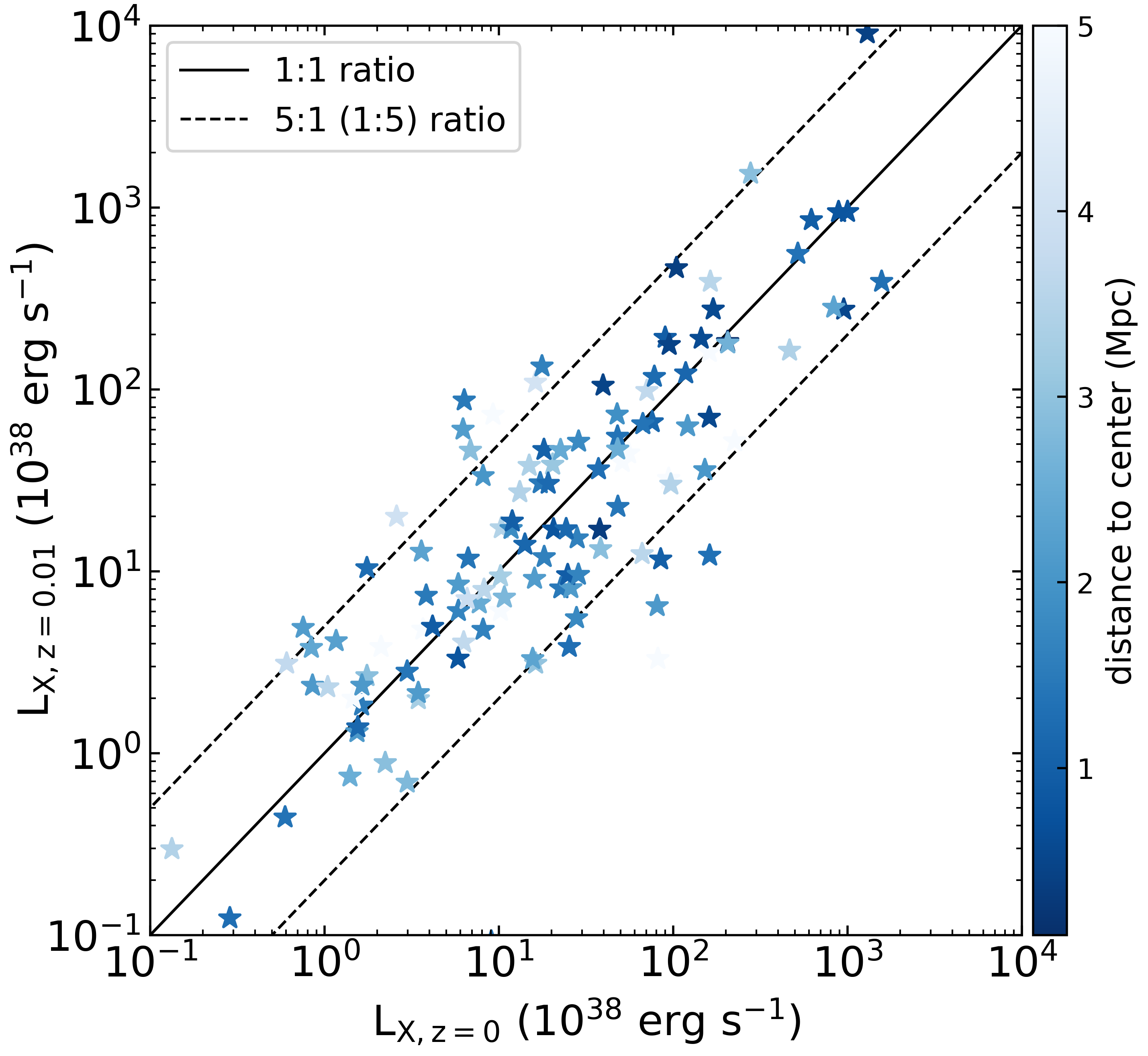}
\includegraphics[scale=0.4, angle=0]{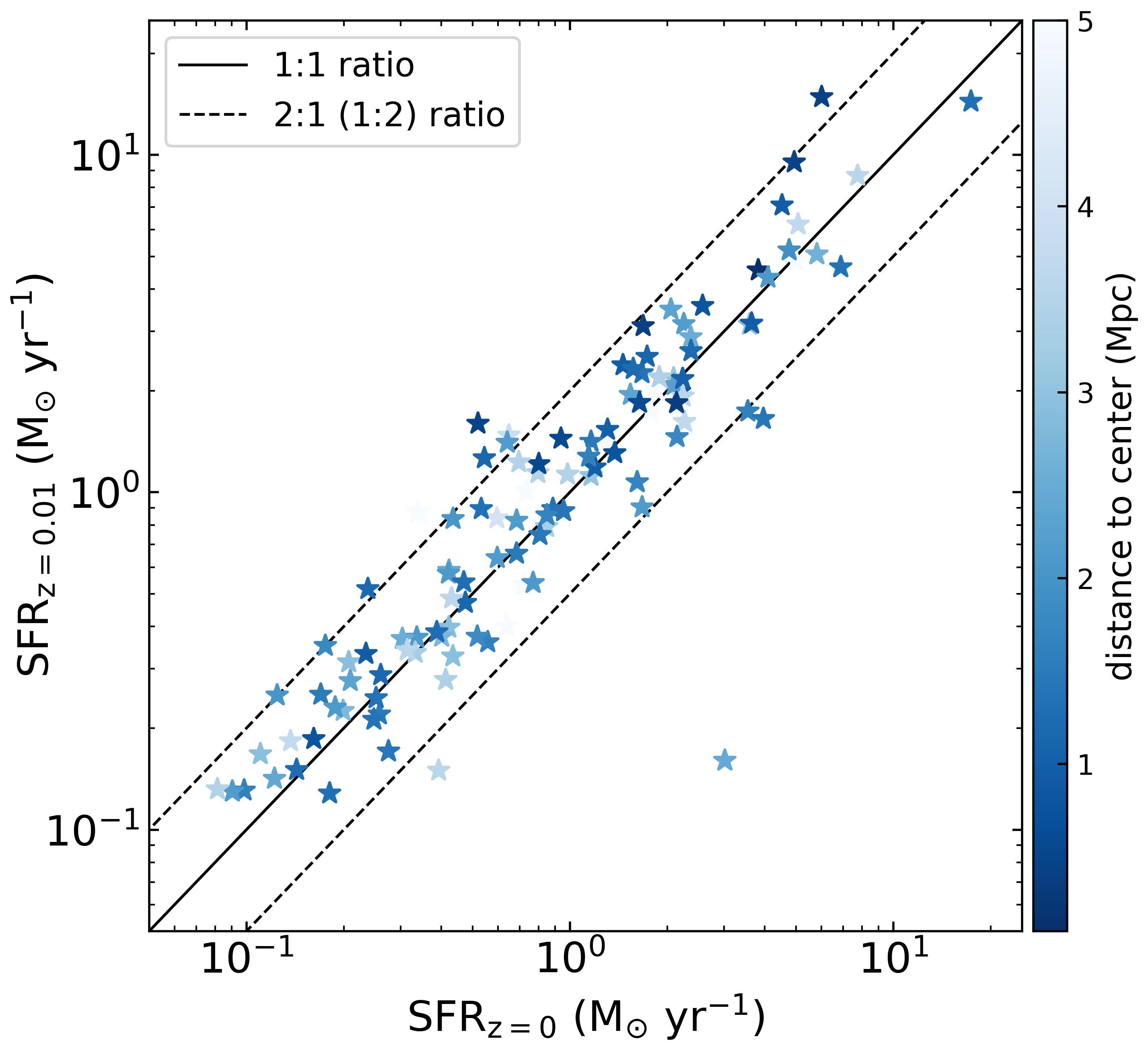}
\hspace{0.1cm}
\includegraphics[scale=0.4, angle=0]{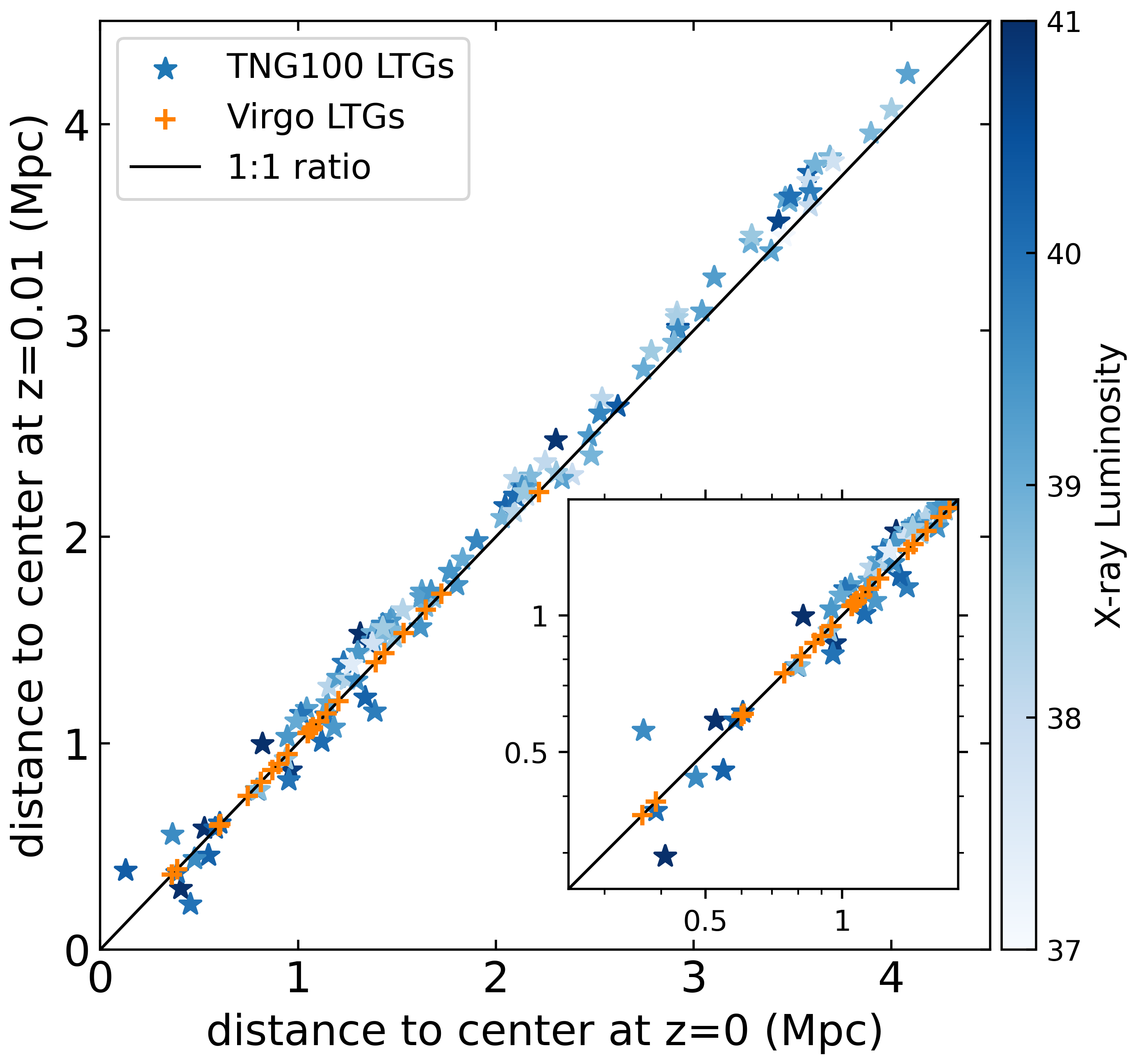}
\caption{The evolution of cluster late-type member galaxies in the TNG100 cosmological simulation. 
{\it Left:} The comparison of 0.5--2 keV extraplanar X-ray luminosity between $z = 0.01$ progenitor galaxies and $z = 0$ galaxies. The Simulated LTGs are shown in blue stars color-coded with the present-day distance to cluster center. The solid line represents the 1:1 ratio and two dashed lines represent the 5:1 (1:5) ratio.
{\it Middle:} The comparison of instantaneous star formation rate between $z = 0.01$ and $z = 0$, showing that more galaxies have a higher SFR at the earlier epoch.  
{\it Right:} The comparison of distance to the cluster center between galaxies at redshift $z = 0.01$ and $z = 0$, showing the systematic infall. LTGs are shown in solid stars color-coded with their 0.5--2 keV extraplanar X-ray luminosity. The projected distance of our Virgo edge-on LTGs are marked by orange crosses at the 1:1 line. A zoom-in view of the inner 1.8 Mpc is shown by the insert.
}
\label{fig:TNGcompare}
\end{figure*}

The hot gas cells of a given simulated LTG are defined as those having a temperature $\gtrsim 10^{5.5}$ K.
The X-ray luminosity of hot gas in each simulated LTG is then obtained by integrating the 0.5--2 keV emission from all hot gas cells within the region of interest. 
For consistency, we assume a single-temperature apec model to generate the X-ray emission using the PyAtomDB code \citep{Foster2020}, considering the temperature and metallicity of each gas cell.
To roughly match the observed X-ray corona regions of the Virgo LTGs, we measure the 0.5--2 keV luminosity in the vertical range of 1--5 kpc on both sides of the disk, with a length of 20 kpc along the major axis, effectively excluding the diffuse X-ray emission from the galactic disk and being consistent with the Virgo edge-on LTGs. 
The orientation of the rotation axis of the disk plane is derived according to the angular momentum distribution of the star particles.
To characterize the TNG100 LTGs, we define their stellar mass as the total mass of all bound star particles, and the SFR as the sum of instantaneous star formation rate of all gas cells. 
We have tested that using the averaged SFR over 100 Myr does not affect the results.
We have also tested an alternative definition which consider all the particles/cells within twice of the half stellar-mass radius and found consistent results. 

The TNG100 LTGs are plotted as open grey boxes in Figure \ref{fig:lxTNG}, which together span a similar range in both SFR and stellar mass with the real galaxies. The TGN100 LTGs exhibit a large scatter (up to two orders of magnitude for a given SFR or $M_*$) in the corona X-ray luminosity, but the median value for a given SFR or $M_*$ interval (shown by a dashed line) shows a more-or-less positive trend, similar to that seen among the real galaxies. 
However, most of the field LTGs with a significant detection of corona X-ray emission, as well as the stacked bin of high-SFR Virgo LTGs, are located well below the median trend of the TNG100 LTGs in the $L\rm_X$ vs. SFR plot (top left panel). 
While for the $L\rm_X$ vs. $M_*$ plot (top right panel), the difference is less significant, although most field LTGs and the stacked bin of high-mass Virgo LTGs also appear to lie below the median trend of the TNG100 LTGs.
In terms of the specific X-ray luminosity versus specific SFR (bottom panel), the Virgo LTGs also lie significantly below the median trend of the simulated LTGs.
Therefore, the simulated cluster LTGs appear to have a systematically higher hot gas content than the Virgo LTGs.
This might be related to the systematically lower number of star-forming galaxies than the actual case of Virgo, as pointed out in the above.

To shed light on the likely cause of this difference, it is instructive to examine the potentially time-dependent (hence position-dependent as the galaxies move inside the cluster) evolution of the X-ray coronae around the simulated cluster LTGs. 
Thus we track the same set of LTGs when they were at an earlier epoch of $z = 0.01$, or $\sim$100 Myr ago, which is about the sound crossing time of a putative hot gas corona with a size of 10 kpc, i.e., the timescale on which the content of the corona may evolve significantly. 
The progenitor of each present-day subhalo is tracked using the SubLink merger trees \citep{Rodriguez-Gomez2015}.
We further add 4 galaxies with SFR$(z=0.01) > 0.1\rm ~M_{\odot}~yr^{-1}$ but SFR$(z=0) < 0.1\rm ~M_{\odot}~yr^{-1}$, to include those galaxies which have their SFR reduced from the earlier epoch.
The corona X-ray luminosity of these progenitors, $L\rm_{X,z=0.01}$, is similarly calculated and compared to the present-day $L\rm_{X}$ in the left panel of Figure~\ref{fig:TNGcompare}. 
Remarkably, most simulated LTGs show a significant variation (by a factor of up to 5 in both directions) in their corona X-ray luminosity, indicating that the amount of extraplanar hot gas in these galaxies does evolve substantially. 

However, there is no systematic increase or decrease for the LTGs as a whole, which is at odds with the {\it naive} expectation that ram pressure would tend to remove the hot gas corona.
Indeed, by comparing the galaxy position relative to the cluster center (defined as the position of the parent halo that is linked to the brightest central galaxy at z=0) between the two epochs (right panel of Figure~\ref{fig:TNGcompare}), there is clearly a substantial inward displacement ($\sim$100 kpc on average) for most simulated LTGs, which should generally move the galaxies into a regime more susceptible to ram pressure. 
One possibility is that some other mechanism(s) operate to compensate for the RPS effect. 
The middle panel plots the variation of the instantaneous SFR between the two epochs, which shows that more LTGs ($\sim$ 63\%) have a higher SFR at $z = 0.01$ than at $z = 0$, consistent with the expectation that RPS progressively removes disk gas and consequently suppresses global star formation at a later epoch.  
In the meantime, ram pressure might also be able to enhance localized SFR in some galaxies at $z = 0.01$, preferentially at the leading side of the infall.
If star formation thus triggered could effectively supply extraplanar hot gas by massive star winds and SNe, this may counteract the more long-lasting effect of RPS, such that the simulated cluster LTGs as a whole shows no systematic variation in their corona X-ray luminosity. 
Of course, this simple picture neglects many other physical processes, such as AGN activities triggered by ram pressure and the subsequent feedback, tidal interactions, etc., as outlined at the beginning of this Section, which may also contribute to increasing or decreasing the hot coronal X-ray luminosity. 
We defer a thorough exploration of various processes in TNG and other cosmological simulations to a future work.



\section{Summary} \label{sec:summary}
We have presented a systematic study of the diffuse hot gas coronae around 21 highly edge-on late-type galaxies in the Virgo cluster, using archival {\it Chandra} observations. The main results are as follows: 

\begin{itemize}
\item Significant extraplanar diffuse X-ray (0.5--2 keV) emission is detected in only three Virgo LTGs, all with a star formation rate $\gtrsim 1\rm~M_\odot~yr^{-1}$ and a stellar mass $\gtrsim 10^{10}\rm~M_{\odot}$). The remaining 18 LTGs, most with SFR $\sim 0.1-10\rm~M_\odot~yr^{-1}$ and stellar mass $0.3-3 \times 10^{10}\rm~M_{\odot}$, do not have significant extraplanar X-ray emission (S/N $<$ 3).

\item For the individually undetected LTGs, we constrain their average corona X-ray emission with a stacking analysis, dividing the whole sample into two subsets of low- and high-SFR, low- and high-stellar mass, or low- and high-sSFR. Only the high-SFR bin yields a significant detection, which has a value of $\sim 3\times10^{38}\rm~erg~s^{-1}$.  
The apparent paucity of truly diffuse hot gas in these LTGs might be understood as efficient ram pressure stripping by the hot ICM. 
However, the stacked signals of the Virgo LTGs, as well as the X-ray luminosity of the three individually detected LTGs, appear to fit into the empirical $L\rm_X - SFR$ and $L{\rm_X} - M_*$ relations among a sample of highly inclined disk galaxies in the field, suggesting the internal feedback effects can also effectively remove the hot gas coronae for sampled galaxies.
This situation is similar to what was previously found among a sample of low-to-intermediate mass Virgo ETGs, whose average diffuse X-ray luminosity per stellar mass is consistent with field ETGs in the same mass range. 

\item A comparison sample of simulated cluster LTGs, identified in the TNG100 cosmological simulation, shows a systematically higher extraplanar X-ray luminosity than the Virgo edge-on LTGs in terms of the stacked signals. 
The simulated cluster LTGs also show no systematic increase or decrease in the extraplanar X-ray luminosity, between an earlier epoch of $z = 0.01$ and the present epoch, suggesting that in the TNG100 simulation, ram pressure stripping is not the sole dominant process that controls the hot gas content of cluster LTGs.  

\end{itemize}

In future work, it is desirable to study a larger sample of observed galaxies in nearby clusters, such as those afforded by {\it Chandra} and the on-going eROSITA all-sky survey, to better probe the effect of RPS and other environmental effects on the star formation and global gas content of cluster galaxies, which in turn would be useful to calibrate next-generation cosmological simulations.
Future missions with the potential to provide accurate abundance measurements for the hot gas corona, e.g., the recently launched XRISM \citep{Tashiro2022}, the planned HUBS \citep{Bregman2023} and LEM \citep{Kraft2022}, would promise to shed light on the difference between accreted IGM and internally supplied hot gas.

\begin{acknowledgments}
This research has made use of data and software provided by the {\it Chandra X-ray Observatory}. The list of Chandra datasets is contained in~\dataset[Chandra Data Collection (CDC) 175]{https://doi.org/10.25574/cdc.175}.
M.H. is supported by the National Natural Science Foundation of China (12203001) and the fellowship of China National Postdoctoral Program for Innovation Talents (grant BX2021016). 
L.H. and Z.H. and Z.L. acknowledge support by the National Key Research and Development Program of China (NO.2022YFF0503402) and National Natural Science Foundation of China (grant 12225302).
C.J. and W.F. acknowledge support from the Smithsonian Institution, the Chandra High Resolution Camera Project through NASA contract NAS8-03060. 
W.F. acknowledges support from NASA Grants 80NSSC19K0116, GO1-22132X, and GO9-20109X.
L.C.H. was supported by the National Science Foundation of China (11721303, 11991052, 12011540375) and the China Manned Space Project (CMS-CSST-2021-A04, CMS-CSST-2021-A06). 
\end{acknowledgments}



\appendix
The SDSS $gri$-color composite image and {\it Chandra}/ACIS 0.5--2 keV flux image of the 18 galaxies without significant X-ray emission from a putative hot gas corona are shown in Figure~\ref{fig:nondetection}.

\begin{figure*}\centering
\subfloat{
\includegraphics[scale=0.25, angle=0]{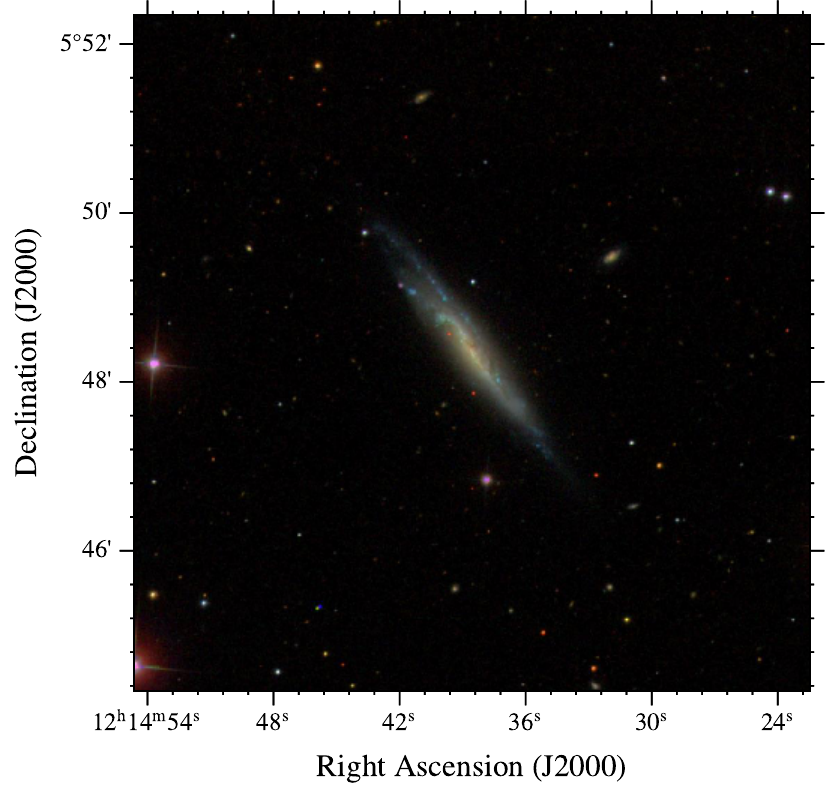}
\includegraphics[scale=0.255, angle=0]{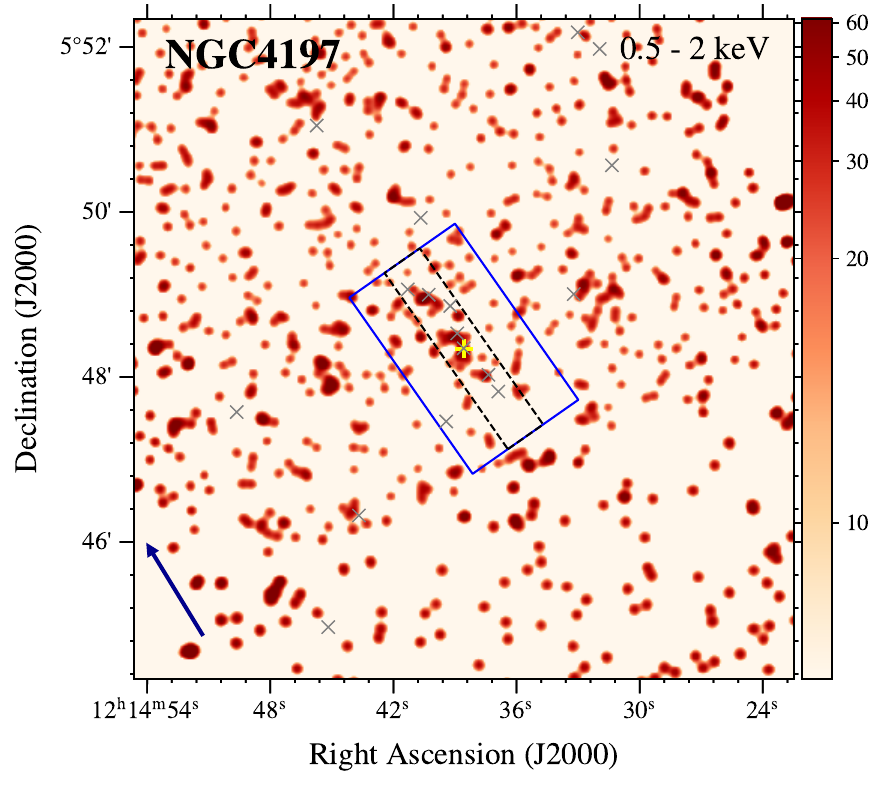}
\includegraphics[scale=0.25, angle=0]{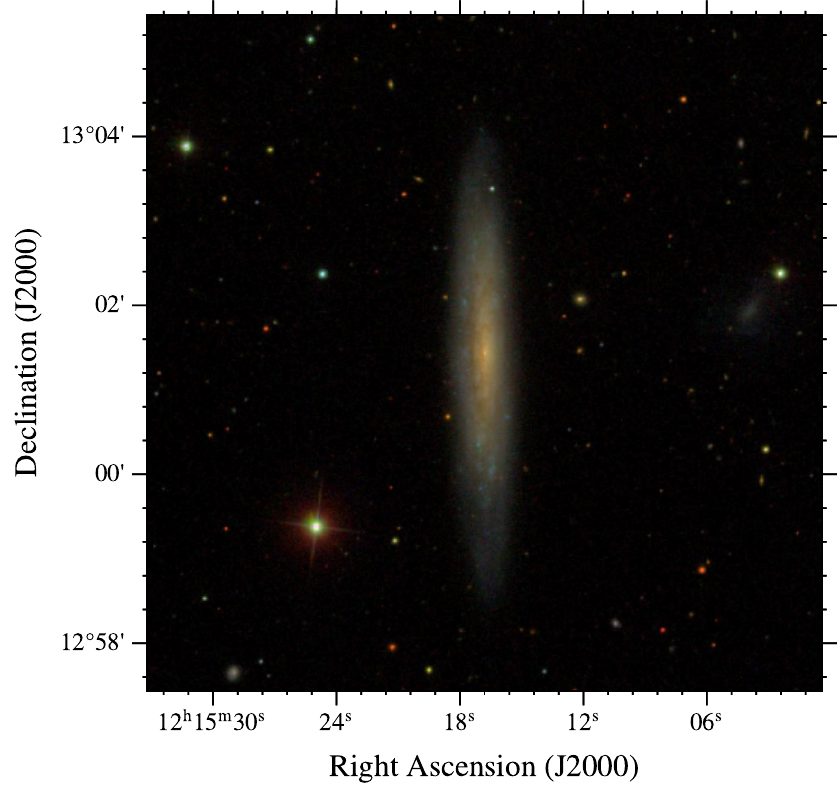}
\includegraphics[scale=0.255, angle=0]{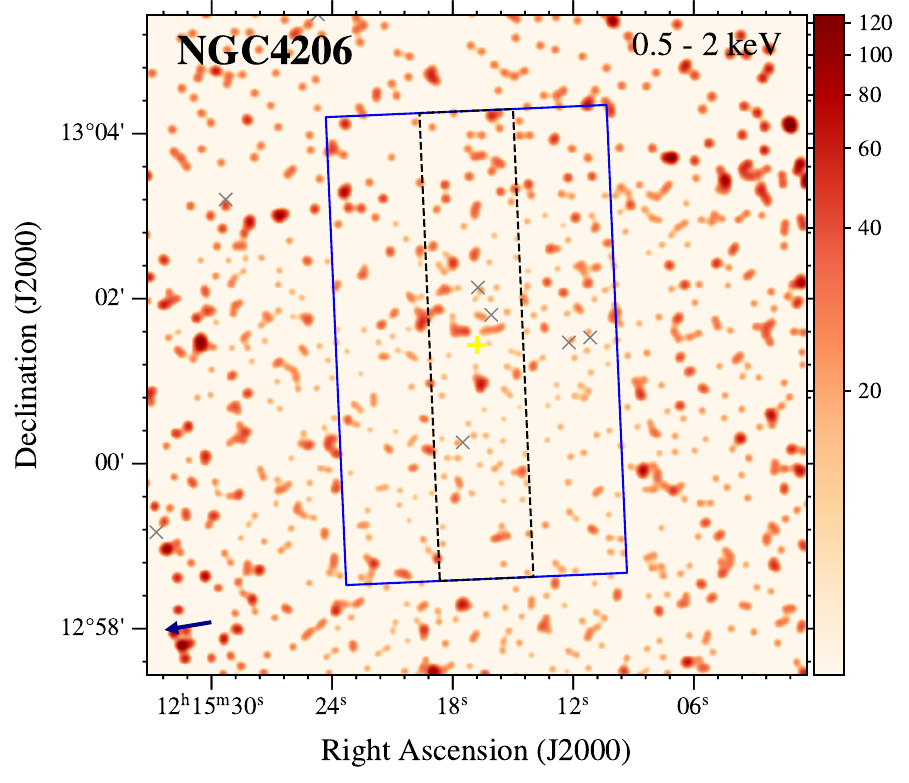}
}
\vspace{-0.35cm}
\qquad 
\subfloat{
\includegraphics[scale=0.25, angle=0]{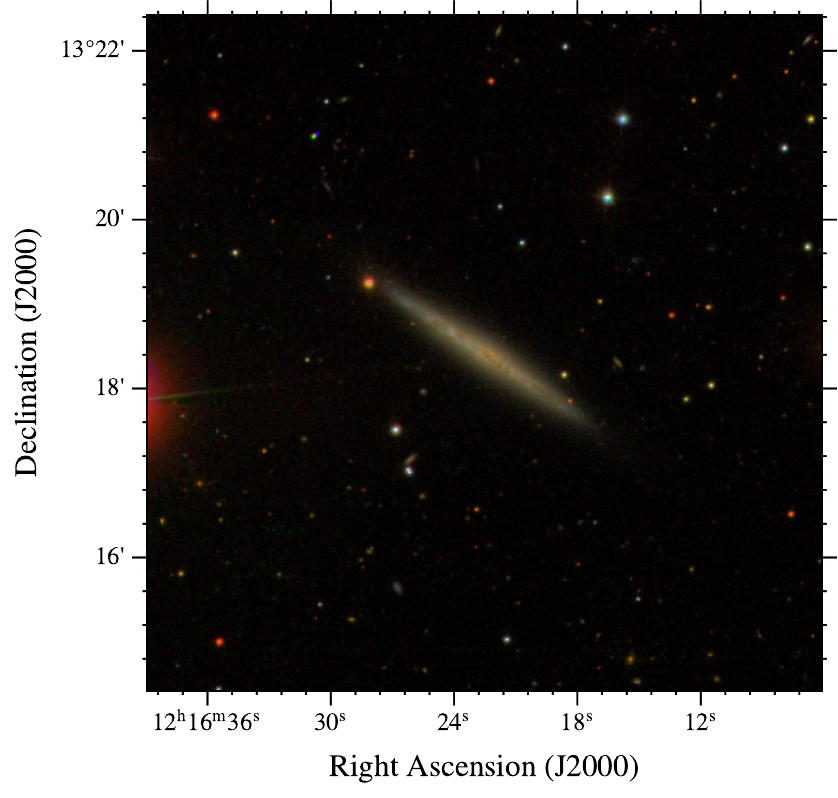}
\includegraphics[scale=0.255, angle=0]{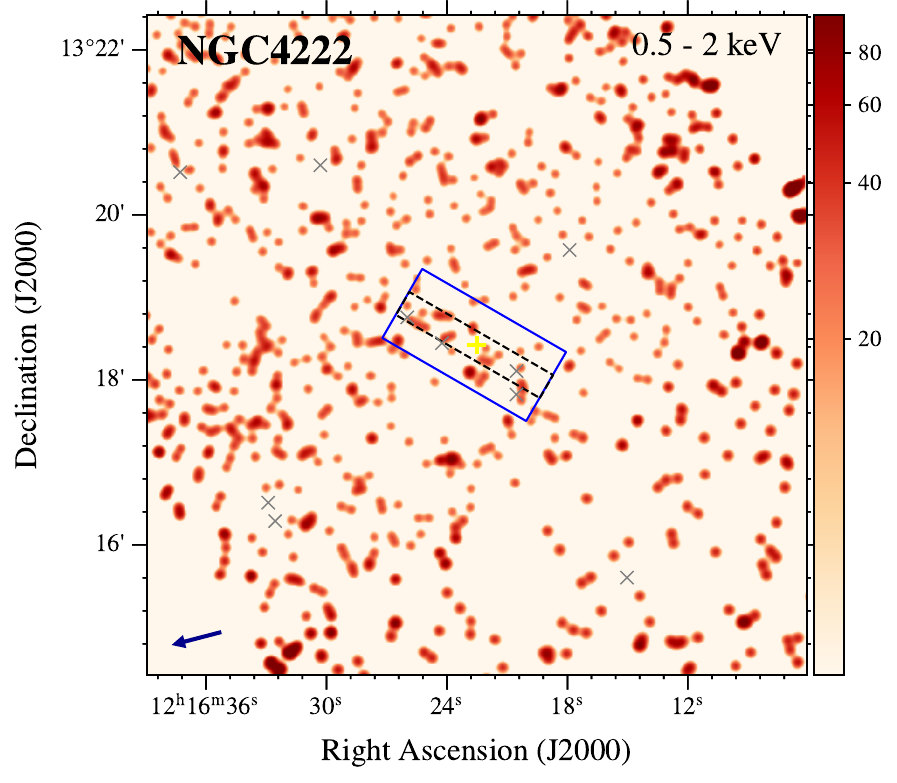}
\includegraphics[scale=0.25, angle=0]{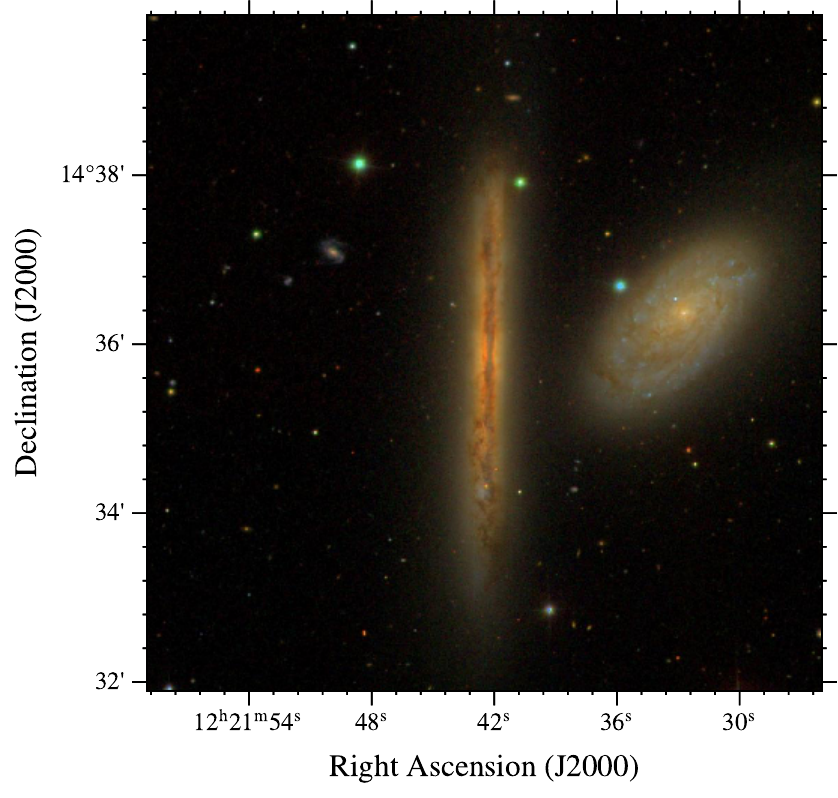}
\includegraphics[scale=0.255, angle=0]{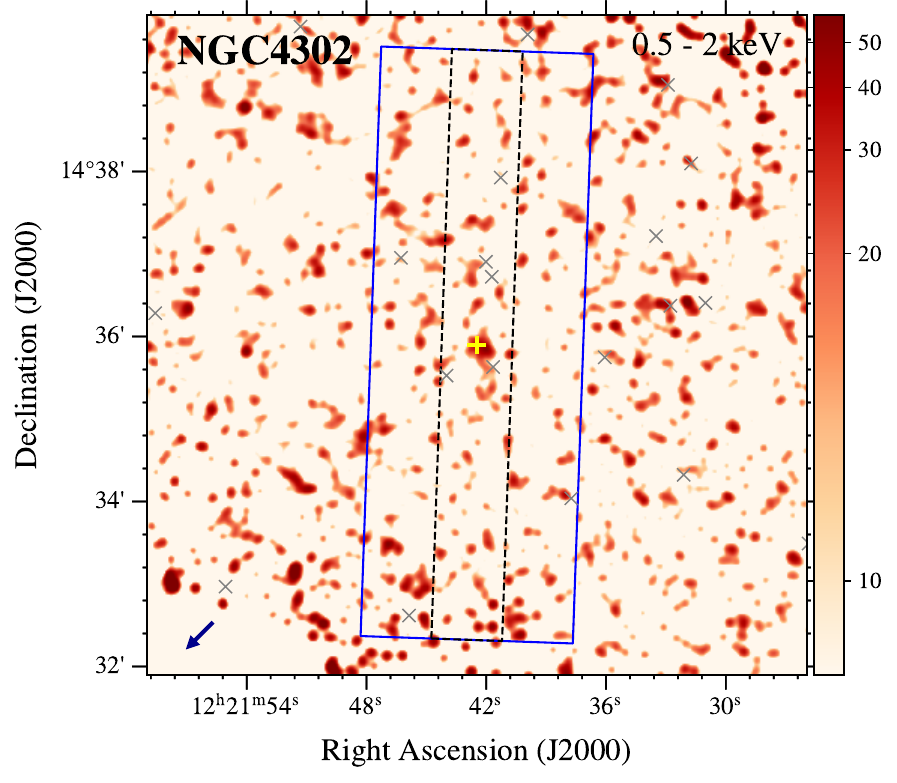}
}
\vspace{-0.35cm}
\qquad 
\subfloat{
\includegraphics[scale=0.25, angle=0]{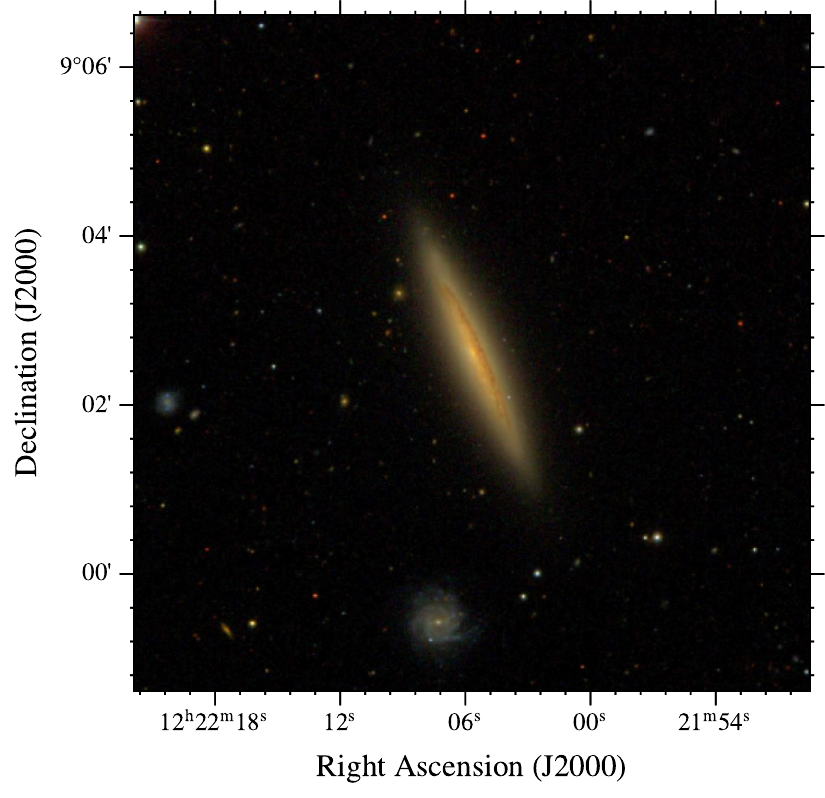}
\includegraphics[scale=0.255, angle=0]{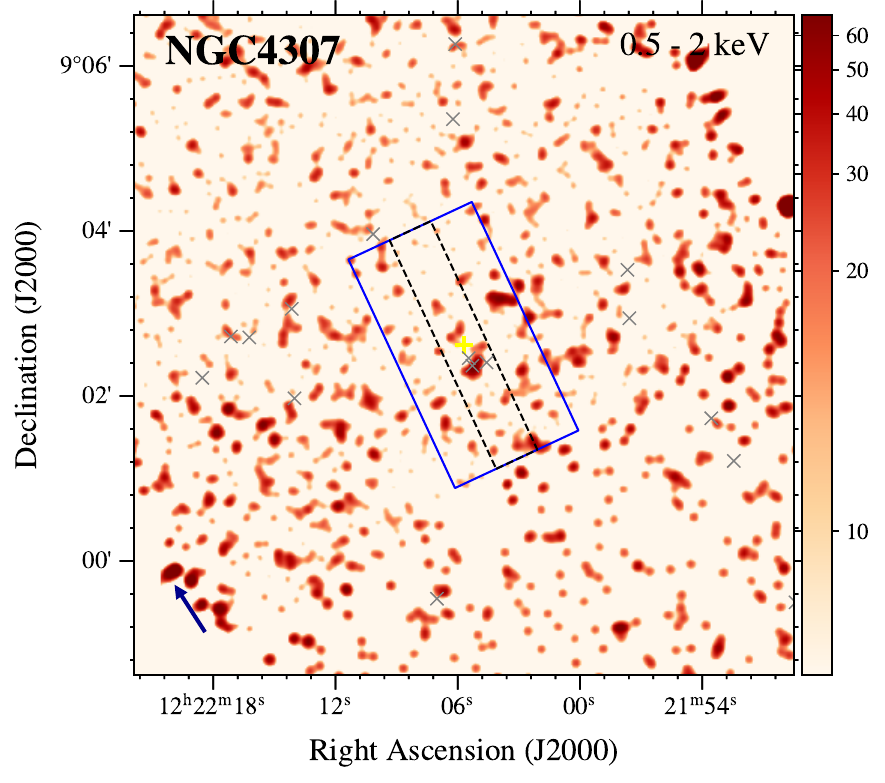}
\includegraphics[scale=0.25, angle=0]{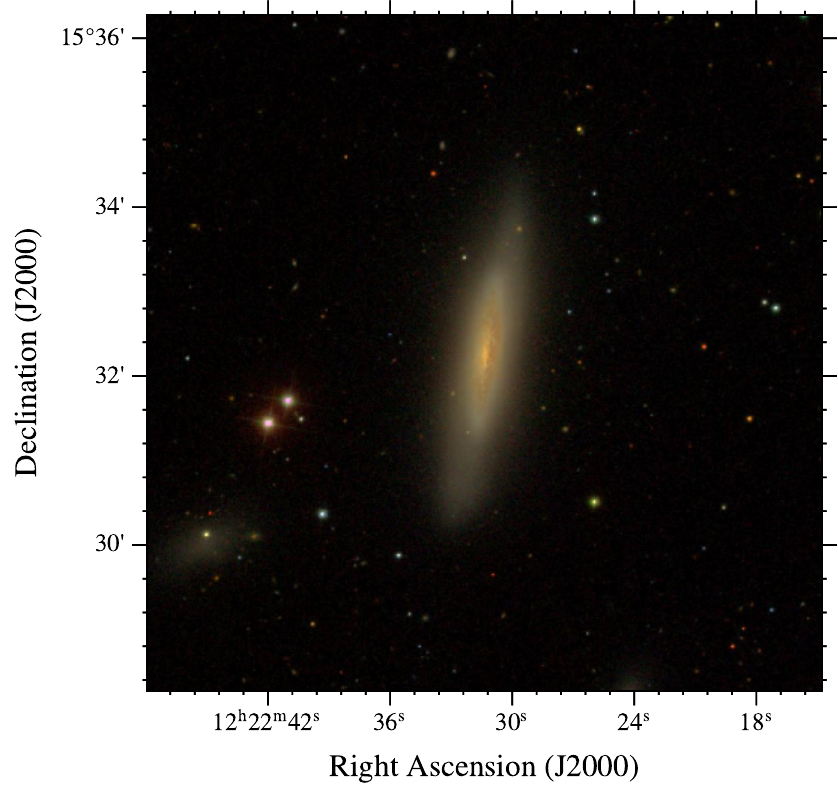}
\includegraphics[scale=0.255, angle=0]{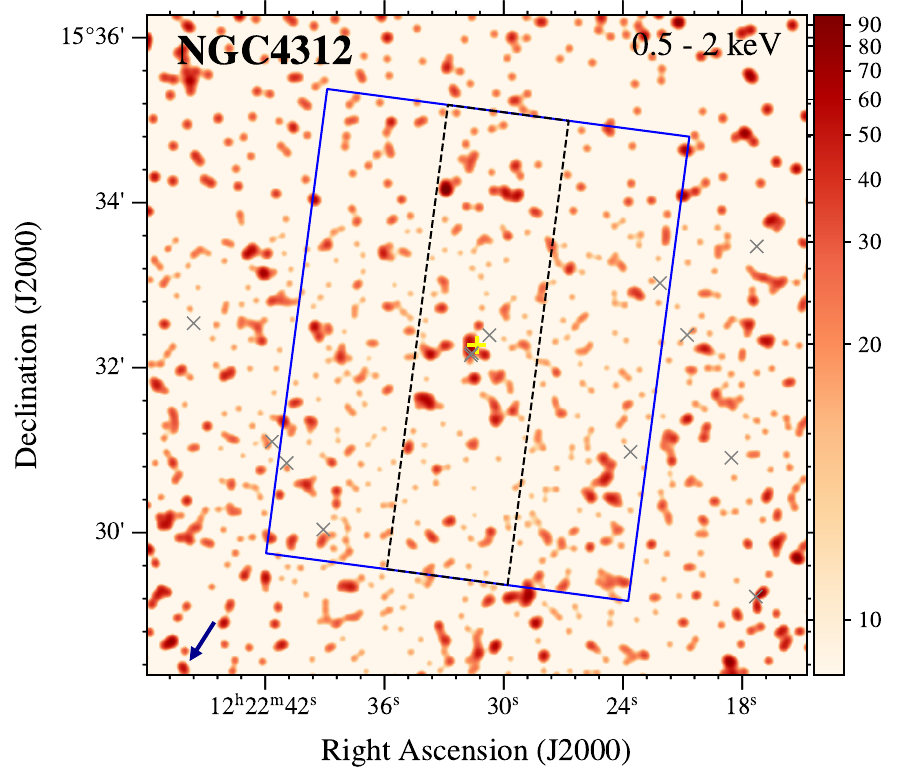}
}
\vspace{-0.35cm}
\qquad 
\subfloat{
\includegraphics[scale=0.25, angle=0]{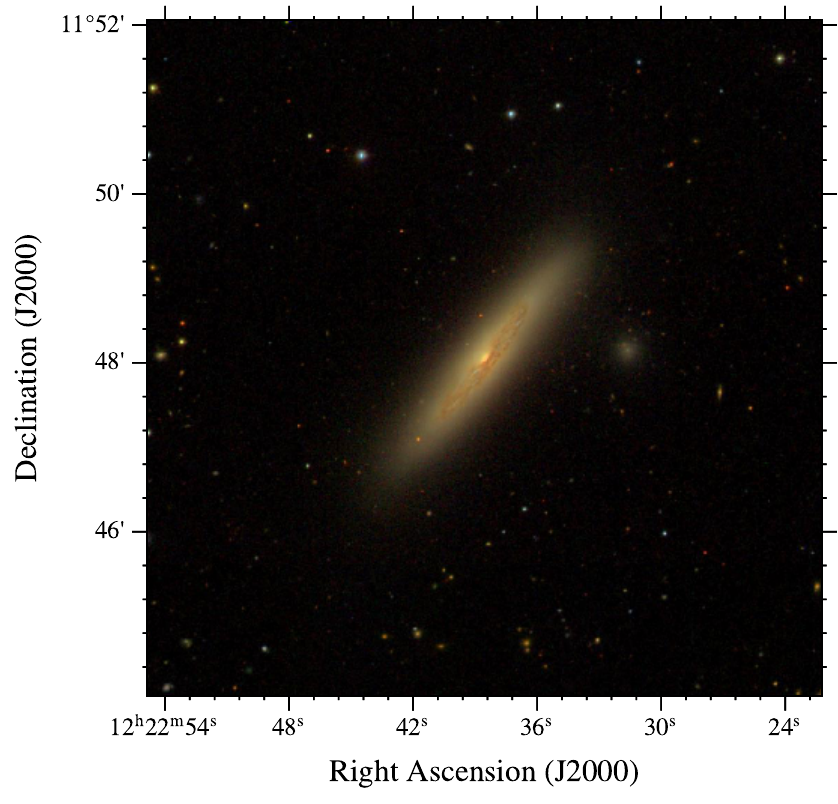}
\includegraphics[scale=0.255, angle=0]{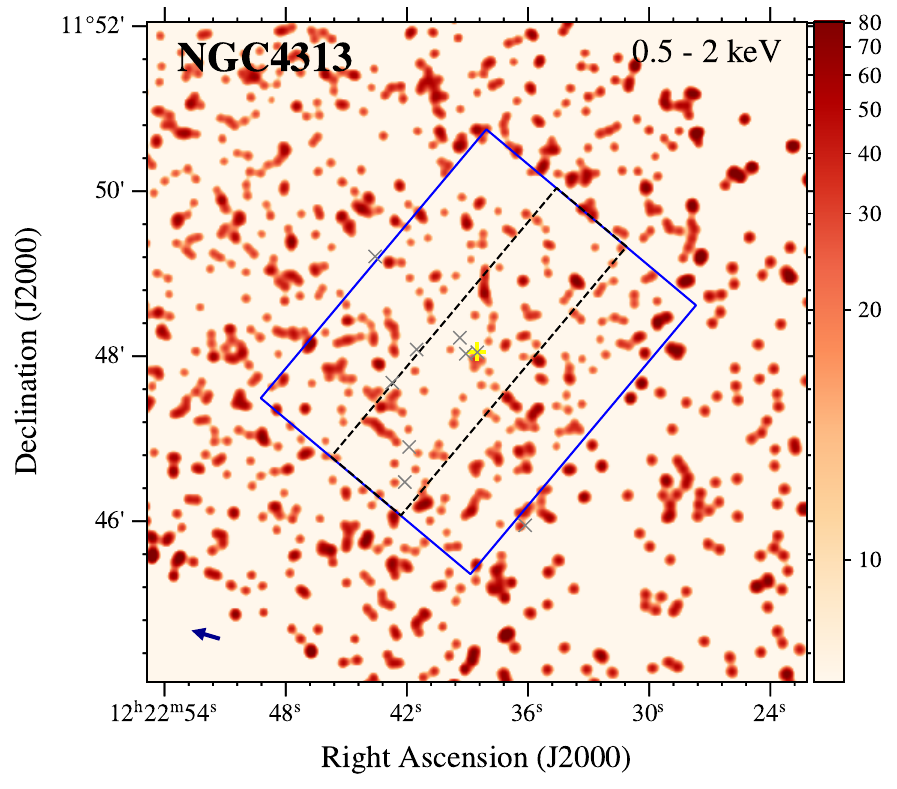}
\includegraphics[scale=0.25, angle=0]{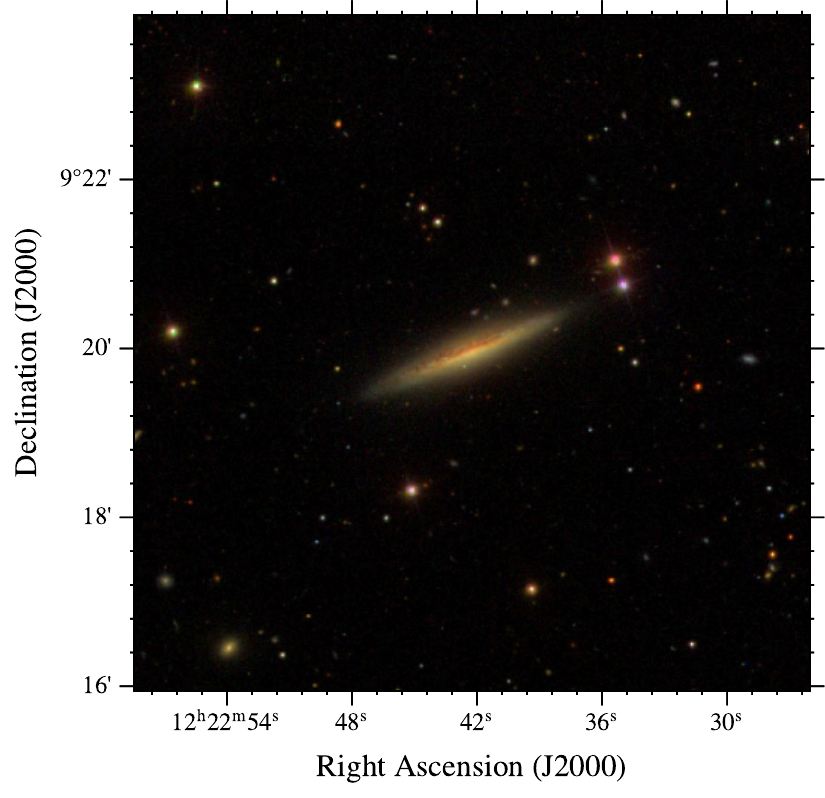}
\includegraphics[scale=0.255, angle=0]{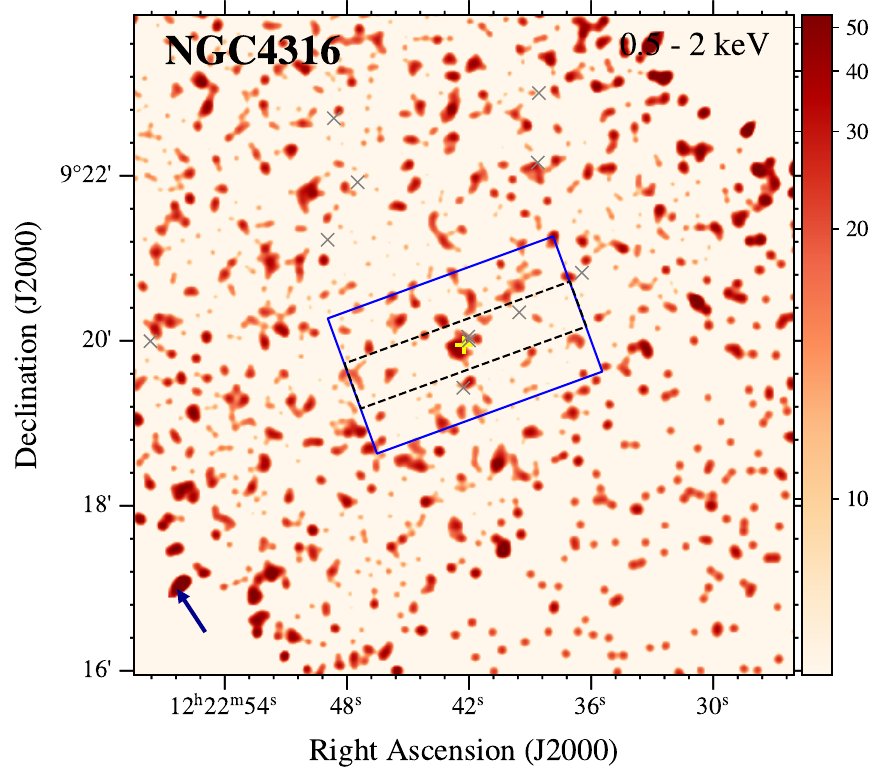}
}
\vspace{-0.35cm}
\qquad 
\subfloat{
\includegraphics[scale=0.25, angle=0]{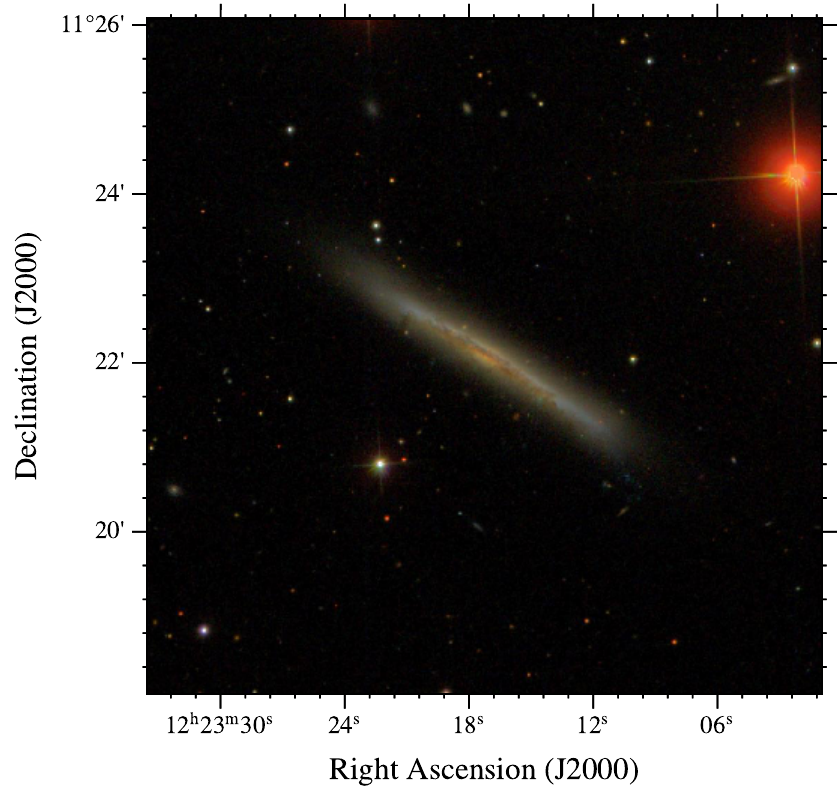}
\includegraphics[scale=0.255, angle=0]{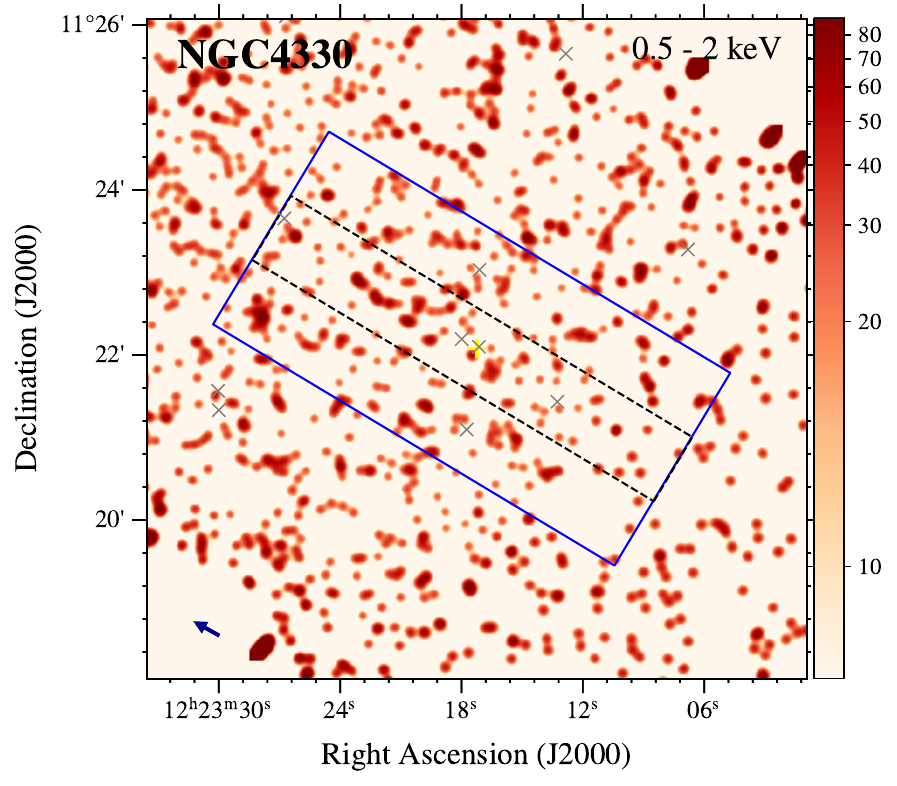}
\includegraphics[scale=0.25, angle=0]{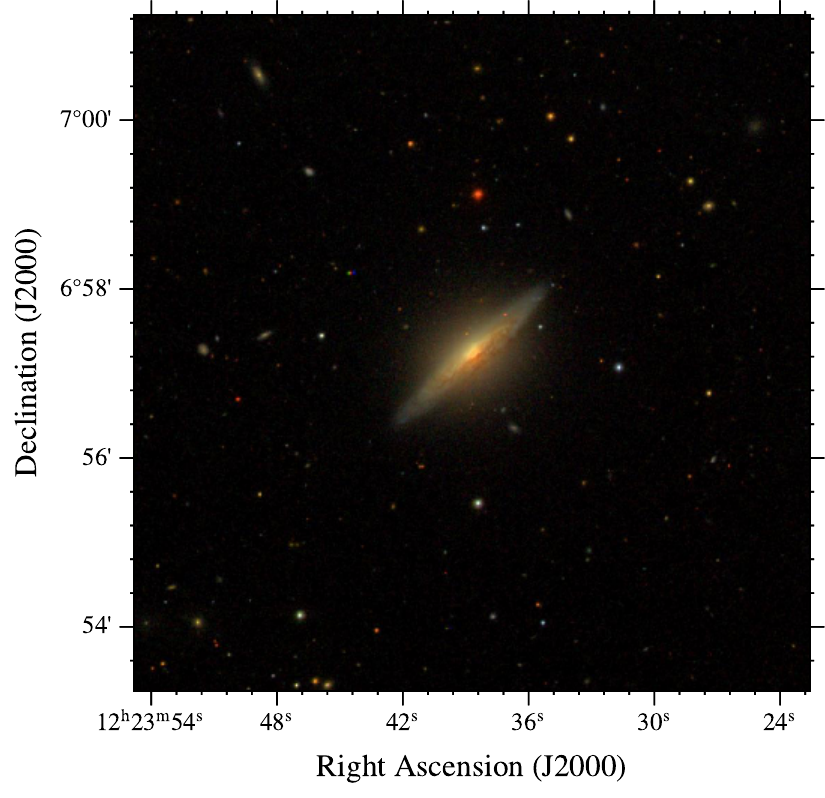}
\includegraphics[scale=0.255, angle=0]{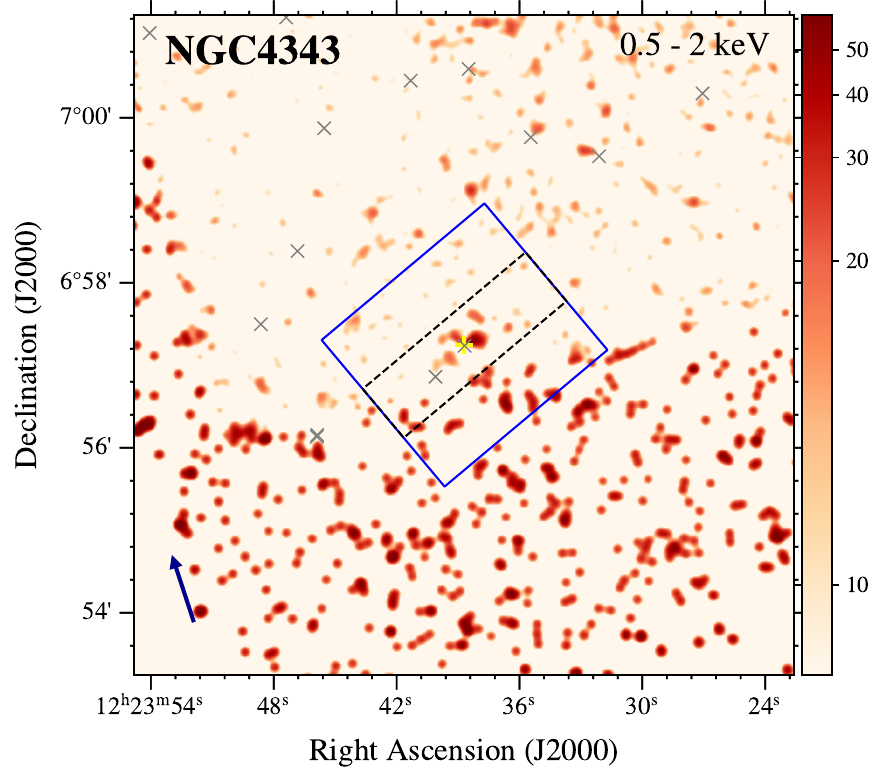}
}
\vspace{-0.35cm}
\qquad 
\subfloat{
\includegraphics[scale=0.25, angle=0]{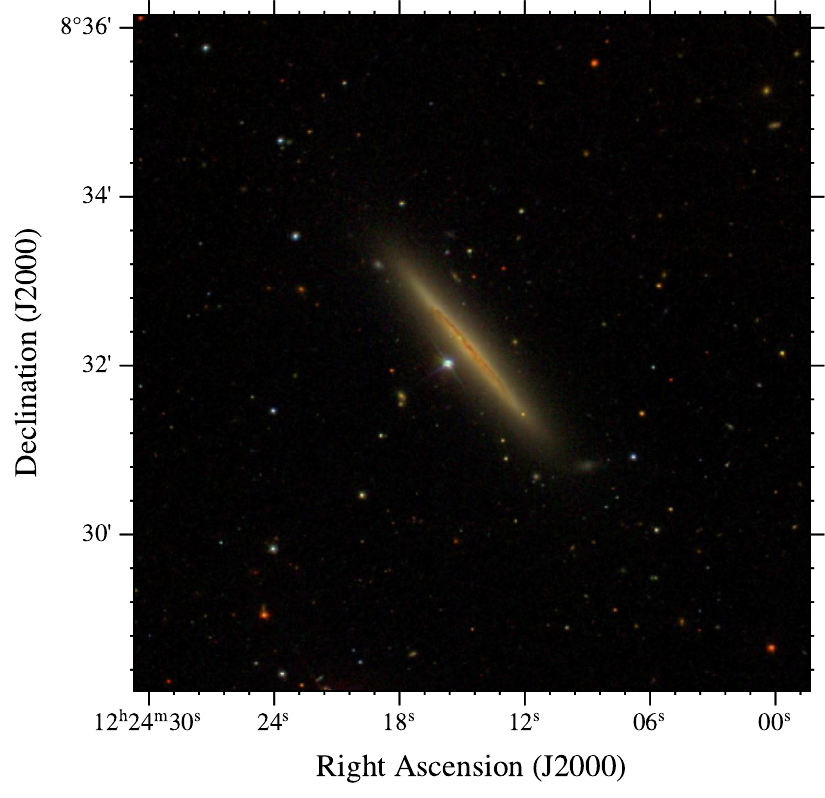}
\includegraphics[scale=0.255, angle=0]{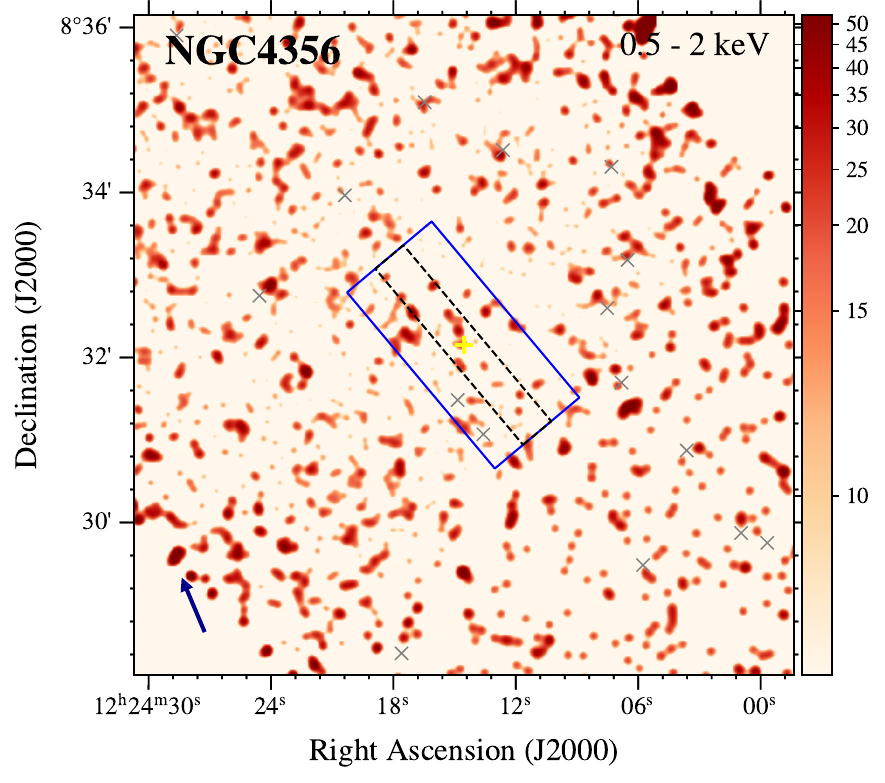}
\includegraphics[scale=0.25, angle=0]{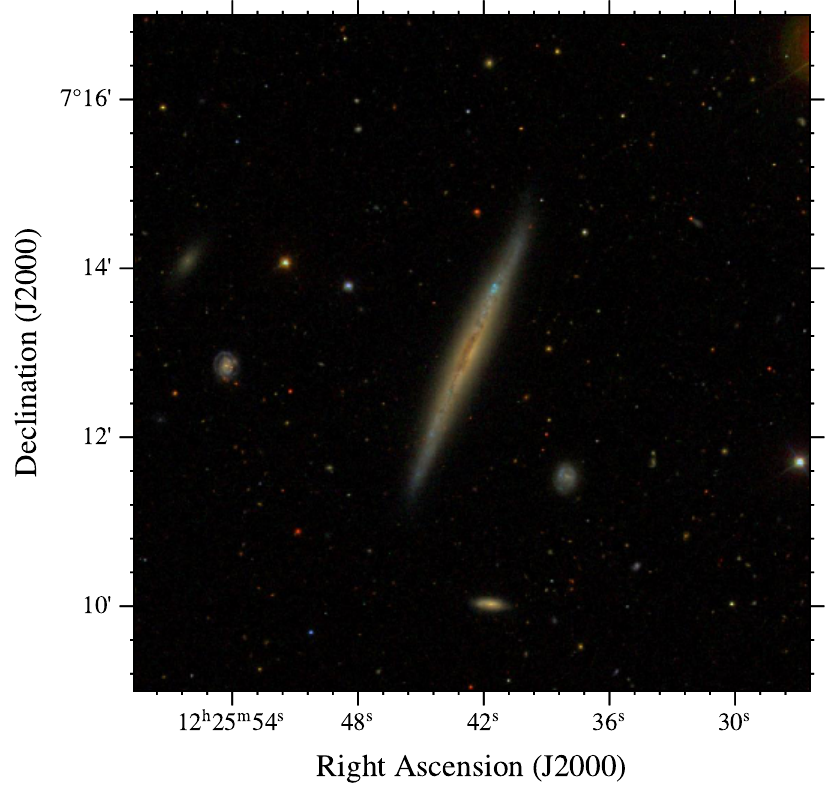}
\includegraphics[scale=0.255, angle=0]{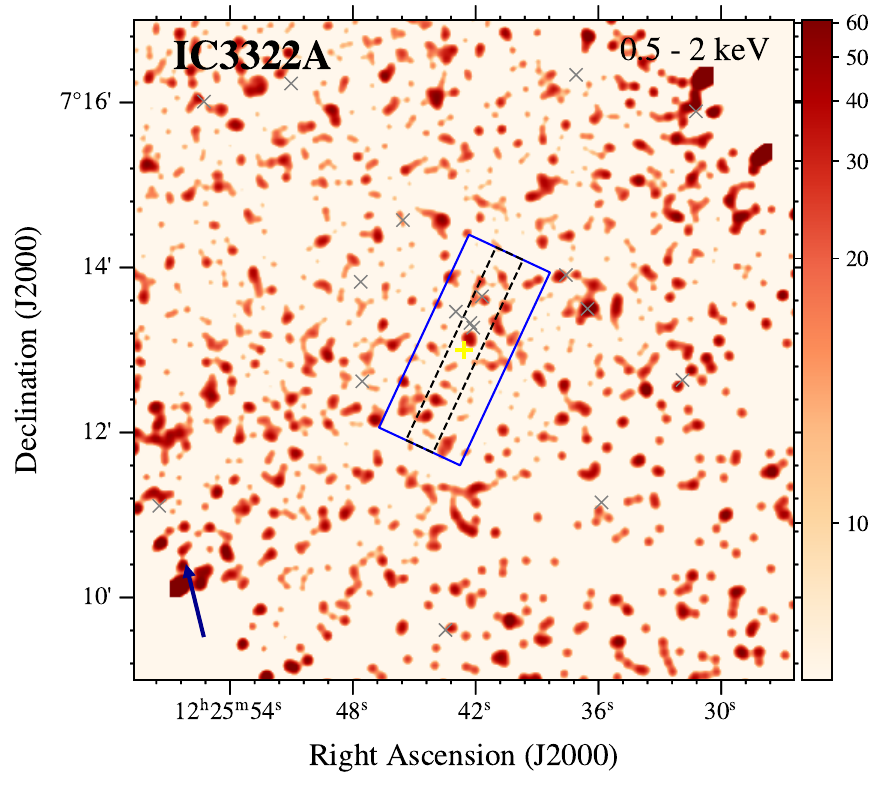}
}
\caption{Similar to Figure \ref{fig:image}, but for the rest of Virgo edge-on LTGs without a significant detection of X-ray hot gas corona emission. The left and right panels of each galaxy show the SDSS $gri$-color composite image and the X-ray flux map in 0.5--2 keV band, respectively. The images have a size of $8' \times 8'$.
}
\label{fig:nondetection}
\end{figure*} 

\begin{figure*}
  \ContinuedFloat 
\centering
\subfloat{
\includegraphics[scale=0.25, angle=0]{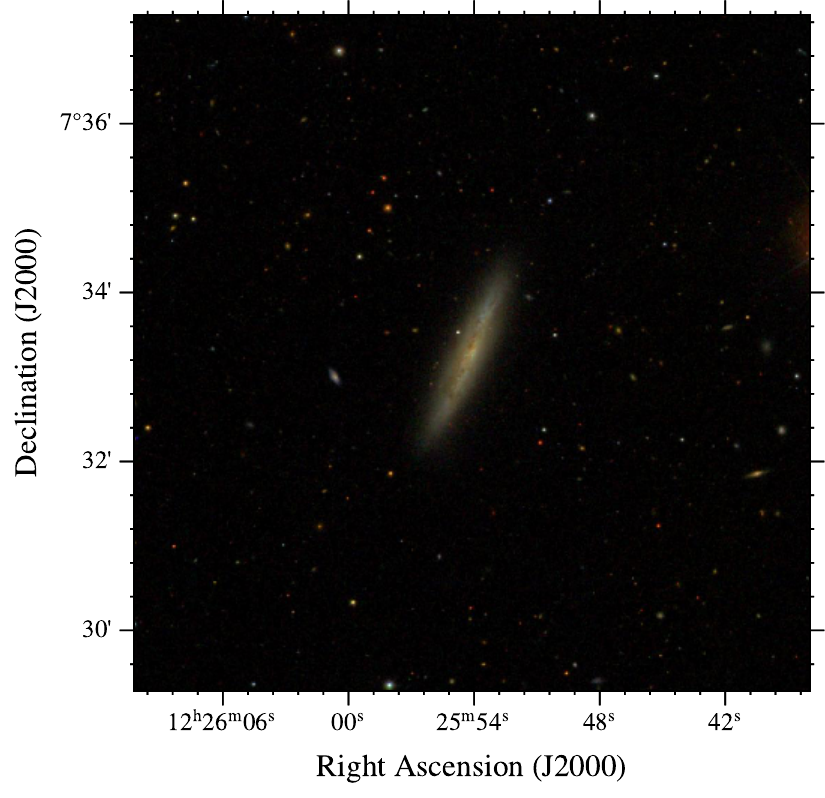}
\includegraphics[scale=0.255, angle=0]{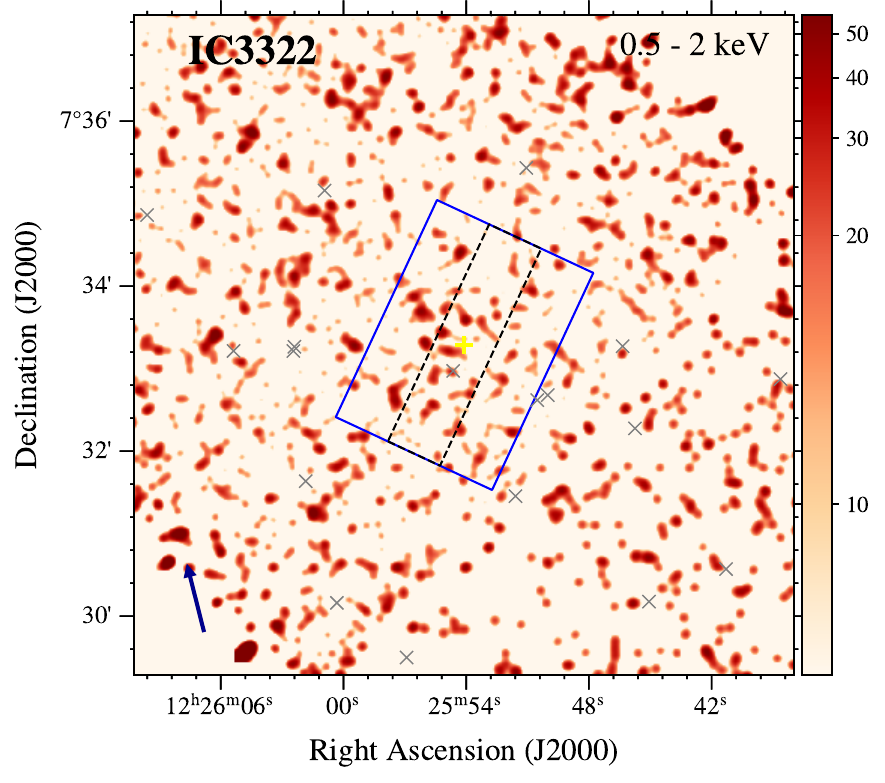}
\includegraphics[scale=0.25, angle=0]{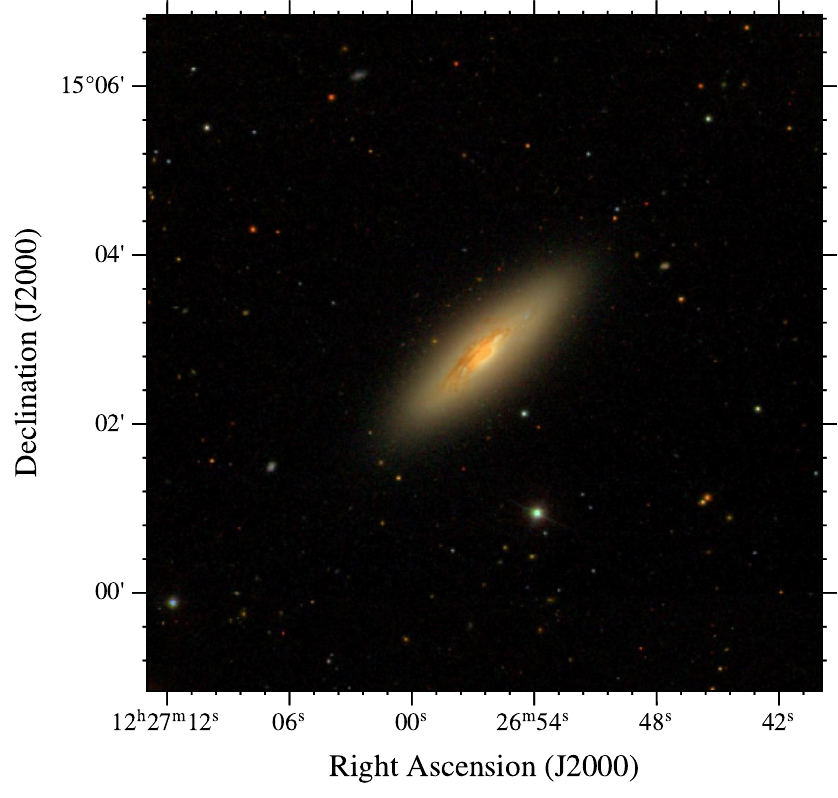}
\includegraphics[scale=0.255, angle=0]{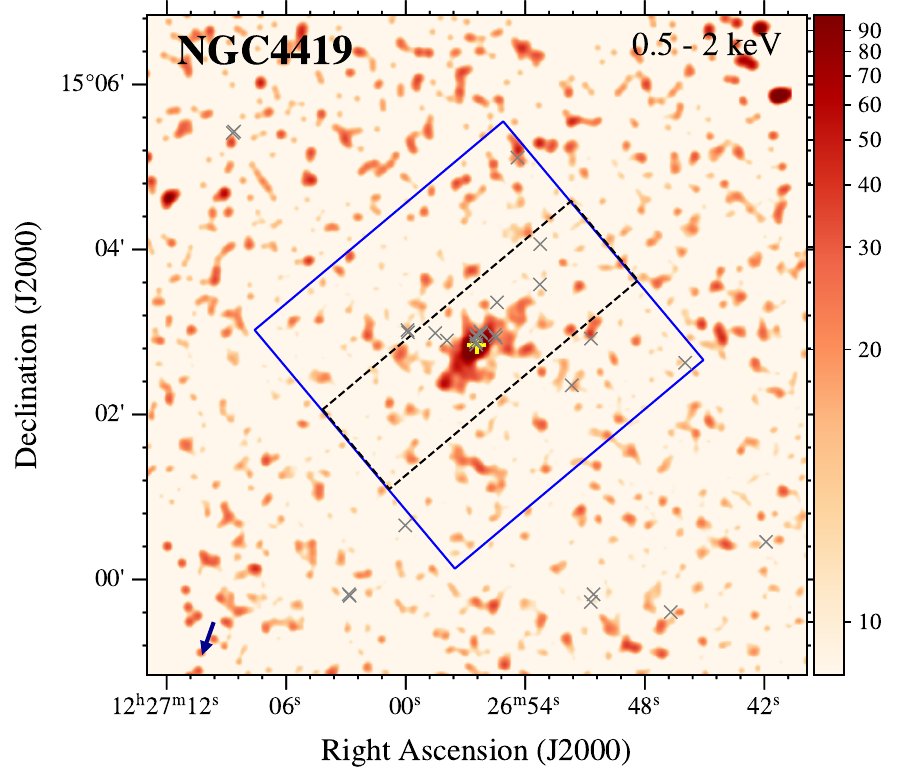}
}
\vspace{-0.35cm}
\qquad 
\subfloat{
\includegraphics[scale=0.25, angle=0]{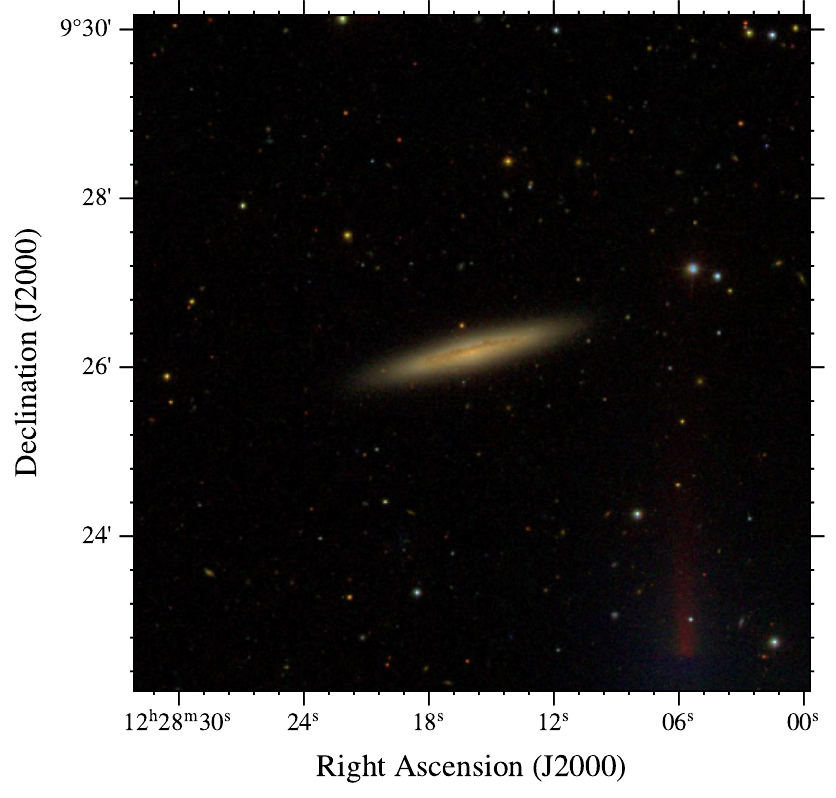}
\includegraphics[scale=0.255, angle=0]{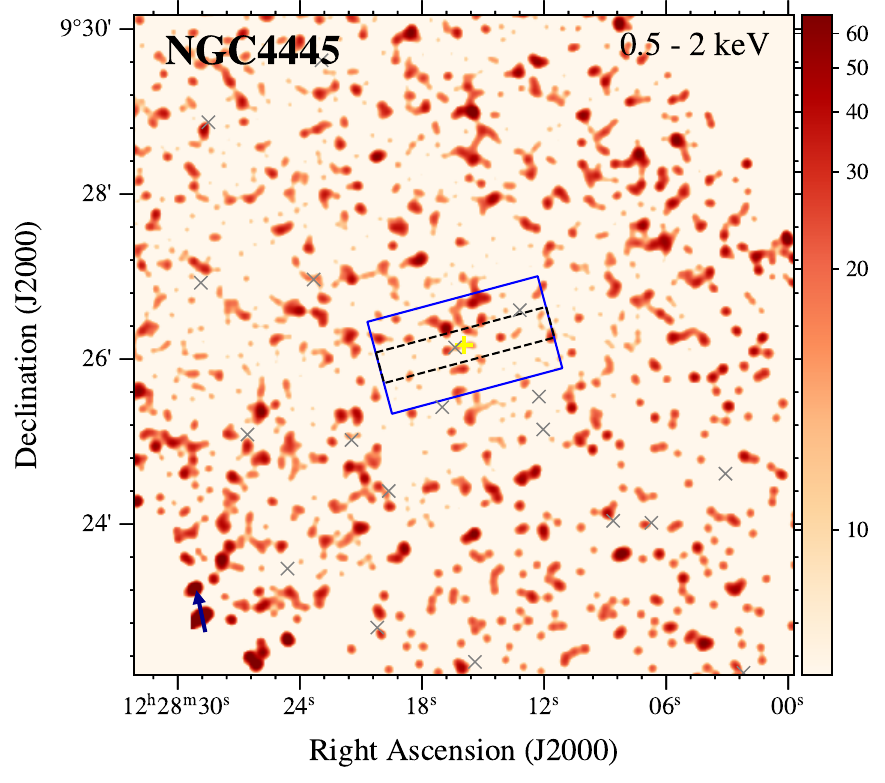}
\includegraphics[scale=0.25, angle=0]{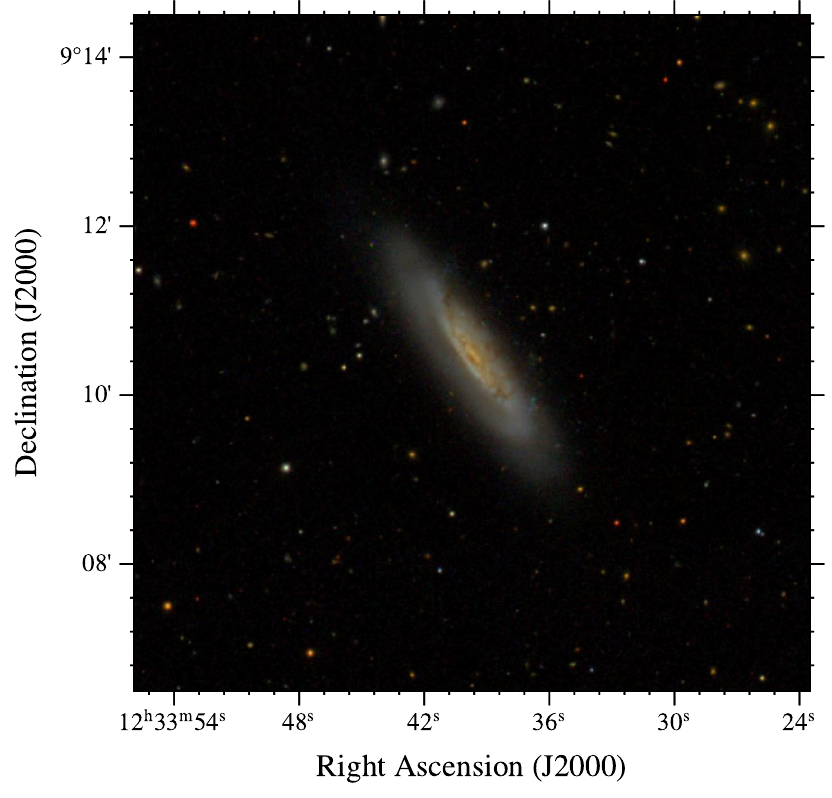}
\includegraphics[scale=0.255, angle=0]{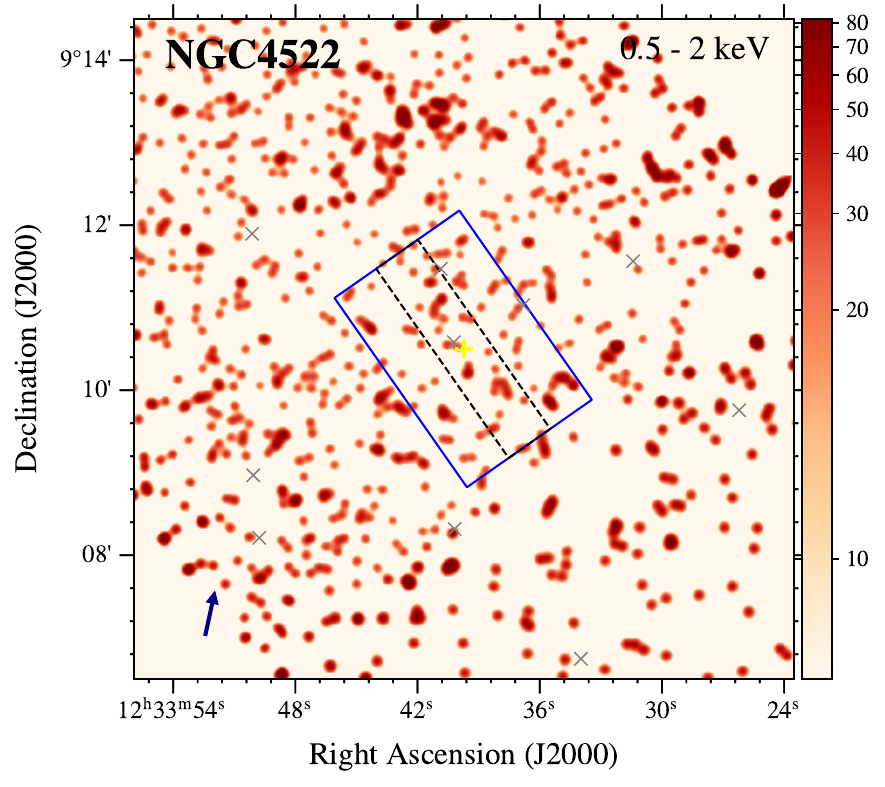}
}
\vspace{-0.35cm}
\qquad 
\subfloat{
\includegraphics[scale=0.25, angle=0]{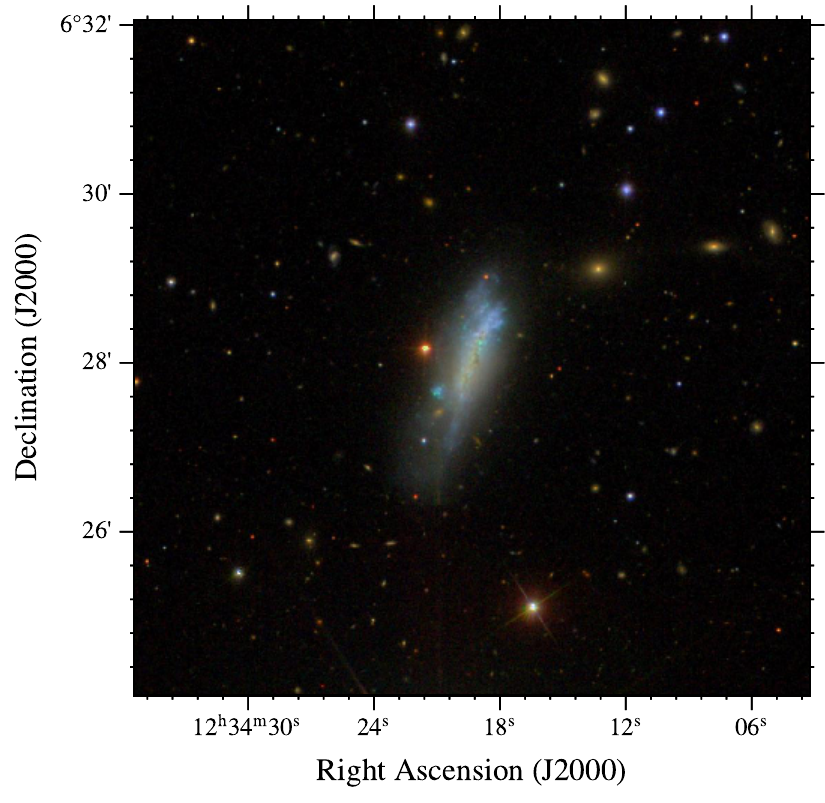}
\includegraphics[scale=0.255, angle=0]{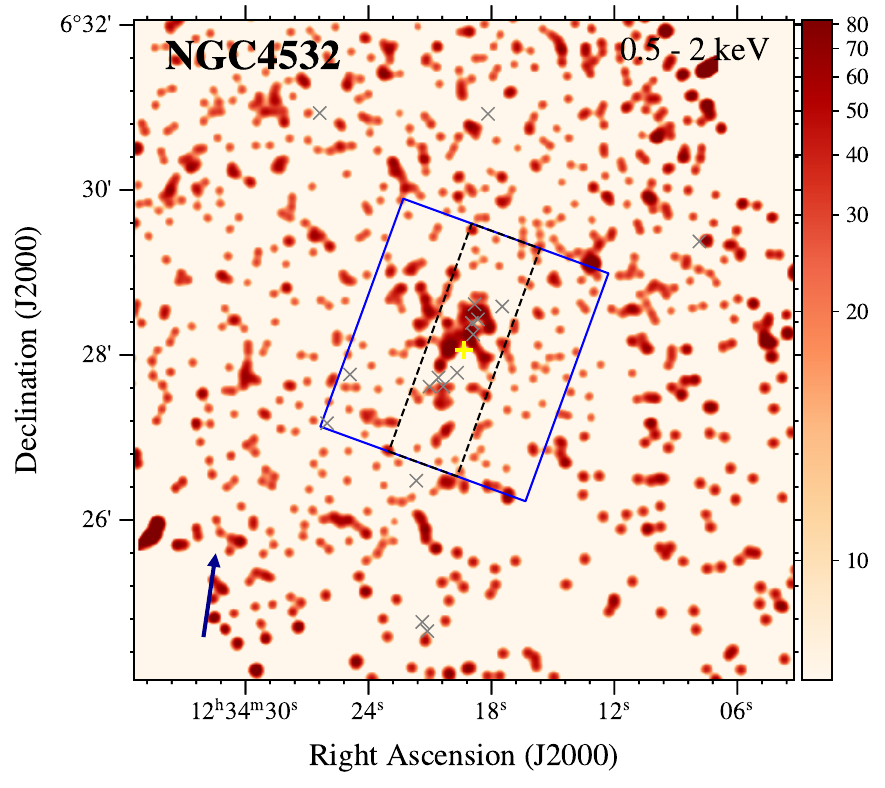}
\includegraphics[scale=0.25, angle=0]{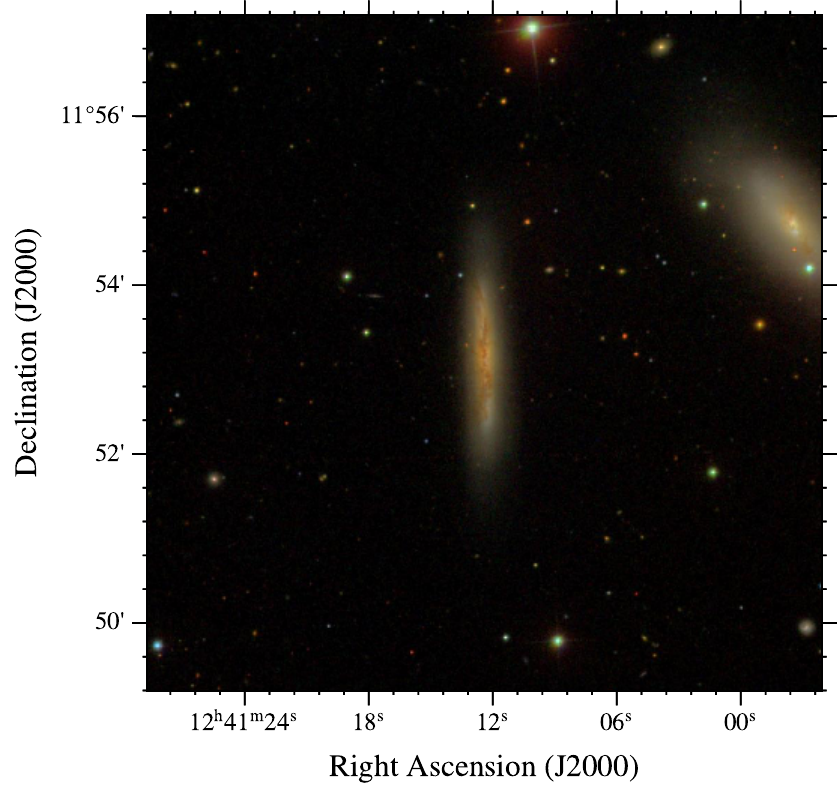}
\includegraphics[scale=0.255, angle=0]{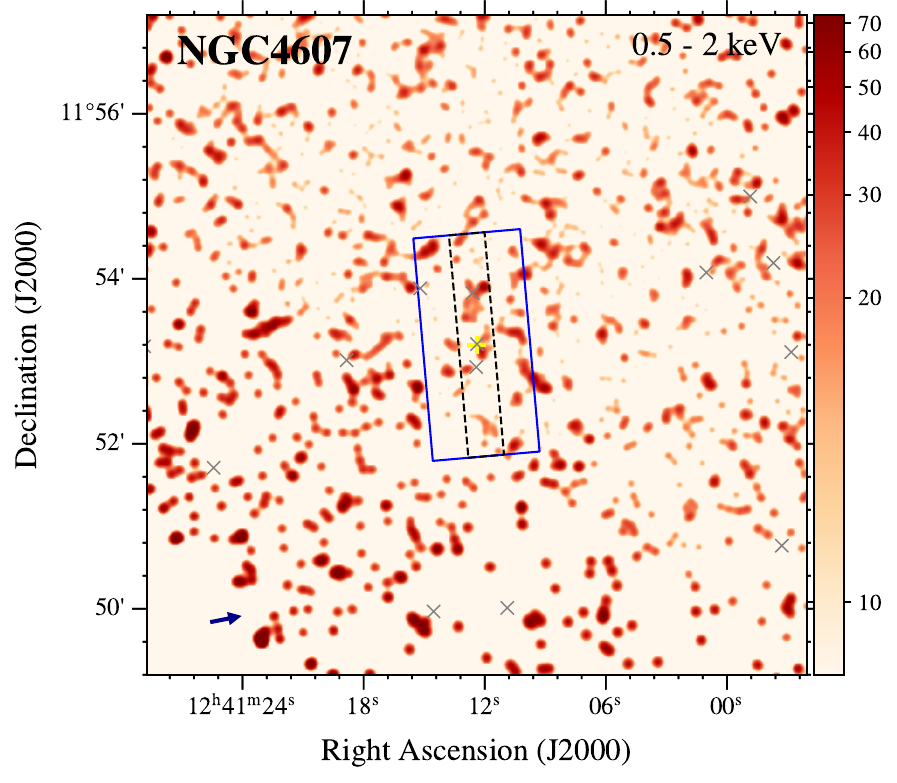}
}
\caption{--continued}
\end{figure*}

\bibliography{VirgoeLTG}

\begin{thebibliography}{}
\expandafter\ifx\csname natexlab\endcsname\relax\def\natexlab#1{#1}\fi
\providecommand{\url}[1]{\href{#1}{#1}}
\providecommand{\dodoi}[1]{doi:~\href{http://doi.org/#1}{\nolinkurl{#1}}}
\providecommand{\doeprint}[1]{\href{http://ascl.net/#1}{\nolinkurl{http://ascl.net/#1}}}
\providecommand{\doarXiv}[1]{\href{https://arxiv.org/abs/#1}{\nolinkurl{https://arxiv.org/abs/#1}}}

\bibitem[{{Anderson} \& {Bregman}(2011)}]{Anderson2011}
{Anderson}, M.~E., \& {Bregman}, J.~N. 2011, \apj, 737, 22, \dodoi{10.1088/0004-637X/737/1/22}

\bibitem[{{Anderson} {et~al.}(2016){Anderson}, {Churazov}, \& {Bregman}}]{Anderson2016}
{Anderson}, M.~E., {Churazov}, E., \& {Bregman}, J.~N. 2016, \mnras, 455, 227, \dodoi{10.1093/mnras/stv2314}

\bibitem[{{Bekki}(2009)}]{Bekki2009}
{Bekki}, K. 2009, \mnras, 399, 2221, \dodoi{10.1111/j.1365-2966.2009.15431.x}

\bibitem[{{Benson}(2010)}]{Benson2010}
{Benson}, A.~J. 2010, \physrep, 495, 33, \dodoi{10.1016/j.physrep.2010.06.001}

\bibitem[{{Benson} {et~al.}(2000){Benson}, {Bower}, {Frenk}, \& {White}}]{Benson2000}
{Benson}, A.~J., {Bower}, R.~G., {Frenk}, C.~S., \& {White}, S.~D.~M. 2000, \mnras, 314, 557, \dodoi{10.1046/j.1365-8711.2000.03362.x}

\bibitem[{{Biller} {et~al.}(2004){Biller}, {Jones}, {Forman}, {Kraft}, \& {Ensslin}}]{Biller2004}
{Biller}, B.~A., {Jones}, C., {Forman}, W.~R., {Kraft}, R., \& {Ensslin}, T. 2004, \apj, 613, 238, \dodoi{10.1086/423020}

\bibitem[{{Bogd{\'a}n} {et~al.}(2013{\natexlab{a}}){Bogd{\'a}n}, {Forman}, {Kraft}, \& {Jones}}]{Bogdan2013a}
{Bogd{\'a}n}, {\'A}., {Forman}, W.~R., {Kraft}, R.~P., \& {Jones}, C. 2013{\natexlab{a}}, \apj, 772, 98, \dodoi{10.1088/0004-637X/772/2/98}

\bibitem[{{Bogd{\'a}n} {et~al.}(2013{\natexlab{b}}){Bogd{\'a}n}, {Forman}, {Vogelsberger}, {Bourdin}, {Sijacki}, {Mazzotta}, {Kraft}, {Jones}, {Gilfanov}, {Churazov}, \& {David}}]{Bogdan2013b}
{Bogd{\'a}n}, {\'A}., {Forman}, W.~R., {Vogelsberger}, M., {et~al.} 2013{\natexlab{b}}, \apj, 772, 97, \dodoi{10.1088/0004-637X/772/2/97}

\bibitem[{{Bogd{\'a}n} {et~al.}(2015){Bogd{\'a}n}, {Vogelsberger}, {Kraft}, {Hernquist}, {Gilfanov}, {Torrey}, {Churazov}, {Genel}, {Forman}, {Murray}, {Vikhlinin}, {Jones}, \& {B{\"o}hringer}}]{Bogdan2015}
{Bogd{\'a}n}, {\'A}., {Vogelsberger}, M., {Kraft}, R.~P., {et~al.} 2015, \apj, 804, 72, \dodoi{10.1088/0004-637X/804/1/72}

\bibitem[{{B{\"o}hringer} {et~al.}(1994){B{\"o}hringer}, {Briel}, {Schwarz}, {Voges}, {Hartner}, \& {Tr{\"u}mper}}]{Bohringer1994}
{B{\"o}hringer}, H., {Briel}, U.~G., {Schwarz}, R.~A., {et~al.} 1994, \nat, 368, 828, \dodoi{10.1038/368828a0}

\bibitem[{{Boselli} {et~al.}(2022){Boselli}, {Fossati}, \& {Sun}}]{Boselli2022}
{Boselli}, A., {Fossati}, M., \& {Sun}, M. 2022, \aapr, 30, 3, \dodoi{10.1007/s00159-022-00140-3}

\bibitem[{{Bregman} {et~al.}(2023){Bregman}, {Cen}, {Chen}, {Cui}, {Fang}, {Guo}, {Hodges-Kluck}, {Huang}, {Ho}, {Ji}, {Ji}, {Kang}, {Lai}, {Li}, {Li}, {Li}, {Li}, {Li}, {Li}, {Liang}, {Liu}, {Liu}, {Lu}, {Mao}, {Ponti}, {Qu}, {Shan}, {Shao}, {Shi}, {Shu}, {Sun}, {Sun}, {Tong}, {Wang}, {Wang}, {Wang}, {Wang}, {Wang}, {Wang}, {Wang}, {Xu}, {Xu}, {Xu}, {Xu}, {Xu}, {Xue}, {Yang}, {Yuan}, {Zhang}, {Zhang}, {Zhang}, {Zhao}, {Zhou}, \& {Zhou}}]{Bregman2023}
{Bregman}, J., {Cen}, R., {Chen}, Y., {et~al.} 2023, Science China Physics, Mechanics, and Astronomy, 66, 299513, \dodoi{10.1007/s11433-023-2149-y}

\bibitem[{{Bregman}(1980)}]{Bregman1980}
{Bregman}, J.~N. 1980, \apj, 236, 577, \dodoi{10.1086/157776}

\bibitem[{{Brown} \& {Bregman}(2000)}]{Brown2000}
{Brown}, B.~A., \& {Bregman}, J.~N. 2000, \apj, 539, 592, \dodoi{10.1086/309240}

\bibitem[{{Chadayammuri} {et~al.}(2022){Chadayammuri}, {Bogd{\'a}n}, {Oppenheimer}, {Kraft}, {Forman}, \& {Jones}}]{Chadayammuri2022}
{Chadayammuri}, U., {Bogd{\'a}n}, {\'A}., {Oppenheimer}, B.~D., {et~al.} 2022, \apjl, 936, L15, \dodoi{10.3847/2041-8213/ac8936}

\bibitem[{{Chevalier} \& {Clegg}(1985)}]{Chevalier1985}
{Chevalier}, R.~A., \& {Clegg}, A.~W. 1985, \nat, 317, 44, \dodoi{10.1038/317044a0}

\bibitem[{{Cluver} {et~al.}(2017){Cluver}, {Jarrett}, {Dale}, {Smith}, {August}, \& {Brown}}]{Cluver2017}
{Cluver}, M.~E., {Jarrett}, T.~H., {Dale}, D.~A., {et~al.} 2017, \apj, 850, 68, \dodoi{10.3847/1538-4357/aa92c7}

\bibitem[{{Cluver} {et~al.}(2014){Cluver}, {Jarrett}, {Hopkins}, {Driver}, {Liske}, {Gunawardhana}, {Taylor}, {Robotham}, {Alpaslan}, {Baldry}, {Brown}, {Peacock}, {Popescu}, {Tuffs}, {Bauer}, {Bland-Hawthorn}, {Colless}, {Holwerda}, {Lara-L{\'o}pez}, {Leschinski}, {L{\'o}pez-S{\'a}nchez}, {Norberg}, {Owers}, {Wang}, \& {Wilkins}}]{Cluver2014}
{Cluver}, M.~E., {Jarrett}, T.~H., {Hopkins}, A.~M., {et~al.} 2014, \apj, 782, 90, \dodoi{10.1088/0004-637X/782/2/90}

\bibitem[{{Colbert} {et~al.}(1998){Colbert}, {Baum}, {O'Dea}, \& {Veilleux}}]{Colbert1998}
{Colbert}, E. J.~M., {Baum}, S.~A., {O'Dea}, C.~P., \& {Veilleux}, S. 1998, \apj, 496, 786, \dodoi{10.1086/305417}

\bibitem[{{Comparat} {et~al.}(2022){Comparat}, {Truong}, {Merloni}, {Pillepich}, {Ponti}, {Driver}, {Bellstedt}, {Liske}, {Aird}, {Br{\"u}ggen}, {Bulbul}, {Davies}, {Villalba}, {Georgakakis}, {Haberl}, {Liu}, {Maitra}, {Nandra}, {Popesso}, {Predehl}, {Robotham}, {Salvato}, {Thorne}, \& {Zhang}}]{Comparat2022}
{Comparat}, J., {Truong}, N., {Merloni}, A., {et~al.} 2022, \aap, 666, A156, \dodoi{10.1051/0004-6361/202243101}

\bibitem[{{Cox}(2005)}]{Cox2005}
{Cox}, D.~P. 2005, \araa, 43, 337, \dodoi{10.1146/annurev.astro.43.072103.150615}

\bibitem[{{Cox} \& {Smith}(1974)}]{Cox1974}
{Cox}, D.~P., \& {Smith}, B.~W. 1974, \apjl, 189, L105, \dodoi{10.1086/181476}

\bibitem[{{Crain} {et~al.}(2010){Crain}, {McCarthy}, {Frenk}, {Theuns}, \& {Schaye}}]{Crain2010}
{Crain}, R.~A., {McCarthy}, I.~G., {Frenk}, C.~S., {Theuns}, T., \& {Schaye}, J. 2010, \mnras, 407, 1403, \dodoi{10.1111/j.1365-2966.2010.16985.x}

\bibitem[{{Dai} {et~al.}(2012){Dai}, {Anderson}, {Bregman}, \& {Miller}}]{Dai2012}
{Dai}, X., {Anderson}, M.~E., {Bregman}, J.~N., \& {Miller}, J.~M. 2012, \apj, 755, 107, \dodoi{10.1088/0004-637X/755/2/107}

\bibitem[{{Ehlert} {et~al.}(2013){Ehlert}, {Werner}, {Simionescu}, {Allen}, {Kenney}, {Million}, \& {Finoguenov}}]{Ehlert2013}
{Ehlert}, S., {Werner}, N., {Simionescu}, A., {et~al.} 2013, \mnras, 430, 2401, \dodoi{10.1093/mnras/stt060}

\bibitem[{{Foster} \& {Heuer}(2020)}]{Foster2020}
{Foster}, A.~R., \& {Heuer}, K. 2020, Atoms, 8, 49, \dodoi{10.3390/atoms8030049}

\bibitem[{{Grimes} {et~al.}(2005){Grimes}, {Heckman}, {Strickland}, \& {Ptak}}]{Grimes2005}
{Grimes}, J.~P., {Heckman}, T., {Strickland}, D., \& {Ptak}, A. 2005, \apj, 628, 187, \dodoi{10.1086/430692}

\bibitem[{{Gunn} \& {Gott}(1972)}]{Gunn1972}
{Gunn}, J.~E., \& {Gott}, J.~Richard, I. 1972, \apj, 176, 1, \dodoi{10.1086/151605}

\bibitem[{{Hou} {et~al.}(2021){Hou}, {Li}, {Jones}, {Forman}, \& {Su}}]{Hou2021}
{Hou}, M., {Li}, Z., {Jones}, C., {Forman}, W., \& {Su}, Y. 2021, \apj, 919, 141, \dodoi{10.3847/1538-4357/ac1344}

\bibitem[{{Hou} {et~al.}(2017){Hou}, {Li}, {Peng}, \& {Liu}}]{Hou2017}
{Hou}, M., {Li}, Z., {Peng}, E.~W., \& {Liu}, C. 2017, \apj, 846, 126, \dodoi{10.3847/1538-4357/aa8635}

\bibitem[{{Hubble}(1926)}]{Hubble1926}
{Hubble}, E.~P. 1926, \apj, 64, 321, \dodoi{10.1086/143018}

\bibitem[{{Huertas-Company} {et~al.}(2019){Huertas-Company}, {Rodriguez-Gomez}, {Nelson}, {Pillepich}, {Bottrell}, {Bernardi}, {Dom{\'\i}nguez-S{\'a}nchez}, {Genel}, {Pakmor}, {Snyder}, \& {Vogelsberger}}]{Huertas-Company2019}
{Huertas-Company}, M., {Rodriguez-Gomez}, V., {Nelson}, D., {et~al.} 2019, \mnras, 489, 1859, \dodoi{10.1093/mnras/stz2191}

\bibitem[{{Iwasawa} {et~al.}(2003){Iwasawa}, {Wilson}, {Fabian}, \& {Young}}]{Iwasawa2003}
{Iwasawa}, K., {Wilson}, A.~S., {Fabian}, A.~C., \& {Young}, A.~J. 2003, \mnras, 345, 369, \dodoi{10.1046/j.1365-8711.2003.06857.x}

\bibitem[{{Jarrett} {et~al.}(2000){Jarrett}, {Chester}, {Cutri}, {Schneider}, {Skrutskie}, \& {Huchra}}]{Jarrett2000}
{Jarrett}, T.~H., {Chester}, T., {Cutri}, R., {et~al.} 2000, \aj, 119, 2498, \dodoi{10.1086/301330}

\bibitem[{{Jiang} {et~al.}(2019){Jiang}, {Li}, {Fang}, \& {Wang}}]{Jiang2019}
{Jiang}, X., {Li}, J., {Fang}, T., \& {Wang}, Q.~D. 2019, \apj, 885, 38, \dodoi{10.3847/1538-4357/ab44b4}

\bibitem[{{Jin} {et~al.}(2019){Jin}, {Hou}, {Zhu}, \& {Li}}]{Jin2019}
{Jin}, X., {Hou}, M., {Zhu}, Z., \& {Li}, Z. 2019, \apj, 876, 53, \dodoi{10.3847/1538-4357/ab064f}

\bibitem[{{Kim} {et~al.}(2014){Kim}, {Rey}, {Jerjen}, {Lisker}, {Sung}, {Lee}, {Chung}, {Pak}, {Yi}, \& {Lee}}]{Kim2014}
{Kim}, S., {Rey}, S.-C., {Jerjen}, H., {et~al.} 2014, \apjs, 215, 22, \dodoi{10.1088/0067-0049/215/2/22}

\bibitem[{{Kraft} {et~al.}(2022){Kraft}, {Markevitch}, {Kilbourne}, {Adams}, {Akamatsu}, {Ayromlou}, {Bandler}, {Barbera}, {Bennett}, {Bhardwaj}, {Biffi}, {Bodewits}, {Bogdan}, {Bonamente}, {Borgani}, {Branduardi-Raymont}, {Bregman}, {Burchett}, {Cann}, {Carter}, {Chakraborty}, {Churazov}, {Crain}, {Cumbee}, {Dave}, {DiPirro}, {Dolag}, {Bertrand Doriese}, {Drake}, {Dunn}, {Eckart}, {Eckert}, {Ettori}, {Forman}, {Galeazzi}, {Gall}, {Gatuzz}, {Hell}, {Hodges-Kluck}, {Jackman}, {Jahromi}, {Jennings}, {Jones}, {Kaaret}, {Kavanagh}, {Kelley}, {Khabibullin}, {Kim}, {Koutroumpa}, {Kovacs}, {Kuntz}, {Lau}, {Lee}, {Leutenegger}, {Lin}, {Lisse}, {Lo Cicero}, {Lovisari}, {McCammon}, {McEntee}, {Mernier}, {Miller}, {Nagai}, {Negro}, {Nelson}, {Ness}, {Nulsen}, {Ogorzalek}, {Oppenheimer}, {Oskinova}, {Patnaude}, {Pfeifle}, {Pillepich}, {Plucinsky}, {Pooley}, {Porter}, {Randall}, {Rasia}, {Raymond}, {Ruszkowski}, {Sakai}, {Sarkar}, {Sasaki}, {Sato}, {Schellenberger}, {Schaye}, {Simionescu}, {Smith}, {Steiner}, {Stern},
  {Su}, {Sun}, {Tremblay}, {Truong}, {Tutt}, {Ursino}, {Veilleux}, {Vikhlinin}, {Vladutescu-Zopp}, {Vogelsberger}, {Walker}, {Weaver}, {Weigt}, {Werk}, {Werner}, {Wolk}, {Zhang}, {Zhang}, {Zhuravleva}, \& {ZuHone}}]{Kraft2022}
{Kraft}, R., {Markevitch}, M., {Kilbourne}, C., {et~al.} 2022, arXiv e-prints, arXiv:2211.09827, \dodoi{10.48550/arXiv.2211.09827}

\bibitem[{{Kraft} {et~al.}(2011){Kraft}, {Forman}, {Jones}, {Nulsen}, {Hardcastle}, {Raychaudhury}, {Evans}, {Sivakoff}, \& {Sarazin}}]{Kraft2011}
{Kraft}, R.~P., {Forman}, W.~R., {Jones}, C., {et~al.} 2011, \apj, 727, 41, \dodoi{10.1088/0004-637X/727/1/41}

\bibitem[{{Kraft} {et~al.}(2017){Kraft}, {Roediger}, {Machacek}, {Forman}, {Nulsen}, {Jones}, {Churazov}, {Randall}, {Su}, \& {Sheardown}}]{Kraft2017}
{Kraft}, R.~P., {Roediger}, E., {Machacek}, M., {et~al.} 2017, \apj, 848, 27, \dodoi{10.3847/1538-4357/aa8a6e}

\bibitem[{{Li} {et~al.}(2016){Li}, {Bregman}, {Wang}, {Crain}, \& {Anderson}}]{Li2016}
{Li}, J.-T., {Bregman}, J.~N., {Wang}, Q.~D., {Crain}, R.~A., \& {Anderson}, M.~E. 2016, \apj, 830, 134, \dodoi{10.3847/0004-637X/830/2/134}

\bibitem[{{Li} {et~al.}(2018){Li}, {Bregman}, {Wang}, {Crain}, \& {Anderson}}]{Li2018}
---. 2018, \apjl, 855, L24, \dodoi{10.3847/2041-8213/aab2af}

\bibitem[{{Li} {et~al.}(2017){Li}, {Bregman}, {Wang}, {Crain}, {Anderson}, \& {Zhang}}]{LiJT2017}
{Li}, J.-T., {Bregman}, J.~N., {Wang}, Q.~D., {et~al.} 2017, \apjs, 233, 20, \dodoi{10.3847/1538-4365/aa96fc}

\bibitem[{{Li} {et~al.}(2014){Li}, {Crain}, \& {Wang}}]{Li2014}
{Li}, J.-T., {Crain}, R.~A., \& {Wang}, Q.~D. 2014, \mnras, 440, 859, \dodoi{10.1093/mnras/stu329}

\bibitem[{{Li} {et~al.}(2008){Li}, {Li}, {Wang}, {Irwin}, \& {Rossa}}]{Li2008}
{Li}, J.-T., {Li}, Z., {Wang}, Q.~D., {Irwin}, J.~A., \& {Rossa}, J. 2008, \mnras, 390, 59, \dodoi{10.1111/j.1365-2966.2008.13749.x}

\bibitem[{{Li} \& {Wang}(2013{\natexlab{a}})}]{Li2013a}
{Li}, J.-T., \& {Wang}, Q.~D. 2013{\natexlab{a}}, \mnras, 428, 2085, \dodoi{10.1093/mnras/sts183}

\bibitem[{{Li} \& {Wang}(2013{\natexlab{b}})}]{Li2013b}
---. 2013{\natexlab{b}}, \mnras, 435, 3071, \dodoi{10.1093/mnras/stt1501}

\bibitem[{{Li} {et~al.}(2007){Li}, {Wang}, \& {Hameed}}]{Li2007}
{Li}, Z., {Wang}, Q.~D., \& {Hameed}, S. 2007, \mnras, 376, 960, \dodoi{10.1111/j.1365-2966.2007.11513.x}

\bibitem[{{Li} {et~al.}(2006){Li}, {Wang}, {Irwin}, \& {Chaves}}]{Li2006}
{Li}, Z., {Wang}, Q.~D., {Irwin}, J.~A., \& {Chaves}, T. 2006, \mnras, 371, 147, \dodoi{10.1111/j.1365-2966.2006.10682.x}

\bibitem[{{Li-M.} {et~al.}(2017){Li-M.}, {Bryan}, \& {Ostriker}}]{LiM2017}
{Li-M.}, M., {Bryan}, G.~L., \& {Ostriker}, J.~P. 2017, \apj, 841, 101, \dodoi{10.3847/1538-4357/aa7263}

\bibitem[{{Machacek} {et~al.}(2006){Machacek}, {Jones}, {Forman}, \& {Nulsen}}]{Machacek2006}
{Machacek}, M., {Jones}, C., {Forman}, W.~R., \& {Nulsen}, P. 2006, \apj, 644, 155, \dodoi{10.1086/503350}

\bibitem[{{Machacek} {et~al.}(2004){Machacek}, {Jones}, \& {Forman}}]{Machacek2004}
{Machacek}, M.~E., {Jones}, C., \& {Forman}, W.~R. 2004, \apj, 610, 183, \dodoi{10.1086/421448}

\bibitem[{{Matt} {et~al.}(1994){Matt}, {Piro}, {Antonelli}, {Fink}, {Meurs}, \& {Perola}}]{Matt1994}
{Matt}, G., {Piro}, L., {Antonelli}, L.~A., {et~al.} 1994, \aap, 292, L13

\bibitem[{{Mei} {et~al.}(2007){Mei}, {Blakeslee}, {C{\^o}t{\'e}}, {Tonry}, {West}, {Ferrarese}, {Jord{\'a}n}, {Peng}, {Anthony}, \& {Merritt}}]{Mei2007}
{Mei}, S., {Blakeslee}, J.~P., {C{\^o}t{\'e}}, P., {et~al.} 2007, \apj, 655, 144, \dodoi{10.1086/509598}

\bibitem[{{Nehlig} {et~al.}(2016){Nehlig}, {Vollmer}, \& {Braine}}]{Nehlig2016}
{Nehlig}, F., {Vollmer}, B., \& {Braine}, J. 2016, \aap, 587, A108, \dodoi{10.1051/0004-6361/201527021}

\bibitem[{{Nelson} {et~al.}(2019){Nelson}, {Springel}, {Pillepich}, {Rodriguez-Gomez}, {Torrey}, {Genel}, {Vogelsberger}, {Pakmor}, {Marinacci}, {Weinberger}, {Kelley}, {Lovell}, {Diemer}, \& {Hernquist}}]{Nelson2019}
{Nelson}, D., {Springel}, V., {Pillepich}, A., {et~al.} 2019, Computational Astrophysics and Cosmology, 6, 2, \dodoi{10.1186/s40668-019-0028-x}

\bibitem[{{Oppenheimer} {et~al.}(2020){Oppenheimer}, {Bogd{\'a}n}, {Crain}, {ZuHone}, {Forman}, {Schaye}, {Wijers}, {Davies}, {Jones}, {Kraft}, \& {Ghirardini}}]{Oppenheimer2020}
{Oppenheimer}, B.~D., {Bogd{\'a}n}, {\'A}., {Crain}, R.~A., {et~al.} 2020, \apjl, 893, L24, \dodoi{10.3847/2041-8213/ab846f}

\bibitem[{{Paggi} {et~al.}(2017){Paggi}, {Kim}, {Anderson}, {Burke}, {D'Abrusco}, {Fabbiano}, {Fruscione}, {Gokas}, {Lauer}, {McCollough}, {Morgan}, {Mossman}, {O'Sullivan}, {Trinchieri}, {Vrtilek}, {Pellegrini}, {Romanowsky}, \& {Brodie}}]{Paggi2017}
{Paggi}, A., {Kim}, D.-W., {Anderson}, C., {et~al.} 2017, \apj, 844, 5, \dodoi{10.3847/1538-4357/aa7897}

\bibitem[{{Pedersen} {et~al.}(2006){Pedersen}, {Rasmussen}, {Sommer-Larsen}, {Toft}, {Benson}, \& {Bower}}]{Pedersen2006}
{Pedersen}, K., {Rasmussen}, J., {Sommer-Larsen}, J., {et~al.} 2006, \na, 11, 465, \dodoi{10.1016/j.newast.2005.11.004}

\bibitem[{{Pillepich} {et~al.}(2018{\natexlab{a}}){Pillepich}, {Springel}, {Nelson}, {Genel}, {Naiman}, {Pakmor}, {Hernquist}, {Torrey}, {Vogelsberger}, {Weinberger}, \& {Marinacci}}]{Pillepich2018}
{Pillepich}, A., {Springel}, V., {Nelson}, D., {et~al.} 2018{\natexlab{a}}, \mnras, 473, 4077, \dodoi{10.1093/mnras/stx2656}

\bibitem[{{Pillepich} {et~al.}(2018{\natexlab{b}}){Pillepich}, {Nelson}, {Hernquist}, {Springel}, {Pakmor}, {Torrey}, {Weinberger}, {Genel}, {Naiman}, {Marinacci}, \& {Vogelsberger}}]{Pillepich2018b}
{Pillepich}, A., {Nelson}, D., {Hernquist}, L., {et~al.} 2018{\natexlab{b}}, \mnras, 475, 648, \dodoi{10.1093/mnras/stx3112}

\bibitem[{{Poggianti} {et~al.}(2017){Poggianti}, {Jaff{\'e}}, {Moretti}, {Gullieuszik}, {Radovich}, {Tonnesen}, {Fritz}, {Bettoni}, {Vulcani}, {Fasano}, {Bellhouse}, {Hau}, \& {Omizzolo}}]{Poggianti2017}
{Poggianti}, B.~M., {Jaff{\'e}}, Y.~L., {Moretti}, A., {et~al.} 2017, \nat, 548, 304, \dodoi{10.1038/nature23462}

\bibitem[{{Predehl} {et~al.}(2020){Predehl}, {Sunyaev}, {Becker}, {Brunner}, {Burenin}, {Bykov}, {Cherepashchuk}, {Chugai}, {Churazov}, {Doroshenko}, {Eismont}, {Freyberg}, {Gilfanov}, {Haberl}, {Khabibullin}, {Krivonos}, {Maitra}, {Medvedev}, {Merloni}, {Nandra}, {Nazarov}, {Pavlinsky}, {Ponti}, {Sanders}, {Sasaki}, {Sazonov}, {Strong}, \& {Wilms}}]{Predehl2020}
{Predehl}, P., {Sunyaev}, R.~A., {Becker}, W., {et~al.} 2020, \nat, 588, 227, \dodoi{10.1038/s41586-020-2979-0}

\bibitem[{{Randall} {et~al.}(2008){Randall}, {Nulsen}, {Forman}, {Jones}, {Machacek}, {Murray}, \& {Maughan}}]{Randall2008}
{Randall}, S., {Nulsen}, P., {Forman}, W.~R., {et~al.} 2008, \apj, 688, 208, \dodoi{10.1086/592324}

\bibitem[{{Randall} {et~al.}(2004){Randall}, {Sarazin}, \& {Irwin}}]{Randall2004}
{Randall}, S.~W., {Sarazin}, C.~L., \& {Irwin}, J.~A. 2004, \apj, 600, 729, \dodoi{10.1086/380109}

\bibitem[{{Rasmussen} {et~al.}(2009){Rasmussen}, {Sommer-Larsen}, {Pedersen}, {Toft}, {Benson}, {Bower}, \& {Grove}}]{Rasmussen2009}
{Rasmussen}, J., {Sommer-Larsen}, J., {Pedersen}, K., {et~al.} 2009, \apj, 697, 79, \dodoi{10.1088/0004-637X/697/1/79}

\bibitem[{{Rodriguez-Gomez} {et~al.}(2015){Rodriguez-Gomez}, {Genel}, {Vogelsberger}, {Sijacki}, {Pillepich}, {Sales}, {Torrey}, {Snyder}, {Nelson}, {Springel}, {Ma}, \& {Hernquist}}]{Rodriguez-Gomez2015}
{Rodriguez-Gomez}, V., {Genel}, S., {Vogelsberger}, M., {et~al.} 2015, \mnras, 449, 49, \dodoi{10.1093/mnras/stv264}

\bibitem[{{Roediger} {et~al.}(2015){Roediger}, {Kraft}, {Nulsen}, {Forman}, {Machacek}, {Randall}, {Jones}, {Churazov}, \& {Kokotanekova}}]{Roediger2015}
{Roediger}, E., {Kraft}, R.~P., {Nulsen}, P.~E.~J., {et~al.} 2015, \apj, 806, 103, \dodoi{10.1088/0004-637X/806/1/103}

\bibitem[{{Saintonge} {et~al.}(2016){Saintonge}, {Catinella}, {Cortese}, {Genzel}, {Giovanelli}, {Haynes}, {Janowiecki}, {Kramer}, {Lutz}, {Schiminovich}, {Tacconi}, {Wuyts}, \& {Accurso}}]{Saintonge2016}
{Saintonge}, A., {Catinella}, B., {Cortese}, L., {et~al.} 2016, \mnras, 462, 1749, \dodoi{10.1093/mnras/stw1715}

\bibitem[{{Schaye} {et~al.}(2015){Schaye}, {Crain}, {Bower}, {Furlong}, {Schaller}, {Theuns}, {Dalla Vecchia}, {Frenk}, {McCarthy}, {Helly}, {Jenkins}, {Rosas-Guevara}, {White}, {Baes}, {Booth}, {Camps}, {Navarro}, {Qu}, {Rahmati}, {Sawala}, {Thomas}, \& {Trayford}}]{Schaye2015}
{Schaye}, J., {Crain}, R.~A., {Bower}, R.~G., {et~al.} 2015, \mnras, 446, 521, \dodoi{10.1093/mnras/stu2058}

\bibitem[{{Smith} {et~al.}(2001){Smith}, {Brickhouse}, {Liedahl}, \& {Raymond}}]{Smith2001}
{Smith}, R.~K., {Brickhouse}, N.~S., {Liedahl}, D.~A., \& {Raymond}, J.~C. 2001, \apjl, 556, L91, \dodoi{10.1086/322992}

\bibitem[{{Soria} {et~al.}(2022){Soria}, {Kolehmainen}, {Graham}, {Swartz}, {Yukita}, {Motch}, {Jarrett}, {Miller-Jones}, {Plotkin}, {Maccarone}, {Ferrarese}, {Guest}, \& {Lan{\c{c}}on}}]{Soria2022}
{Soria}, R., {Kolehmainen}, M., {Graham}, A.~W., {et~al.} 2022, \mnras, 512, 3284, \dodoi{10.1093/mnras/stac148}

\bibitem[{{Spitzer}(1956)}]{Spitzer1956}
{Spitzer}, Lyman, J. 1956, \apj, 124, 20, \dodoi{10.1086/146200}

\bibitem[{{Springel} {et~al.}(2018){Springel}, {Pakmor}, {Pillepich}, {Weinberger}, {Nelson}, {Hernquist}, {Vogelsberger}, {Genel}, {Torrey}, {Marinacci}, \& {Naiman}}]{Springel2018}
{Springel}, V., {Pakmor}, R., {Pillepich}, A., {et~al.} 2018, \mnras, 475, 676, \dodoi{10.1093/mnras/stx3304}

\bibitem[{{Strickland} {et~al.}(2004){Strickland}, {Heckman}, {Colbert}, {Hoopes}, \& {Weaver}}]{Strickland2004}
{Strickland}, D.~K., {Heckman}, T.~M., {Colbert}, E. J.~M., {Hoopes}, C.~G., \& {Weaver}, K.~A. 2004, \apjs, 151, 193, \dodoi{10.1086/382214}

\bibitem[{{Strickland} \& {Stevens}(2000)}]{Strickland2000}
{Strickland}, D.~K., \& {Stevens}, I.~R. 2000, \mnras, 314, 511, \dodoi{10.1046/j.1365-8711.2000.03391.x}

\bibitem[{{Su} {et~al.}(2019){Su}, {Kraft}, {Nulsen}, {Jones}, {Maccarone}, {Mernier}, {Lovisari}, {Sheardown}, {Randall}, {Roediger}, {Fish}, {Forman}, \& {Churazov}}]{Su2019}
{Su}, Y., {Kraft}, R.~P., {Nulsen}, P.~E.~J., {et~al.} 2019, \aj, 158, 6, \dodoi{10.3847/1538-3881/ab1d51}

\bibitem[{{Tashiro}(2022)}]{Tashiro2022}
{Tashiro}, M.~S. 2022, International Journal of Modern Physics D, 31, 2230001, \dodoi{10.1142/S0218271822300014}

\bibitem[{{Toft} {et~al.}(2002){Toft}, {Rasmussen}, {Sommer-Larsen}, \& {Pedersen}}]{Toft2002}
{Toft}, S., {Rasmussen}, J., {Sommer-Larsen}, J., \& {Pedersen}, K. 2002, \mnras, 335, 799, \dodoi{10.1046/j.1365-8711.2002.05663.x}

\bibitem[{{Truong} {et~al.}(2021){Truong}, {Pillepich}, {Nelson}, {Werner}, \& {Hernquist}}]{Truong2021}
{Truong}, N., {Pillepich}, A., {Nelson}, D., {Werner}, N., \& {Hernquist}, L. 2021, \mnras, 508, 1563, \dodoi{10.1093/mnras/stab2638}

\bibitem[{{Tsch{\"o}ke} {et~al.}(2001){Tsch{\"o}ke}, {Bomans}, {Hensler}, \& {Junkes}}]{Tschoke2001}
{Tsch{\"o}ke}, D., {Bomans}, D.~J., {Hensler}, G., \& {Junkes}, N. 2001, \aap, 380, 40, \dodoi{10.1051/0004-6361:20011354}

\bibitem[{{T{\"u}llmann} {et~al.}(2006){T{\"u}llmann}, {Pietsch}, {Rossa}, {Breitschwerdt}, \& {Dettmar}}]{Tullmann2006}
{T{\"u}llmann}, R., {Pietsch}, W., {Rossa}, J., {Breitschwerdt}, D., \& {Dettmar}, R.~J. 2006, \aap, 448, 43, \dodoi{10.1051/0004-6361:20052936}

\bibitem[{{Vijayaraghavan} \& {Ricker}(2015)}]{Vijayaraghavan2015}
{Vijayaraghavan}, R., \& {Ricker}, P.~M. 2015, \mnras, 449, 2312, \dodoi{10.1093/mnras/stv476}

\bibitem[{{Vollmer} {et~al.}(2012){Vollmer}, {Wong}, {Braine}, {Chung}, \& {Kenney}}]{Vollmer2012}
{Vollmer}, B., {Wong}, O.~I., {Braine}, J., {Chung}, A., \& {Kenney}, J.~D.~P. 2012, \aap, 543, A33, \dodoi{10.1051/0004-6361/201118690}

\bibitem[{{Vulcani} {et~al.}(2020){Vulcani}, {Poggianti}, {Tonnesen}, {McGee}, {Moretti}, {Fritz}, {Gullieuszik}, {Jaff{\'e}}, {Franchetto}, {Tomi{\v{c}}i{\'c}}, {Mingozzi}, {Bettoni}, \& {Wolter}}]{Vulcani2020}
{Vulcani}, B., {Poggianti}, B.~M., {Tonnesen}, S., {et~al.} 2020, \apj, 899, 98, \dodoi{10.3847/1538-4357/aba4ae}

\bibitem[{{Wang} {et~al.}(2016){Wang}, {Li}, {Jiang}, \& {Fang}}]{Wang2016}
{Wang}, Q.~D., {Li}, J., {Jiang}, X., \& {Fang}, T. 2016, \mnras, 457, 1385, \dodoi{10.1093/mnras/stv2886}

\bibitem[{{We{\.z}gowiec} {et~al.}(2012){We{\.z}gowiec}, {Bomans}, {Ehle}, {Chy{\.z}y}, {Urbanik}, {Braine}, \& {Soida}}]{Wezgowie2012}
{We{\.z}gowiec}, M., {Bomans}, D.~J., {Ehle}, M., {et~al.} 2012, \aap, 544, A99, \dodoi{10.1051/0004-6361/201117652}

\bibitem[{{We{\.z}gowiec} {et~al.}(2011){We{\.z}gowiec}, {Vollmer}, {Ehle}, {Dettmar}, {Bomans}, {Chy{\.z}y}, {Urbanik}, \& {Soida}}]{Wezgowie2011}
{We{\.z}gowiec}, M., {Vollmer}, B., {Ehle}, M., {et~al.} 2011, \aap, 531, A44, \dodoi{10.1051/0004-6361/201016344}

\bibitem[{{White} \& {Frenk}(1991)}]{White1991}
{White}, S. D.~M., \& {Frenk}, C.~S. 1991, \apj, 379, 52, \dodoi{10.1086/170483}

\bibitem[{{White} \& {Rees}(1978)}]{White1978}
{White}, S.~D.~M., \& {Rees}, M.~J. 1978, \mnras, 183, 341, \dodoi{10.1093/mnras/183.3.341}

\bibitem[{{Wood} {et~al.}(2017){Wood}, {Jones}, {Machacek}, {Forman}, {Bogdan}, {Andrade-Santos}, {Kraft}, {Paggi}, \& {Roediger}}]{Wood2017}
{Wood}, R.~A., {Jones}, C., {Machacek}, M.~E., {et~al.} 2017, \apj, 847, 79, \dodoi{10.3847/1538-4357/aa8723}

\bibitem[{{Wright} {et~al.}(2010){Wright}, {Eisenhardt}, {Mainzer}, {Ressler}, {Cutri}, {Jarrett}, {Kirkpatrick}, {Padgett}, {McMillan}, {Skrutskie}, {Stanford}, {Cohen}, {Walker}, {Mather}, {Leisawitz}, {Gautier}, {McLean}, {Benford}, {Lonsdale}, {Blain}, {Mendez}, {Irace}, {Duval}, {Liu}, {Royer}, {Heinrichsen}, {Howard}, {Shannon}, {Kendall}, {Walsh}, {Larsen}, {Cardon}, {Schick}, {Schwalm}, {Abid}, {Fabinsky}, {Naes}, \& {Tsai}}]{Wright2010}
{Wright}, E.~L., {Eisenhardt}, P. R.~M., {Mainzer}, A.~K., {et~al.} 2010, \aj, 140, 1868, \dodoi{10.1088/0004-6256/140/6/1868}

\bibitem[{{Yi} {et~al.}(2021){Yi}, {Wang}, {Shu}, {Fabbiano}, {Pappalardo}, {Wang}, \& {Yu}}]{Yi2021}
{Yi}, H., {Wang}, J., {Shu}, X., {et~al.} 2021, \apj, 908, 156, \dodoi{10.3847/1538-4357/abcec3}

\bibitem[{{Zhu} {et~al.}(2023){Zhu}, {Tonnesen}, \& {Bryan}}]{Zhu2023}
{Zhu}, J., {Tonnesen}, S., \& {Bryan}, G.~L. 2023, arXiv e-prints, arXiv:2309.07037.
\newblock \doarXiv{2309.07037}

\end{thebibliography}
\bibliographystyle{aasjournal}


\end{document}